\documentclass[pdflatex,sn-nature]{sn-jnl}

\usepackage{graphicx}%
\usepackage{multirow}%
\usepackage{amsmath,amssymb,amsfonts}%
\usepackage{amsthm}%
\usepackage{mathrsfs}%
\usepackage[title]{appendix}%
\usepackage{xcolor}%
\usepackage{array}%
\usepackage{textcomp}%
\usepackage{manyfoot}%
\usepackage{booktabs}%
\usepackage{algorithm}%
\usepackage{algorithmicx}%
\usepackage{algpseudocode}%
\usepackage{listings}%
\usepackage{bbm}
\usepackage{lineno}

\usepackage{xcolor}

\begin{document}

\title[MedFormer]{MedFormer: a data-driven model for forecasting the Mediterranean Sea}

\author*[1,3]{\fnm{Italo} \sur{Epicoco}}\email{italo.epicoco@cmcc.it}
\author[1,3]{\fnm{Davide} \sur{Donno}}\email{davide.donno@cmcc.it}
\author[4,5,1]{\fnm{Gabriele} \sur{Accarino}}\email{gabriele.accarino@cmcc.it}
\author[1]{\fnm{Simone} \sur{Norberti}}\email{simone.norberti@cmcc.it}
\author[2]{\fnm{Alessandro} \sur{Grandi}}\email{alessandro.grandi@cmcc.it}
\author[2]{\fnm{Michele} \sur{Giurato}}\email{michele.giurato@cmcc.it}
\author[2]{\fnm{Ronan} \sur{McAdam}}\email{ronan.mcadam@cmcc.it}
\author[1]{\fnm{Donatello} \sur{Elia}}\email{donatello.elia@cmcc.it}
\author[2]{\fnm{Emanuela} \sur{Clementi}}\email{emanuela.clementi@cmcc.it}
\author[1]{\fnm{Paola} \sur{Nassisi}}\email{paola.nassisi@cmcc.it}
\author[2]{\fnm{Enrico} \sur{Scoccimarro}}\email{enrico.scoccimarro@cmcc.it}
\author[1]{\fnm{Giovanni} \sur{Coppini}}\email{giovanni.coppini@cmcc.it}
\author[2]{\fnm{Silvio} \sur{Gualdi}}\email{silvio.gualdi@cmcc.it}
\author[1,3]{\fnm{Giovanni} \sur{Aloisio}}\email{giovanni.aloisio@cmcc.it}
\author[2]{\fnm{Simona} \sur{Masina}}\email{simona.masina@cmcc.it}
\author[1]{\fnm{Giulio} \sur{Boccaletti}}\email{giulio.boccaletti@cmcc.it}
\author[2]{\fnm{Antonio} \sur{Navarra}}\email{antonio.navarra@cmcc.it}

\affil[1]{\orgname{CMCC Foundation - Euro-Mediterranean Center on Climate Change}, \orgaddress{\street{via Marco Biagi, 5}, \city{Lecce}, \postcode{73100}, \country{Italy}}}
\affil[2]{\orgname{CMCC Foundation - Euro-Mediterranean Center on Climate Change}, \orgaddress{\street{via C. Berti Pichat, 6/2}, \city{Bologna}, \postcode{40127}, \country{Italy}}}
\affil[3]{\orgdiv{Department Engineering for Innovation}, \orgname{University of Salento}, \orgaddress{\street{via per Monteroni}, \city{Lecce}, \postcode{73100}, \country{Italy}}}
\affil[4]{\orgdiv{Department of Earth and Environmental Engineering}, \orgname{Columbia University}, \orgaddress{\street{500W 120th St.}, \city{New York}, \postcode{10027}, \state{New York}, \country{U.S.A}}}
\affil[5]{\orgdiv{Learning the Earth With Artificial Intelligence and Physics (LEAP) NSF STC}, \orgname{Columbia University}, \orgaddress{\street{2276 12th Ave, 2nd Floor}, \city{New York}, \postcode{10027}, \state{New York}, \country{U.S.A}}}

\abstract{
Accurate ocean forecasting is essential for supporting a wide range of marine applications. Recent advances in artificial intelligence have highlighted the potential of data-driven models to outperform traditional numerical approaches, particularly in atmospheric weather forecasting. However, extending these methods to ocean systems remains challenging due to their inherently slower dynamics and complex boundary conditions.

In this work, we present MedFormer, a fully data-driven deep learning model specifically designed for medium-range ocean forecasting in the Mediterranean Sea. MedFormer is based on a U-Net architecture augmented with 3D attention mechanisms and operates at a high horizontal resolution of 1/24°. The model is trained on 20 years of daily ocean reanalysis data and fine-tuned with high-resolution operational analyses. It generates 9-day forecasts using an autoregressive strategy. The model leverages both historical ocean states and atmospheric forcings, making it well-suited for operational use.

We benchmark MedFormer against the state-of-the-art Mediterranean Forecasting System (MedFS), developed at Euro-Mediterranean Center on Climate Change (CMCC), using both analysis data and independent observations. The forecast skills, evaluated with the Root Mean Squared Difference and the Anomaly Correlation Coefficient, indicate that MedFormer consistently outperforms MedFS across key 3D ocean variables. These findings underscore the potential of data-driven approaches like MedFormer to complement, or even surpass, traditional numerical ocean forecasting systems in both accuracy and computational efficiency.
}


\maketitle

\section{Introduction}\label{introduction}

Ocean forecasting plays a key role in supporting a wide range of marine activities. Conventional ocean forecasting systems are predominantly based on physics-based numerical models that draw upon the principles of fluid mechanics and thermodynamics. These models simulate future ocean states by solving complex partial differential equations, requiring high-performance computing facilities with hundreds of computational nodes. This process is both time-consuming and energy intensive. Moreover, improving the accuracy of such forecasts remains a significant challenge, as it depends on our ability to accurately represent and encode of the underlying physical processes governing ocean dynamics.

Recent advances in Artificial Intelligence (AI) have introduced data-driven methods that offer a promising alternative to traditional numerical approaches. In particular, deep learning models have demonstrated remarkable success in atmospheric weather forecasting, achieving accuracy comparable to, or even surpassing, that of physics-based models, while operating thousands of times faster. Unlike traditional models, AI-based approaches do not require explicit knowledge of physical processes; instead, they learn complex spatio-temporal relationships directly from large datasets. However, extending these capabilities to ocean forecasting presents specific challenges. The ocean system evolves more slowly than the atmospheric system and it is influenced by a range of complex factors, including bottom topography, lateral continental boundaries, air-sea interactions and intricate water mass transformations. These complexities make it more difficult for data-driven models to fully capture the underlying ocean dynamics. Despite these challenges, several AI-based models for ocean forecasting have recently emerged, such as XiHe~\cite{wang2024xihedatadrivenmodelglobal}, GLONET~\cite{aouni2025glonetmercatorsendtoendneural}, Samudra~\cite{10.1029/2024GL114318}, AI-GOMS~\cite{xiong2023aigomslargeaidrivenglobal}, and ORCA-DL~\cite{guo2024datadrivenglobaloceanmodeling}. However, direct comparisons between AI models and state-of-the-art high-resolution regional dynamical forecasting systems, in an operational setting, are still in their early stages. 

In this work, we introduce MedFormer, an efficient and accurate fully data-driven model for medium-range ocean forecasting. MedFormer is tailored for the Mediterranean Sea, operating at a high horizontal resolution of $1/24^\circ$, and it is among the first ocean forecasting models capable of accurately capturing fine-scale ocean processes at high resolution. Specifically designed and trained for operational deployment, MedFormer produces daily forecasts up to 9 days and is compatible with the existing operational pipeline of the Mediterranean Forecasting System (MedFS)~\cite{os-19-1483-2023}. MedFormer has been pre-trained on daily mean reanalysis from the Euro-Mediterranean Center on Climate Change (CMCC) Mediterranean Physical Reanalysis MEDREA24~\cite{10.3389/feart.2021.702285} dataset available through the Copernicus Marine Service (CMEMS), which employs the $1/4^\circ$ European Centre for Medium-Range Weather Forecasts (ECMWF) ERA5~\cite{10.1002/qj.3803} atmospheric reanalysis available through the Copernicus Climate Change Service (C3S) as forcing, covering the period from 2002 to 2021. Fine-tuning is subsequently performed using CMCC high-resolution Mediterranean analyses at $1/24^\circ$ (MedFS version EAS6~\cite{MedAna}), forced by ECMWF operational atmospheric analysis at $1/10^\circ$ resolution (high resolution product HRES), using 4 years of data spanning from 2018 to 2021.

The loss function used to train MedFormer minimizes the Mean Absolute Error (MAE), incorporating variable- and level-specific weights to prioritize accuracy in the upper ocean layers. MedFormer's skill is evaluated against the CMCC MedFS, using both analysis data and independent observational datasets, including in-situ vertical profiles and satellite data as ground truth. The results consistently show that MedFormer achieves lower forecast errors, demonstrating a clear improvement in forecast skill compared to the dynamical operational system MedFS. 

The remainder of this manuscript is organized as follows: Section~\ref{model} describes the MedFormer architecture; Section~\ref{results} provides results and discussion; we draw our conclusion in Section~\ref{conclusion}. Additional methodological details are provided in Section~\ref{methods}.

\section{Model}\label{model}
\begin{figure}[ht]
    \centering
    \includegraphics[width=0.9\linewidth]{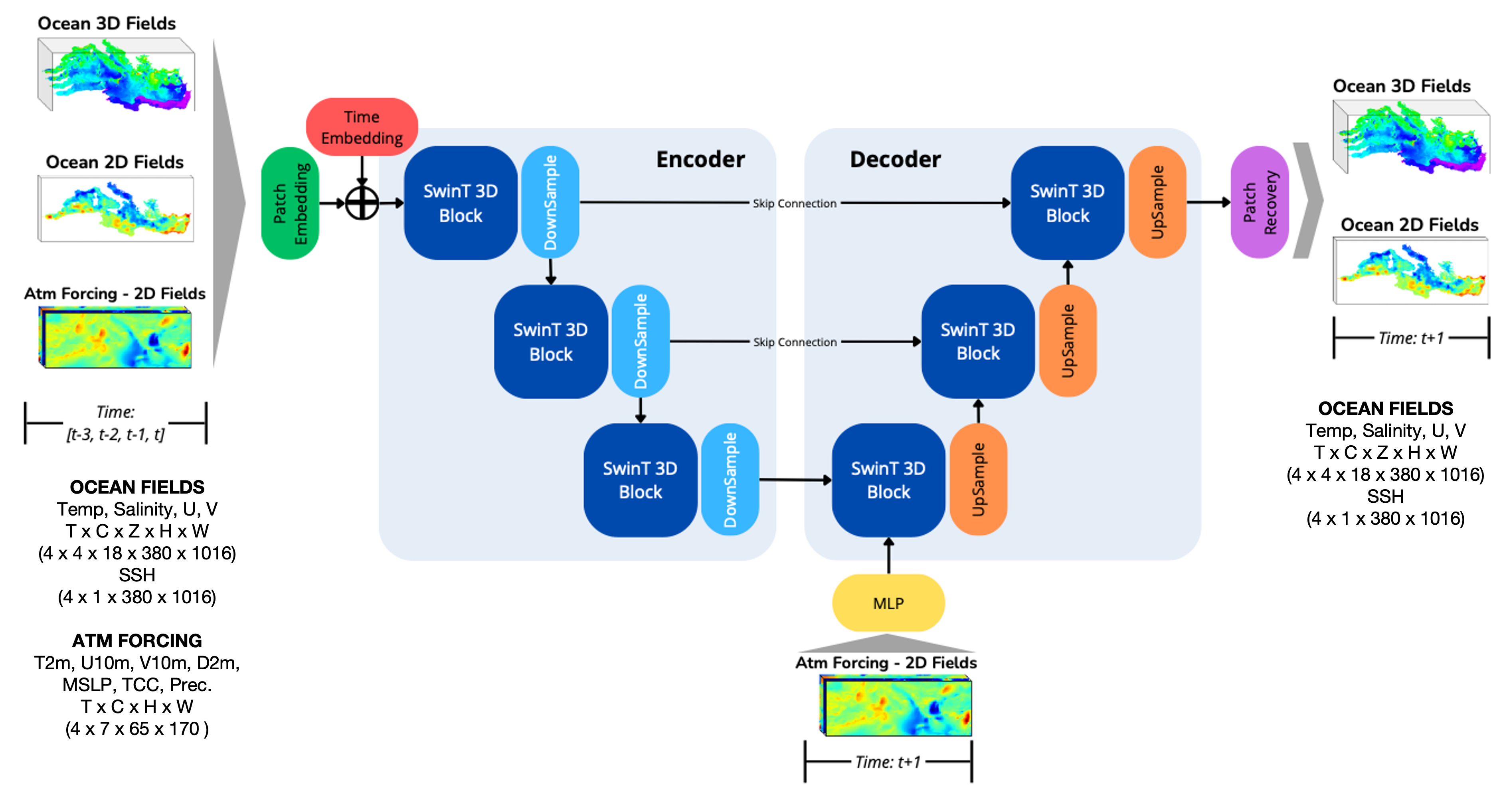}
    \caption{MedFormer architecture relies on a U-Net hierarchical architecture where the encoder and decoder are made of three layer blocks. Each block is a composition of Swin 3D blocks with self attention mechanism. A down-sample layer (in the encoder) or an up-sample layer (in the decoder) are used between Swin 3D blocks}
    \label{fig:medformer_arch}
\end{figure}

The MedFormer architecture is based on a U-Net structure, with both encoder and decoder composed of three layer blocks, as shown in Fig.~\ref{fig:medformer_arch}. Each block contains a sequence of Shifted Windows Attention (Swin) 3D modules, where each module comprises a multi-head attention layer, a normalization layer, a multi layer perceptron (MLP), and an additional normalization layer. The latent space encodes the 3D ocean space using two spatial dimensions, while the third dimension captures the temporal sequence of ocean states. The input consists of a sequence of four consecutive daily mean values ($X_{t-3}, X_{t-2}, X_{t-1}, X_{t}$) of key 3D ocean variables (temperature, salinity, zonal and meridional velocity) and a 2D variable (sea surface height) at a horizontal resolution of $1/24^\circ$. The 3D variables extend to a depth of 971 meters discretized across 18 vertical levels (see Section~\ref{methods} for further details). 
Atmospheric surface variables (2m temperature, 10m zonal and meridional wind velocity, mean sea level pressure, dew point at 2m, 24-hour accumulated precipitation and total cloud cover), along with the day of the year, are used as conditioning inputs. The model forecasts the same set of ocean variables in the future time step ${t+1}$. The forecast is then obtained from the decoder which starts combining the latent representation of the ocean with the atmospheric surface conditions at time ${t+1}$.

MedFormer is designed for operational deployment, producing 1-day daily forecasts using the information of the previous four days as the input sequence. Longer lead-time forecasts are produced through an autoregressive approach, where MedFormer iteratively uses its own predictions as inputs for subsequent time steps. The training strategy is structured in two phases: (i) pre-training phase, in which the model is trained on the MEDREA24 product, which contains daily mean data from 1987 to 2023 and available through the CMEMS, along with ERA5 daily mean surface meteorological data at $1/4^\circ$ resolution. Although the dataset spans 36 years, MedFormer is pre-trained using data from 2000 to 2019, with 2020 and 2021 reserved for validation; (ii) fine-tuning phase, in which model accuracy is  improved by incorporating autoregression directly into the training process. The loss function is defined as the average error accumulated across multiple autoregressive forecast steps. A curriculum learning protocol is used to gradually increase the prediction lead time from 1 to 5 days (see Section~\ref{training} for details). During this phase, training is performed using higher quality datasets: the very high resolution Mediterranean analysis, from the MedFS system~\cite{MedAna}, and surface meteorological forcing from the ECMWF HRES (high resolution) operational analysis at $1/10^\circ$. The years 2018-2020 are used for training, 2021 for validation, and 2022 is reserved for testing the final trained model, for which independent observational data are also available for comparison.

The main characteristics and differences between MedFS and the MEDREA24 systems are provided in Table~\ref{tab:REA_MedFS}.

\begin{table}[ht]
    \centering
    \begin{tabular}{|>{\centering\arraybackslash}p{0.3\linewidth}|>{\centering\arraybackslash}p{0.3\linewidth}|>{\centering\arraybackslash}p{0.3\linewidth}|} \hline  
           & MEDREA24&  MedFS EAS6\\ \hline 
          Grid & 1/24, 141 vertical lev. & 1/24, 141 vertical lev.\\ \hline  
          Modeling System & NEMO v3.6 &  NEMO V3.6 2-way coupled with WW3 \\ \hline 
          Tides & No &  Yes, 8 components\\ \hline 
          Atmospheric Forcing & ECMWF ERA5 (1/4 deg., 1 hour) & ECMWF HRES (1/10 deg., 1-3-6- hours)\\ \hline 
          Rivers & 39. All runoffs from climatology& 39. 38 runoffs from climatolgy except Po river from observations\\ \hline 
          Lateral Open Boundary & Atlantic: nested into the CMCC CGLORS-v5 Reanalysis~\cite{essd-8-679-2016}. Dardanelles Strait: closed & Atlantic: nested into the CMEMS Global Analysis and Forecast system + Dardanelles Strait with daily climatology\\ \hline 
          Assimilated data & CMEMS Reprocessed Sea Level Anomaly, and in-situ temperature and salinity & CMEMS Near Real Time Sea Level Anomaly, and in-situ temperature and salinity\\ \hline
    \end{tabular}
\caption{Characteristics and differences between the Mediterranean Reanalysis and MedFS}
\label{tab:REA_MedFS}
\end{table}

\section{Results and Discussion}\label{results}

\begin{figure}[ht]
    \centering
    \includegraphics[width=0.45\linewidth]{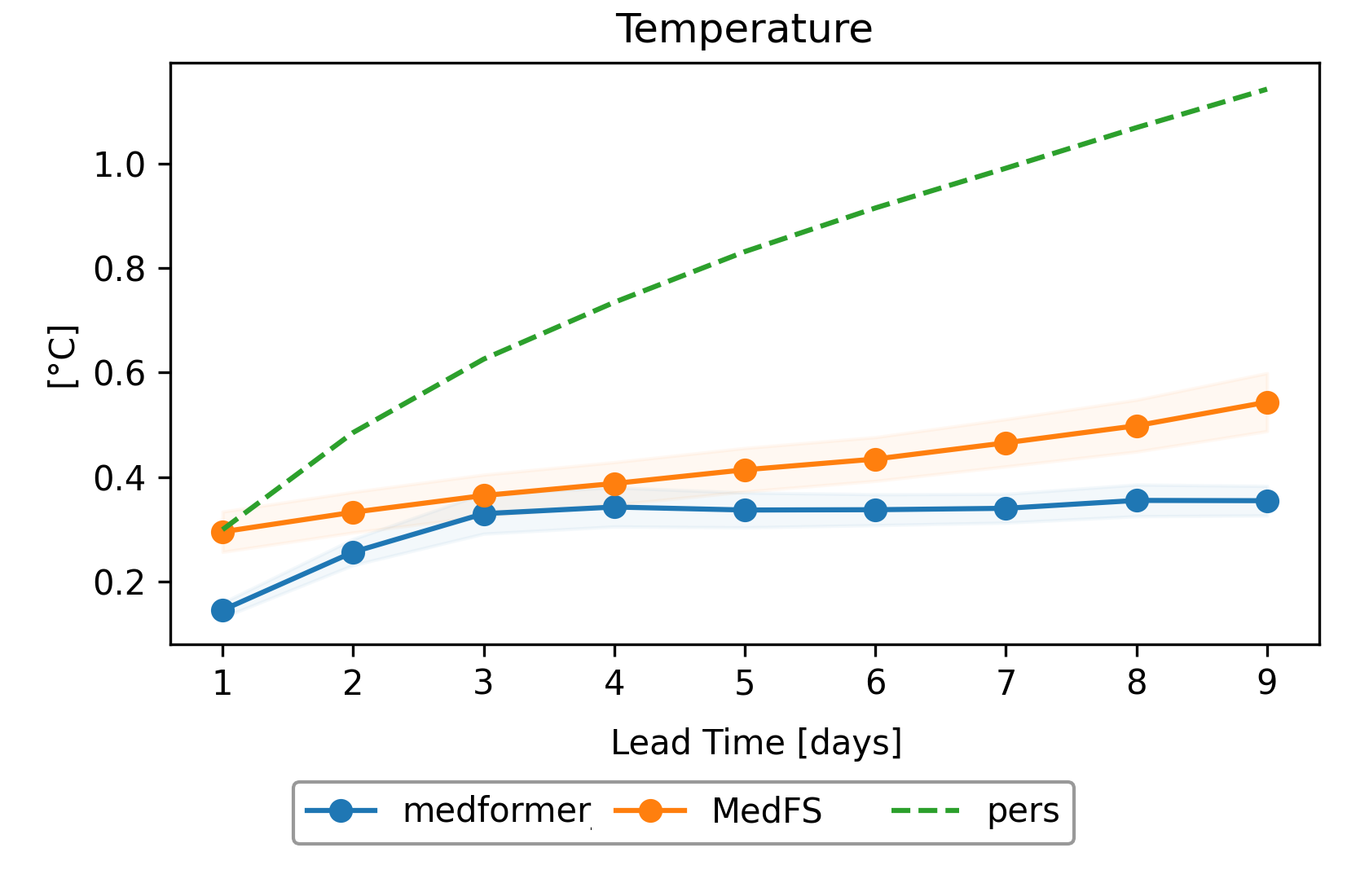}
    \includegraphics[width=0.45\linewidth]{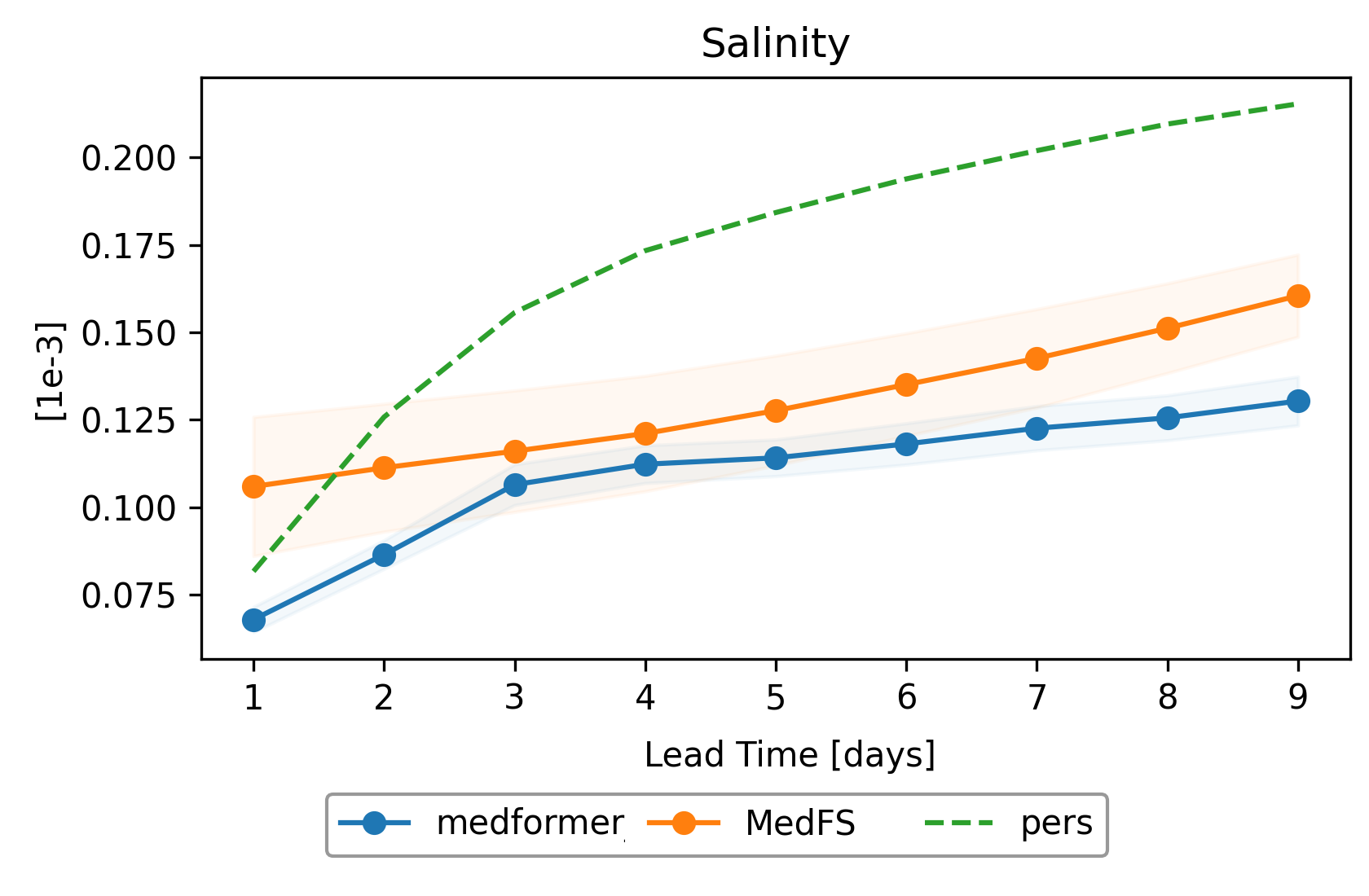}
    \includegraphics[width=0.45\linewidth]{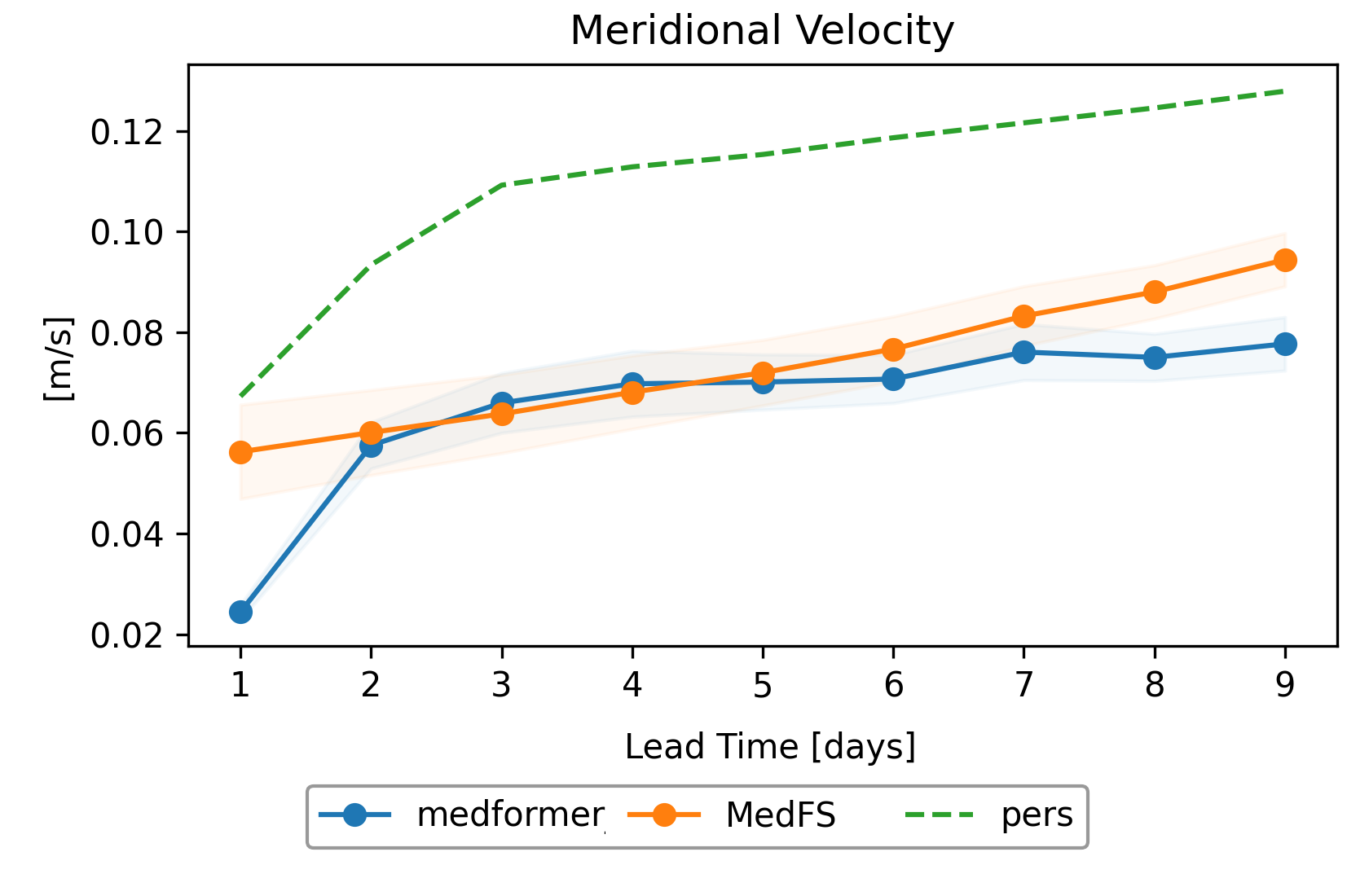}
    \includegraphics[width=0.45\linewidth]{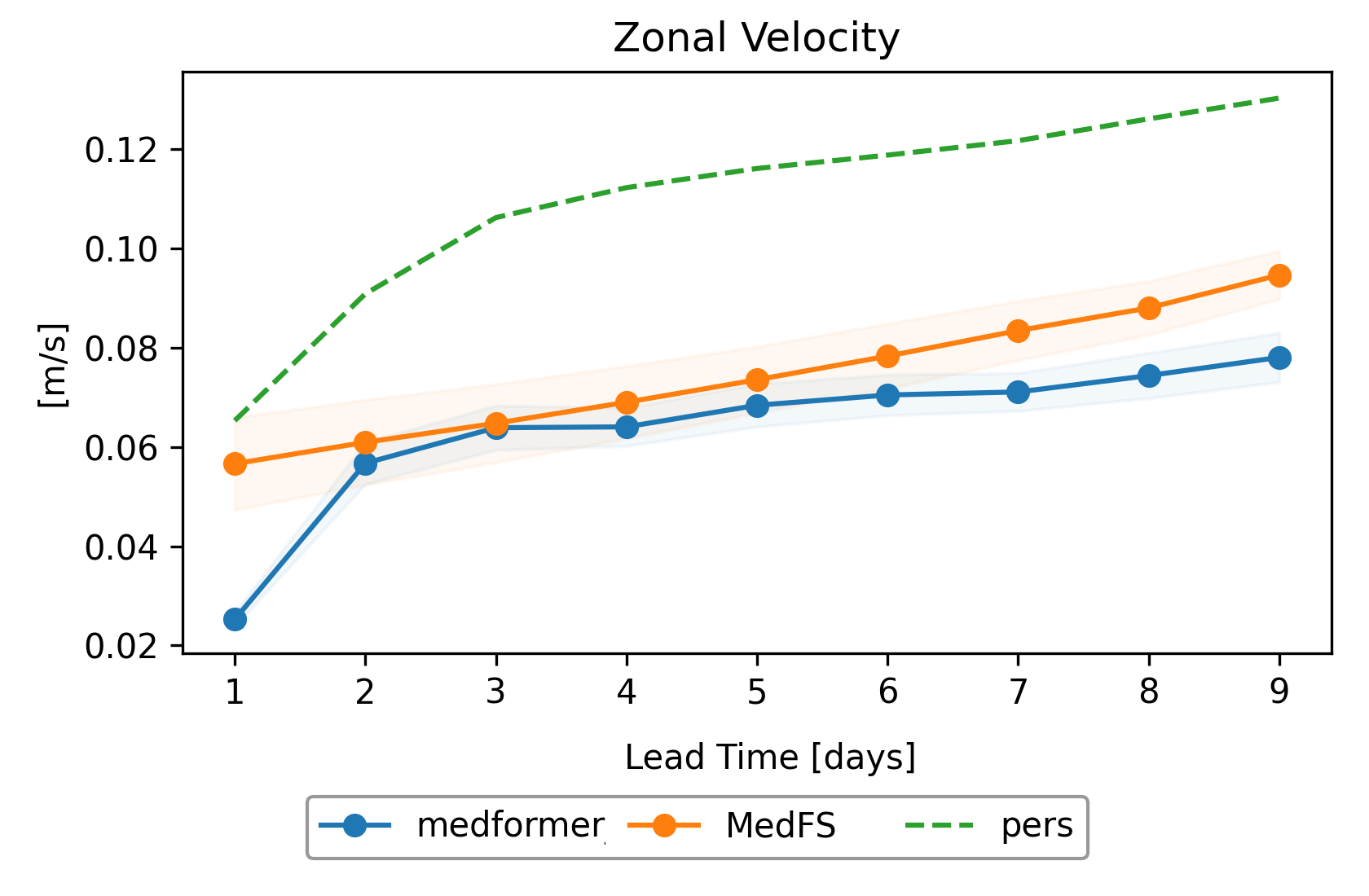}
    \includegraphics[width=0.45\linewidth]{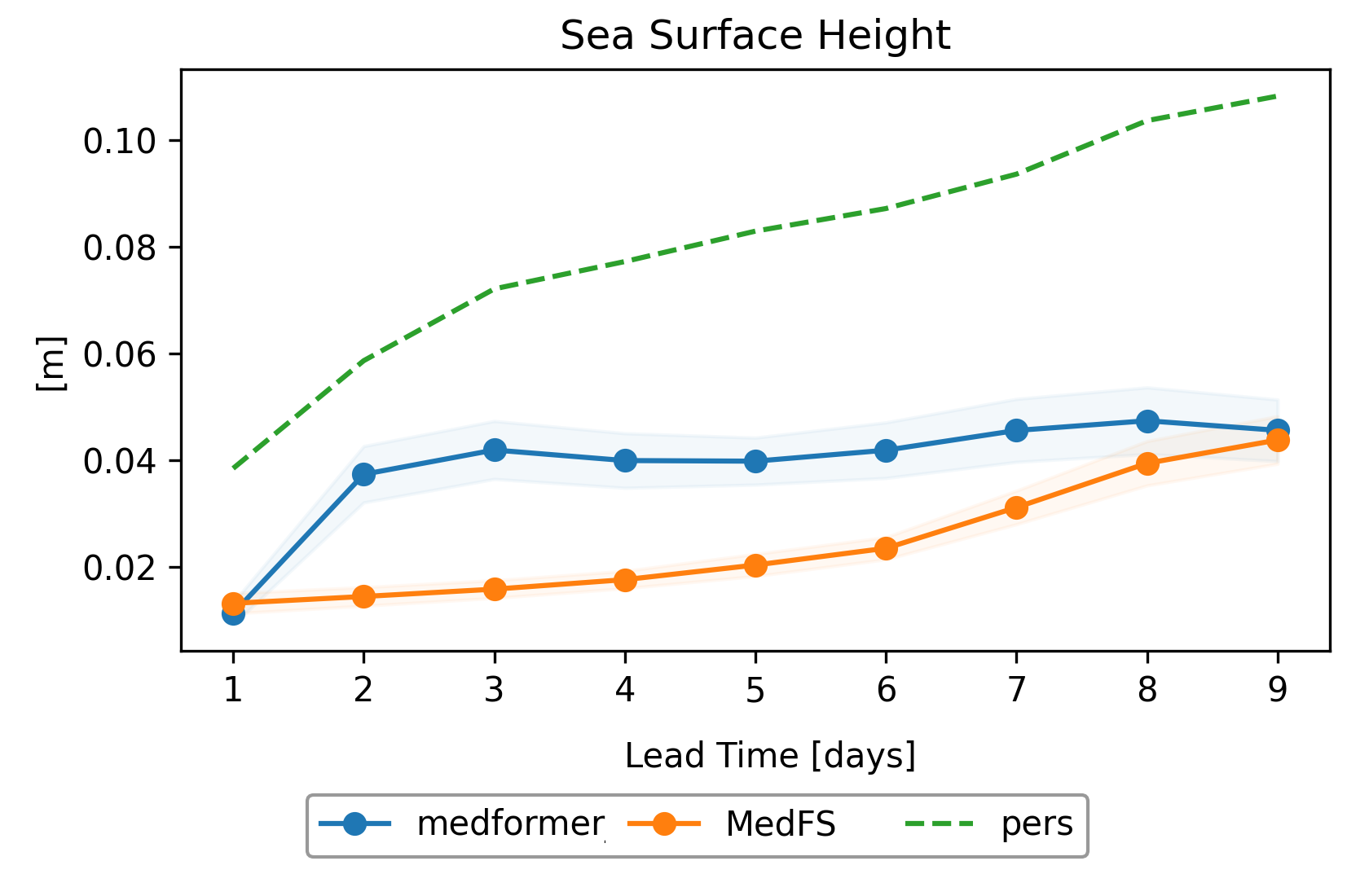}
    
    \caption{MedFormer and MedFS RMSD computed against the Mediterranean Sea Analysis for temperature, salinity, meridional and zonal velocities at the surface and SSH averaged across the 2022 year in the whole Mediterranean basin for all 9 forecast lead times. MedFormer errors are shown in blue, MedFS errors in orange, and the persistence is represented by the dashed green line}
    \label{fig:rmsd-surf}
\end{figure}

In this Section, we evaluate the performances of MedFormer relative to MedFS, using both analysis fields and observational data as reference. The comparison is carried out with forecasts initialized on Tuesdays, aligning with the operational configuration of MedFS, which is initialized from an analysis only once a week, specifically on Tuesdays.
Model skill is assessed using two standard metrics: Root Mean Squared Difference (RMSD) and Anomaly Correlation Coefficient (ACC). 
Figure~\ref{fig:rmsd-surf} depicts the forecast skill of MedFormer in comparison to MedFS. The RMSD values for SSH, temperature, salinity and horizontal velocity components are averaged over the 2022 year, with each forecast cycle providing 9-day forecasts. Both MedFormer and MedFS use the Mediterranean Sea Analysis and the ECMWF HRES atmospheric forecasts as initial conditions. Overall, MedFormer demonstrates a good forecast skill across all 3D ocean variables. The only exception is SSH, for which MedFS yields slightly lower RMSD. Notably, MedFormer shows significant improvements during the early forecast days and exhibits a lower error growth rate over time, even though forecasts are produced in an autoregressive setting. This is evidenced by the shallower slope of the blue line (MedFormer) relative to the orange line (MedFS) in the figure. Both models clearly outperforms the persistence (green dashed line), particularly as the forecast lead time increases. 

\begin{figure}[ht]
    \centering
        \includegraphics[width=0.48\textwidth]{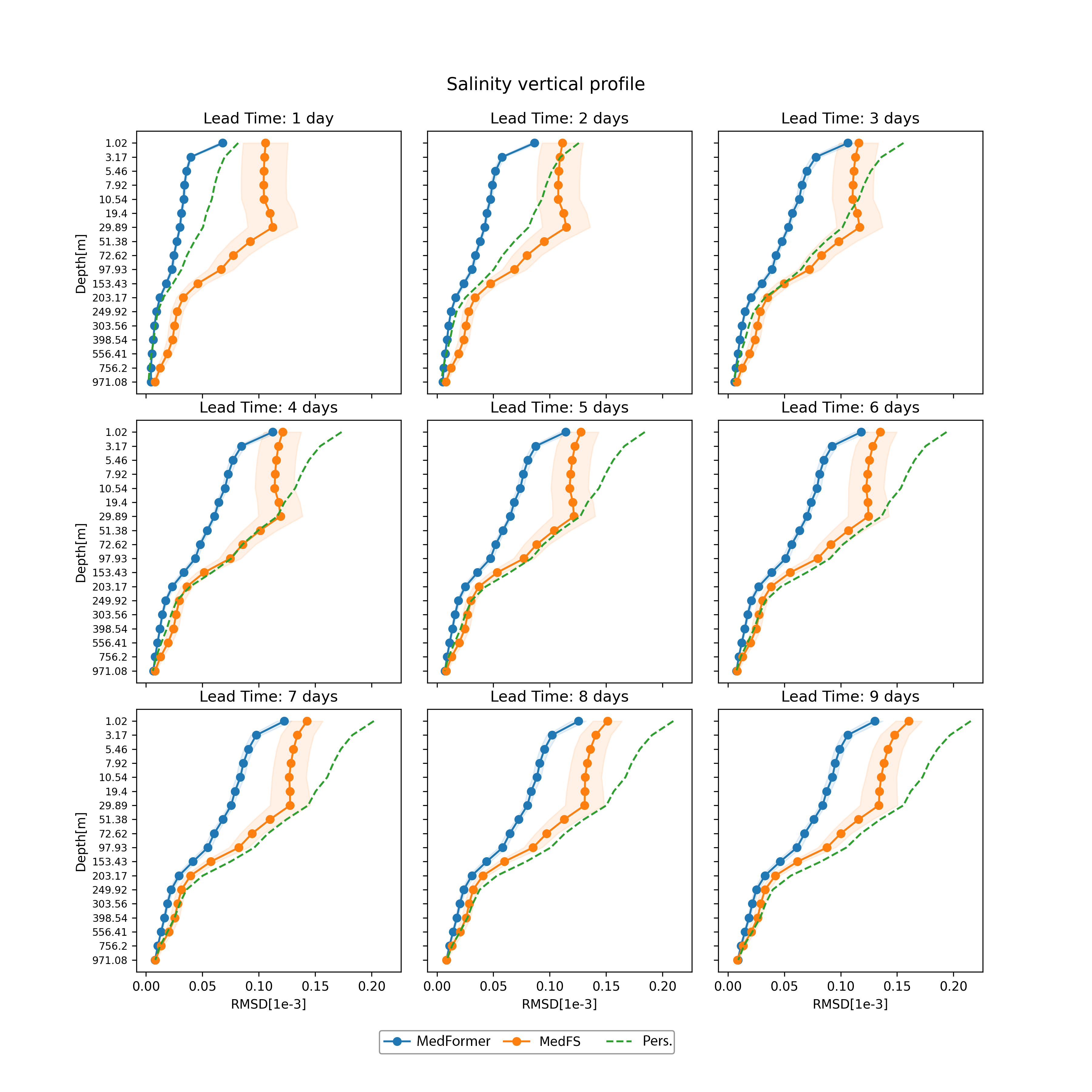} 
        \includegraphics[width=0.48\textwidth]{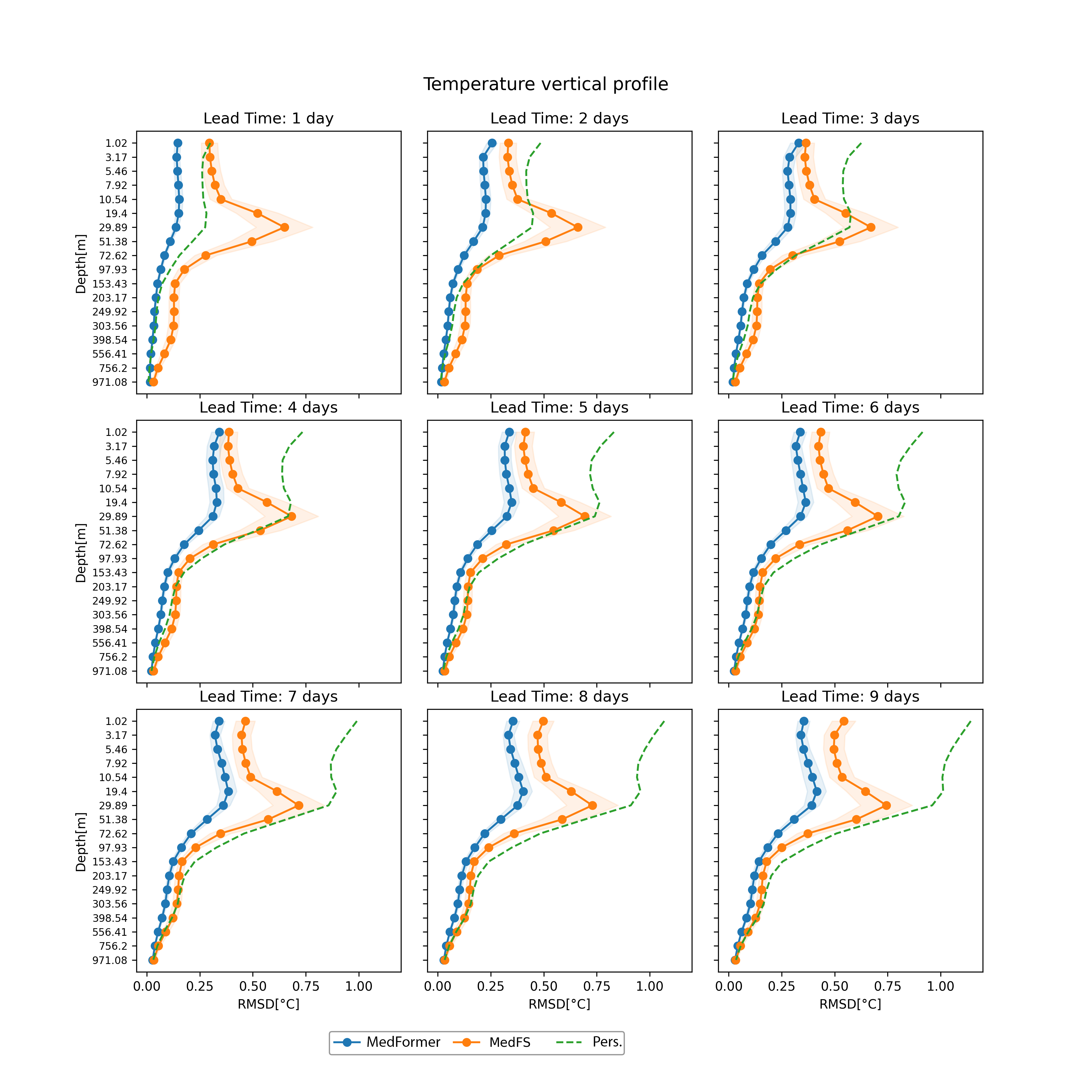}
        
    \caption{RMSD across vertical levels for salinity and temperature, averaged over one year and over the entire Mediterranean basin, for all 9 forecast lead times. MedFormer error is shown in blue, MedFS error in orange and the persistence is indicated by the dashed green line.}
    \label{fig:rmsd-vertical-prof_sal_tem}
\end{figure}

\begin{figure}[ht]
    \centering
        \includegraphics[width=0.48\textwidth]{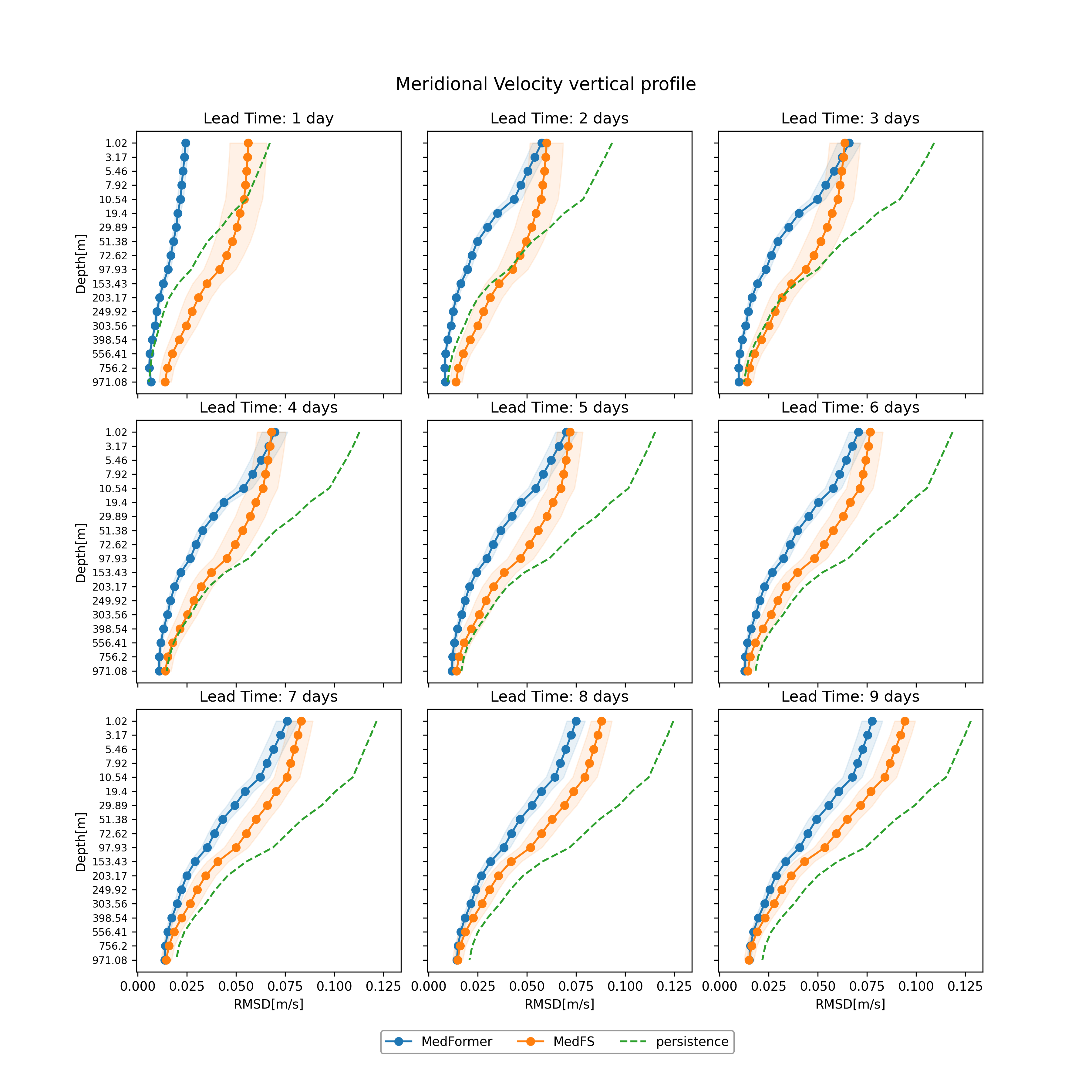} 
        \includegraphics[width=0.48\textwidth]{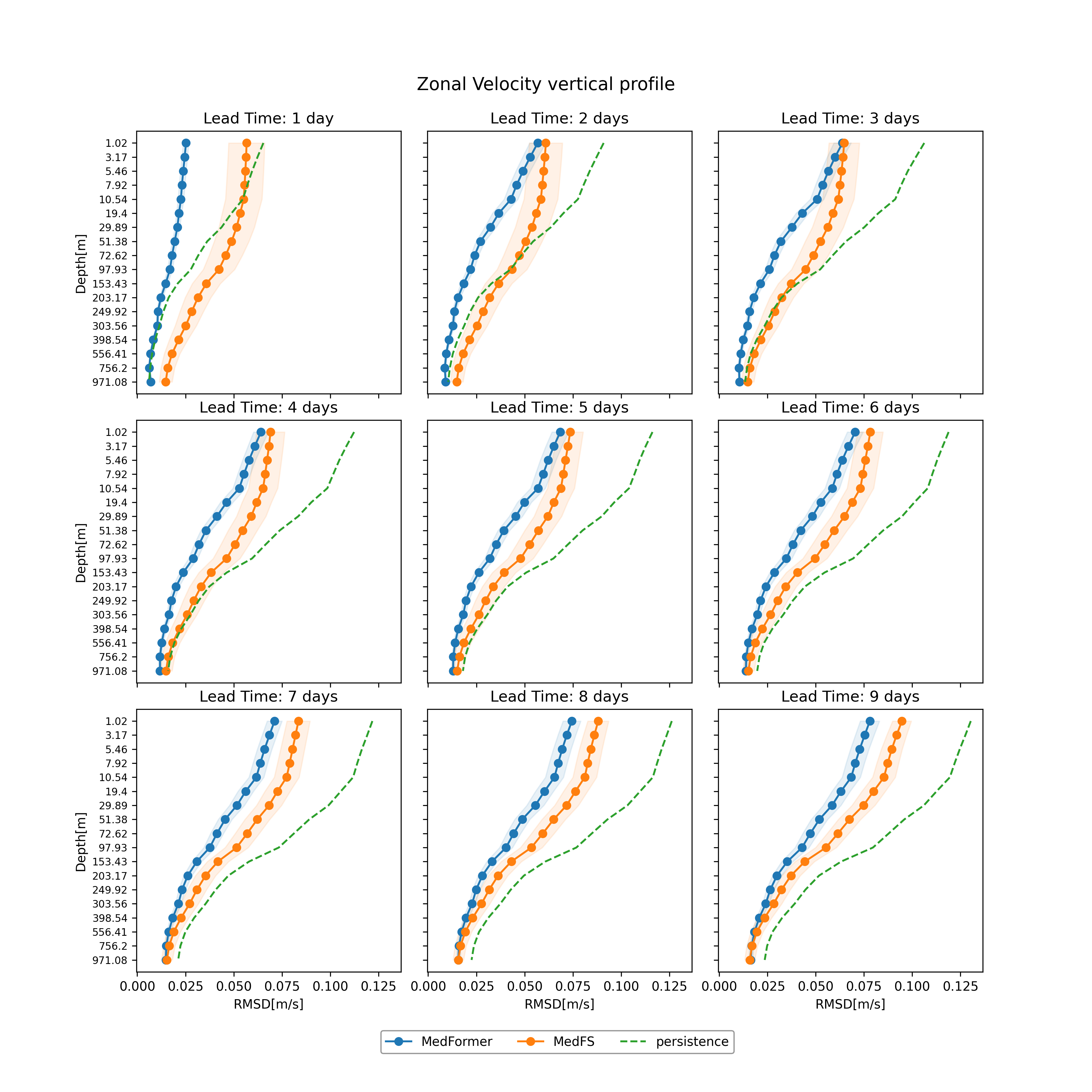}
        
    \caption{RMSD along vertical levels for meridional and zonal velocities, averaged over one year across the entire Mediterranean basin, for all 9 forecast lead times. MedFormer error is shown in blue, MedFS error in orange, and the persistence is indicated by the dashed green line.}
    \label{fig:rmsd-vertical-prof-vel}
\end{figure}

MedFormer's good forecast skill is further confirmed by analyzing RMSD profiles along the vertical levels. Figures~\ref{fig:rmsd-vertical-prof_sal_tem} and \ref{fig:rmsd-vertical-prof-vel} show the RMSD for temperature, salinity, and velocities across $18$ vertical levels down to a depth of $971$m. MedFormer consistently exhibits lower errors than MedFS across all depths and forecast lead times. Notably, it captures the vertical dynamics near the thermocline depth with significantly better accuracy than MedFS. Furthermore, the $95\%$ confidence intervals, represented by the shaded areas in the plots, demonstrate that the performance improvements of MedFormer are statistically significant.     

\begin{figure}[ht]
    \centering
    \includegraphics[width=0.9\linewidth]{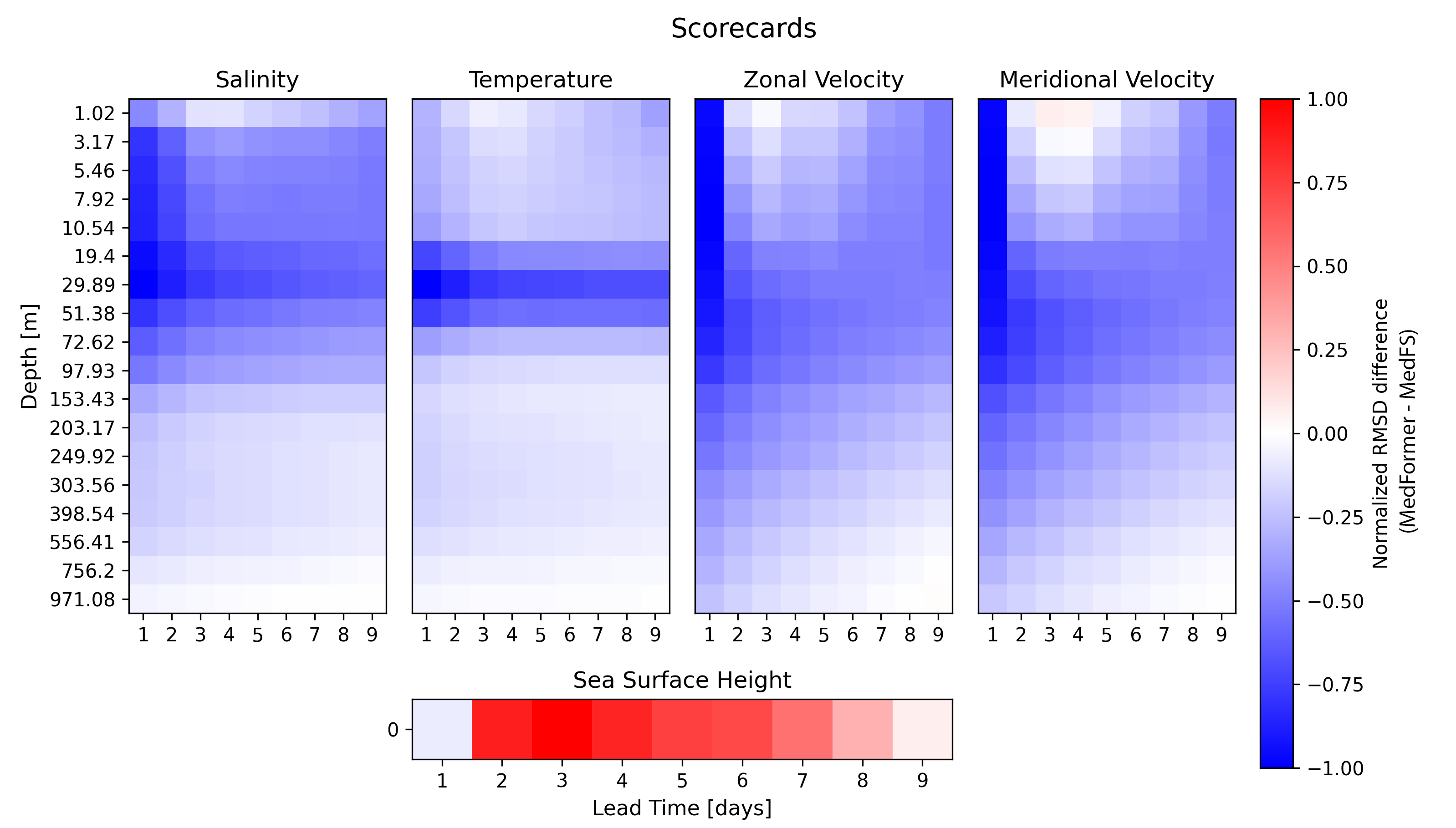}
    \caption{Scorecard comparing the RMSD of MedFormer and MedFS for all the variables considered in the study. A bluish color indicates that MedFormer outperforms MedFS (i.e., lower RMSD), a reddish color indicates that MedFormer performs worse (i.e., higher RMSD) than MedFS.}
    \label{fig:scorecard}
\end{figure}

The scorecards shown in Fig.~\ref{fig:scorecard} provide a comprehensive comparison between MedFormer and MedFS across different vertical levels and forecast lead times. In the scorecards, blue indicates that MedFormer achieves a lower RMSD than MedFS, with the analysis data used as the reference. MedFormer outperforms MedFS across all 3D ocean variables, consistently demonstrating better accuracy at all depths and lead times. However, it shows slightly reduced skill in predicting the SSH beyond the second forecast day.

\begin{figure}[ht]
    \centering
        \includegraphics[width=0.45\textwidth]{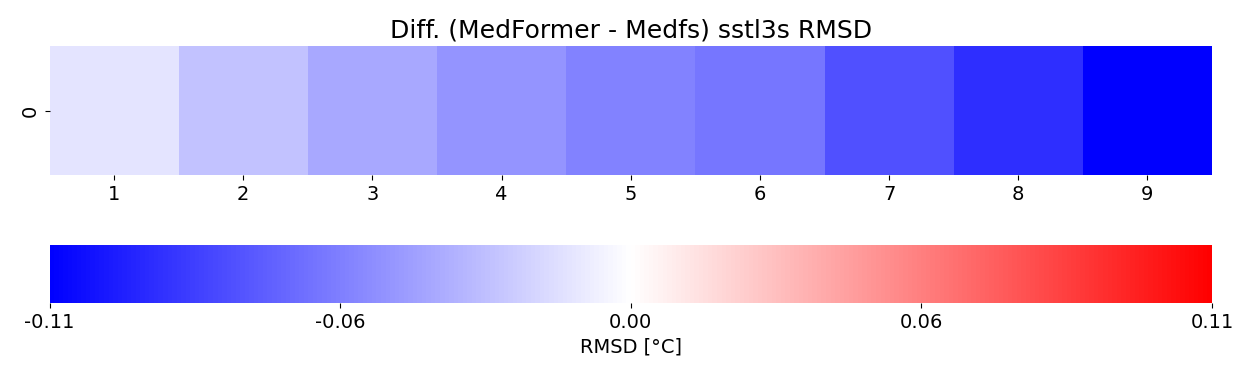} 
        \includegraphics[width=0.45\textwidth]{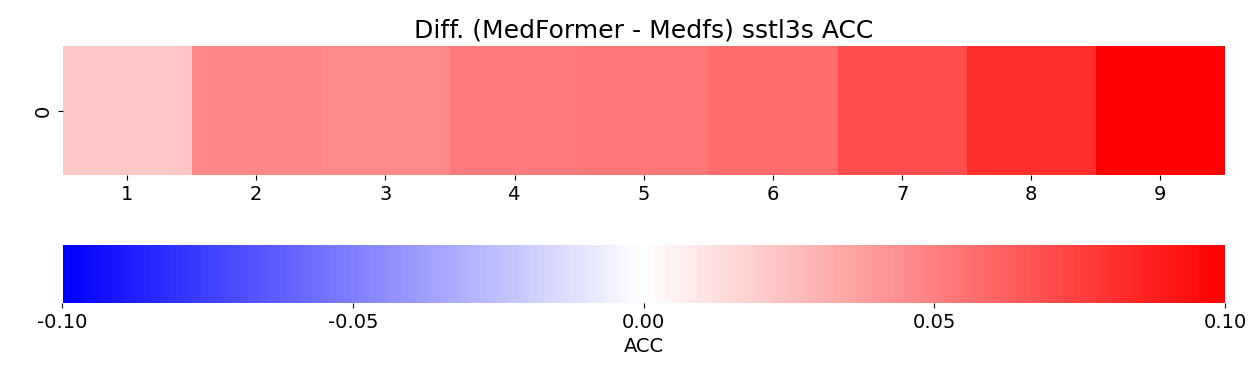} \\
        \includegraphics[width=0.45\textwidth]{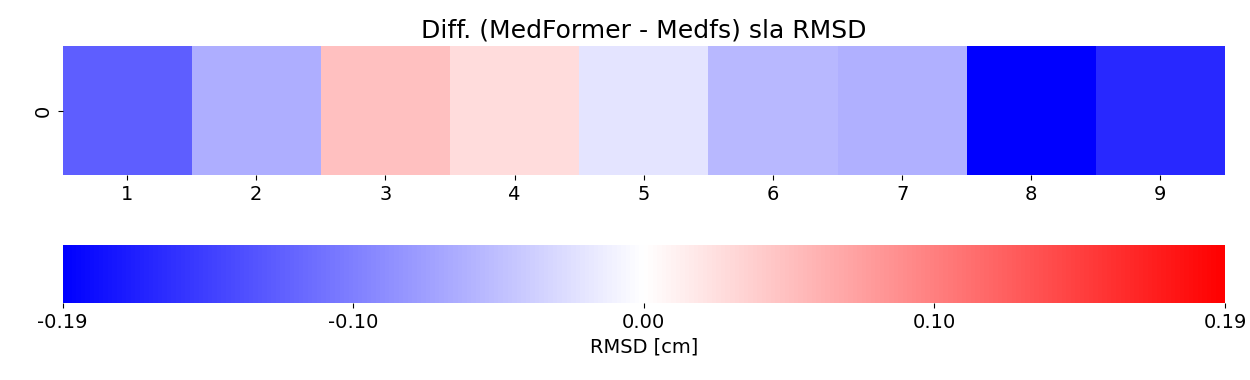} 
        \includegraphics[width=0.45\textwidth]{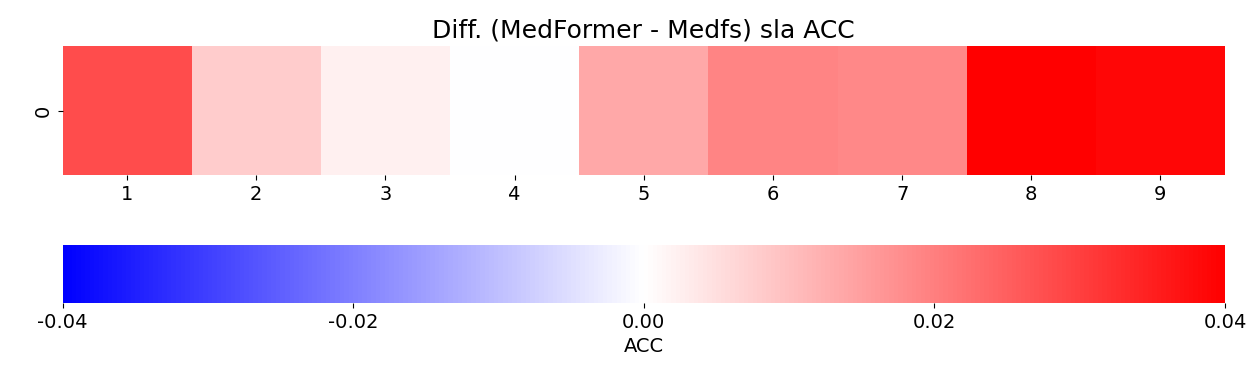}
        
    \caption{Difference in RMSD (left column) and ACC (right column) between MedFormer and MedFS, measured against satellite observations (SST L3s and SLA), averaged over one year across the entire Mediterranean basin for all 9 forecast lead times. For RMSD, blue indicates that MedFormer has lower error than MedFS; for ACC, red indicates that MedFormer exhibits higher skill than MedFS.}
    \label{fig:rmsd-satellite}
\end{figure}

We also evaluated the performance of both models using in-situ and satellite observations as reference to further assess their skills. Notably, MedFormer was not explicitly trained on observational data. 
The observational datasets used as ground truth for surface evaluation include the Copernicus Marine SST L3s product (SST\_MED\_SST\_L3S\_NRT\_OBSERVATIONS\_010\_012) and the along-track Copernicus Marine Sea Level Anomaly (SLA) L3 product (SEALEVEL\_EUR\_PHY\_L3\_NRT\_OBSERVATIONS\_008\_059).     
Figure~\ref{fig:rmsd-satellite} shows the RMSD of MedFormer and MedFS forecasts compared with satellite SST and SLA observations. Both MedFS and MedFormer outperform the persistence. MedFormer exposes superior forecast skill compared to MedFS, with particularly notable improvements in SST accuracy and more moderate gains for SLA.  

\begin{figure}[ht]
    \centering
        \includegraphics[width=0.56\textwidth]{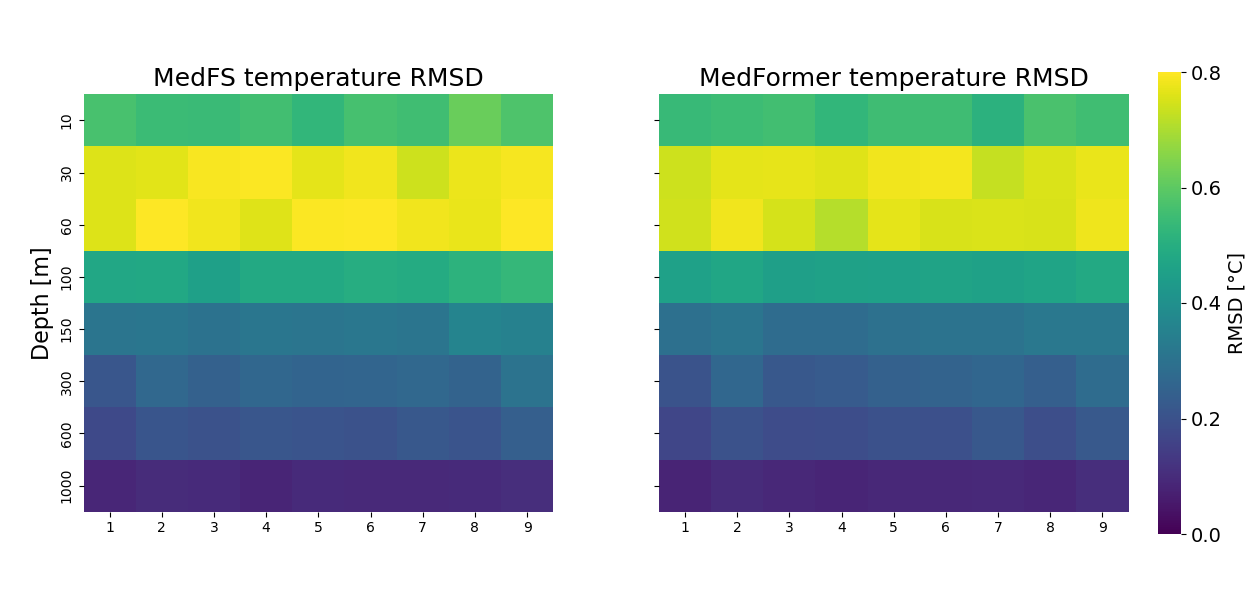} 
        \includegraphics[width=0.34\textwidth]{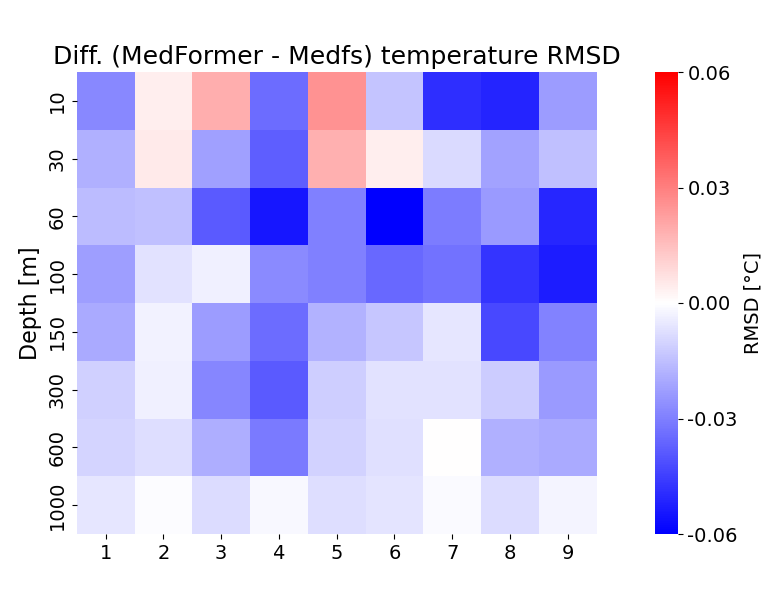}        
    \caption{RMSD of MedFormer and MedFS compared against in-situ temperature observations from Copernicus Marine ARGO vertical profiles, evaluated over 9 forecast lead times. The right panel shows the difference in RMSD between MedFormer and MedFS, with bluish color indicating that MedFormer has lower error than MedFS.}
    \label{fig:rmsd-insitu-tem}
\end{figure}

\begin{figure}[ht]
    \centering
        \includegraphics[width=0.56\textwidth]{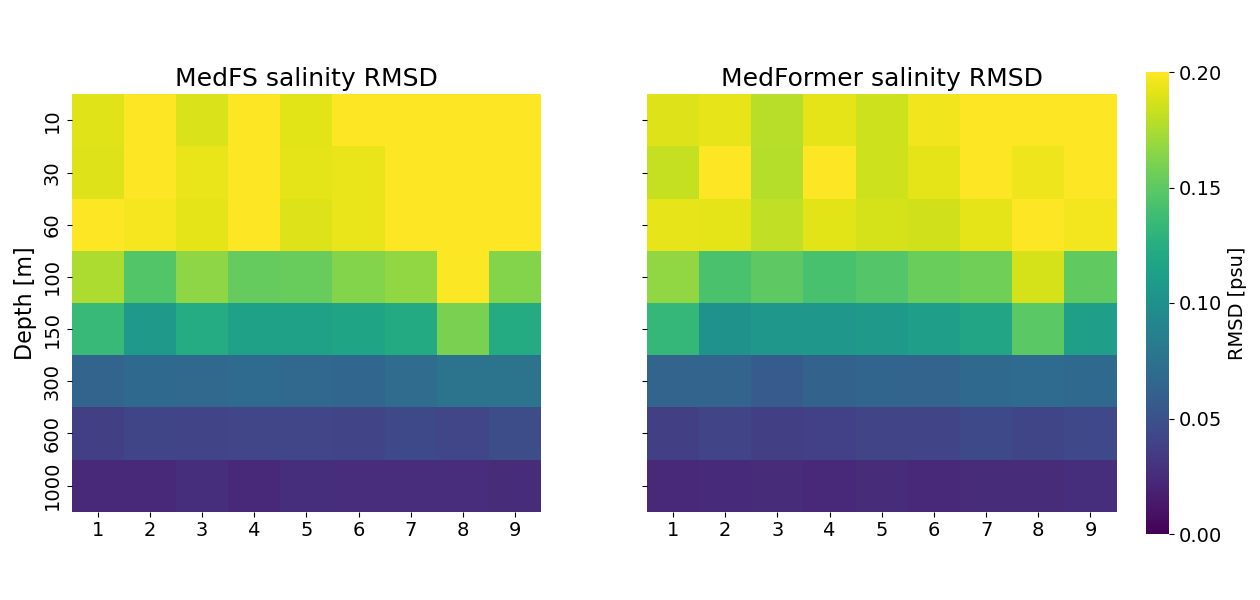} 
        \includegraphics[width=0.34\textwidth]{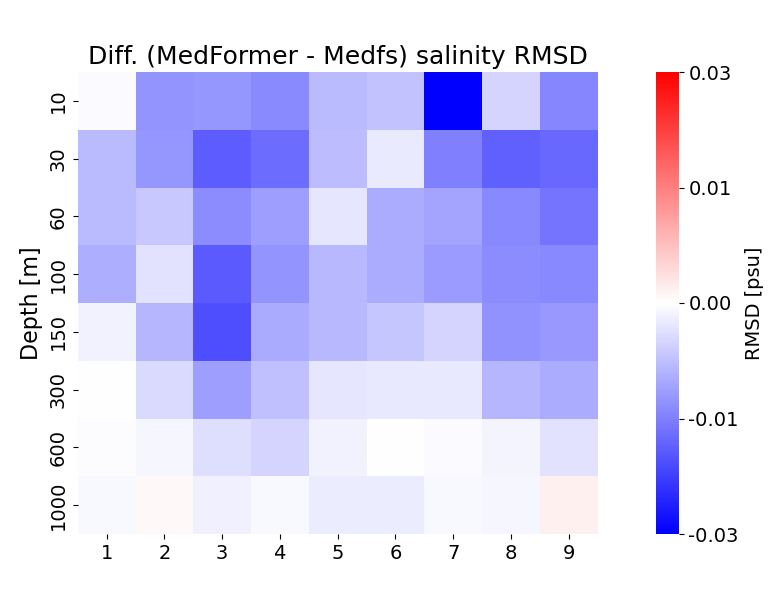} 
        
    \caption{RMSD of MedFormer and MedFS evaluated against in-situ salinity observations from Copernicus Marine ARGO vertical profiles over 9 forecast lead times. The right panel shows the RMSD difference between MedFormer and MedFS, where bluish colors indicate that the MedFormer has lower error than MedFS.}
    \label{fig:rmsd-insitu-sal}
\end{figure}

Figures~\ref{fig:rmsd-insitu-tem} and \ref{fig:rmsd-insitu-sal} depict the results of comparing model forecasts against in-situ observations of temperature and salinity collected along vertical profiles from the Copernicus Marine in-situ ARGO dataset (product INSITU\_GLO\_PHYBGCWAV\_DISCRETE\_MYNRT\_013\_030). Owing to the sparsity and limited availability of the observations, the model skill was assessed by aligning MedFormer's vertical levels with the available data and aggregating the results by vertical layers. Both MedFS and MedFormer exhibit larger errors in the surface and subsurface layers (up to approximately 60m depth), which during the stratified seasons (summer and autumn) typically correspond to the mean level of the thermocline. Across nearly all layers and forecast lead times, MedFormer consistently demonstrates lower RMSD values than MedFS, indicating improved accuracy in capturing the vertical structure of the ocean.

We provide more in-depth evaluations including scorecards, metrics and breakdown of the results in Appendix~\ref{extended_data}.

\section{Conclusion}\label{conclusion}

In this work, we introduced MedFormer, a novel data-driven deep learning model for medium-range ocean forecasting specifically designed for the Mediterranean Sea. Leveraging a UNet architecture enhanced with 3D attention mechanisms, MedFormer effectively captures the complex spatio-temporal dynamics of the ocean at high spatial resolution (1/24°), delivering daily forecasts up to 9 days ahead. Through a carefully structured two-phase training strategy, MedFormer achieves superior performance compared to the state-of-the-art dynamical model, MedFS. 

 Evaluation against both analysis and observational datasets demonstrates that MedFormer consistently outperforms MedFS in forecasting key 3D ocean variables, including temperature, salinity, and horizontal velocities. Its combination of speed, efficiency and accuracy positions MedFormer as a strong candidate for operational deployment within regional ocean forecasting systems.

These results underscore the potential of deep learning approaches in operational oceanography. Future work will aim to extend the forecast lead times, integrate uncertainty quantification, and expand MedFormer's applicability to global ocean domain.

\section{Data and Methods}\label{methods}

The following section describes the methods used for the design and development of the architecture of the deep learning model. MedFormer is a Swin-UNet Transformer 3D tailored to the Mediterranean Sea basin, and trained to forecast the state of the ocean dynamics up to 9 days ahead with high accuracy. 

The Mediterranean Sea dynamics is characterized by four 3-dimensional sea state variables, namely salinity, temperature, meridional and zonal velocities at 18 depth levels (i.e., 1.02m, 3.17m, 5.46m, 7.92m, 10.54m, 19.40m, 29.89m, 51.38m, 72.62m, 97.93m, 153.43m, 203.17m, 249.92m, 303.56m, 398.54m, 556.41m, 756.20m, 971.08m) and one 2-dimensional variable, the Sea Surface Height (SSH).

In addition to the Mediterranean Sea state, seven atmospheric forcings are used as additional inputs to the system: total cloud cover, mean sea level pressure, 2m dewpoint, 2m temperature, 10m u and v components of wind and cumulative precipitation over the last 24 hours.

The input to the MedFormer model is both the sea state and the atmospheric forcing at time $t-3, t-2, t-1, t$, the atmospheric forecast at time $t+1$ is also used as input to the decoder, whereas the Mediterranean Sea state at time $t+1$ is predicted.

\subsection{Datasets}\label{dataset}

 MedFormer works at a daily temporal resolution. The oceanic data sources used for training are the Mediterranean Sea Physics Reanalysis and the Mediterranean Sea Physics Analysis, both datasets are available through the Copernicus Marine Service (CMEMS). The atmospheric forcings are taken from ECMWF ERA5 reanalysis dataset and from ECMWF HRES (high resolution) operational analysis and forecast dataset.

During the first phase of the training we used the Reanalysis dataset from 2000 to 2019 for training and 2020 and 2021 for validation; in the second stage we used the Analysis dataset from 2018 to 2020 for training and 2021 for validation. The model was tested on year 2022.

Both the ocean datasets for reanalysis and analysis consist of daily mean values, with a horizontal resolution of $1/24^\circ$ ($\simeq 4.5 km$) and 141 vertical levels. The region extends from Latitude $30.19^\circ$ to $45.98^\circ$ and from Longitude $-6^\circ$ to $36.29^\circ$, resulting in a shape of $1016 \times 380$ points.

Also the atmospheric fields are averaged over one day but with a different horizontal resolution; ERA5 data have $1/4^\circ$ horizontal resolution, while IFS data for atmospheric analysis are at $1/10^\circ$.

\subsubsection{Preprocessing}\label{preprocessing}

Before starting the training, the datasets is preprocessed. The atmospheric fields are interpolated to the oceanic grid at $1/24^\circ$ using a bilinear interpolation. 

All the ocean and atmospheric fields are normalized using a standard normalization. Each vertical level is treated separately and after the standardization each variable has a statistic distribution with mean $\mu = 0$ and standard deviation $\sigma = 1$. Namely, the following transformation is applied $\bar{X}_{f,l} = \frac{X_{f,l}-\mu_{f,l}}{\sigma_{f,l}}$, where $\mu_{f,l}$ and $\sigma_{f,l}$ are, respectively, the mean and the standard deviation of the variable $f$ at level $l$, while $X_{f,l}$ is the original fields and $\bar{X}_{f,l}$ is its normalized version. 
 
Each 3D variable at each depth level is considered as separate variable and stacked along the channels dimensions: $V \times Z \times H \times W \rightarrow VZ \times H \times W$, where $VZ = 4 \cdot 18 = 72$, while $(H,W)$ denote the lat-lon grid of dimension $1016 \times 380$. The 2D variable (SSH) has shape $1 \times H \times W$, while the atmospheric forcings have shape $7 \times H \times W$ after the interpolation. 

The input to the model includes a sequence of 4 time steps at $t-3, t-2, t-1, t$, which are stacked along a new dimension $T$. To summarize, the input of the model is a 4D data tensor $X_{in} \in \mathbb{R}^{C_{in} \times T \times H \times W}$, where $C_{in} = 7 + 4 \cdot 18 + 1 = 80$, $T=4$, $H=1016$ and $W=380$. Since the model outputs only the ocean state at time $t + 1$, the corresponding output tensor is $X_{out} \in \mathbb{R}^{C_{out} \times 1 \times H \times W}$, where $C_{out} = 1 + 4 \cdot 18 = 73$.

\subsection{Mathematical problem statement}
\label{probelm_sta}

Given a sequence of 4 time steps of the ocean state ${\mathbb{X}^t_{t-3} = \{X^{t-3}, X^{t-2}, X^{t-1},X^t\}}$ and a sequence of 5 time steps of the surface fields of the atmosphere ${\mathbb{A}^{t+1}_{t-3}=\{A^{t-3}, A^{t-2}, A^{t-1}, A^t, A^{t+1}\}}$, MedFormer is meant to forecast the Mediterranean Sea state at the next time step $t+1$, $\hat{X}^{t+1}$. Therefore, the objective is to find a set of parameters $\theta$ that best approximate the relationship between the input sequence and the output, thus providing an accurate forecast for the next day. The regression task to address can be formulated as:

\begin{equation}
\hat{X}^{t+1} = \mathcal{F}_\theta (\mathbb{X}^t_{t-3}, \mathbb{A}^{t+1}_{t-3}) \simeq X^{t+1}  
\end{equation}

The trained model can be used autoregressively to forecast the ocean state at longer time leads. Indeed, we can use the predicted state $\hat{X}^{t+1}$ at time $t+1$ to update the input sequence $\mathbb{X}^{t+1}_{t-2} = \{X^{t-2}, X^{t-1},X^t,\hat{X}^{t+1}\}$ to predict the state at time $t+2$, assuming that the atmospheric fields at time $t+2$ are available from an external data source or from an atmospheric model; ${\hat{X}^{t+2} = \mathcal{F}_\theta (\mathbb{X}^{t+1}_{t-2}, \mathbb{A}^{t+2}_{t-2})}$.
In this regards we can extend the definition of the regression task if we consider the whole chain of forecasts. Namely, given an input sequence of 4 ocean states $\mathbb{X}^t_{t-3}$ and a sequence of $T+4$ time steps of surface atmospheric fields $\mathbb{A}^{t+T}_{t-3} = \{A^{t-3}, A^{t-2}, \dotsc ,A^{t+T}\}$, the MedFormer model is able to forecast a sequence of $T$ time steps of the ocean state in the future.

\begin{equation}
\hat{\mathbb{X}}^{t+T}_{t+1} = \mathcal{F}_\theta (\mathbb{X}^t_{t-3}, \mathbb{A}^{t+T}_{t-3})  
\end{equation}

where $\hat{\mathbb{X}}^{t+T}_{t+1} = \{\hat{X}^{t+1}, \hat{X}^{t+2}, \dotsc ,\hat{X}^{t+T}\}$ is the sequence of predicted states. In this work we chose $T=9$

\subsection{MedFormer architecture}\label{architecture}

\begin{figure}[ht]
    \centering
        \includegraphics[width=0.5\textwidth]{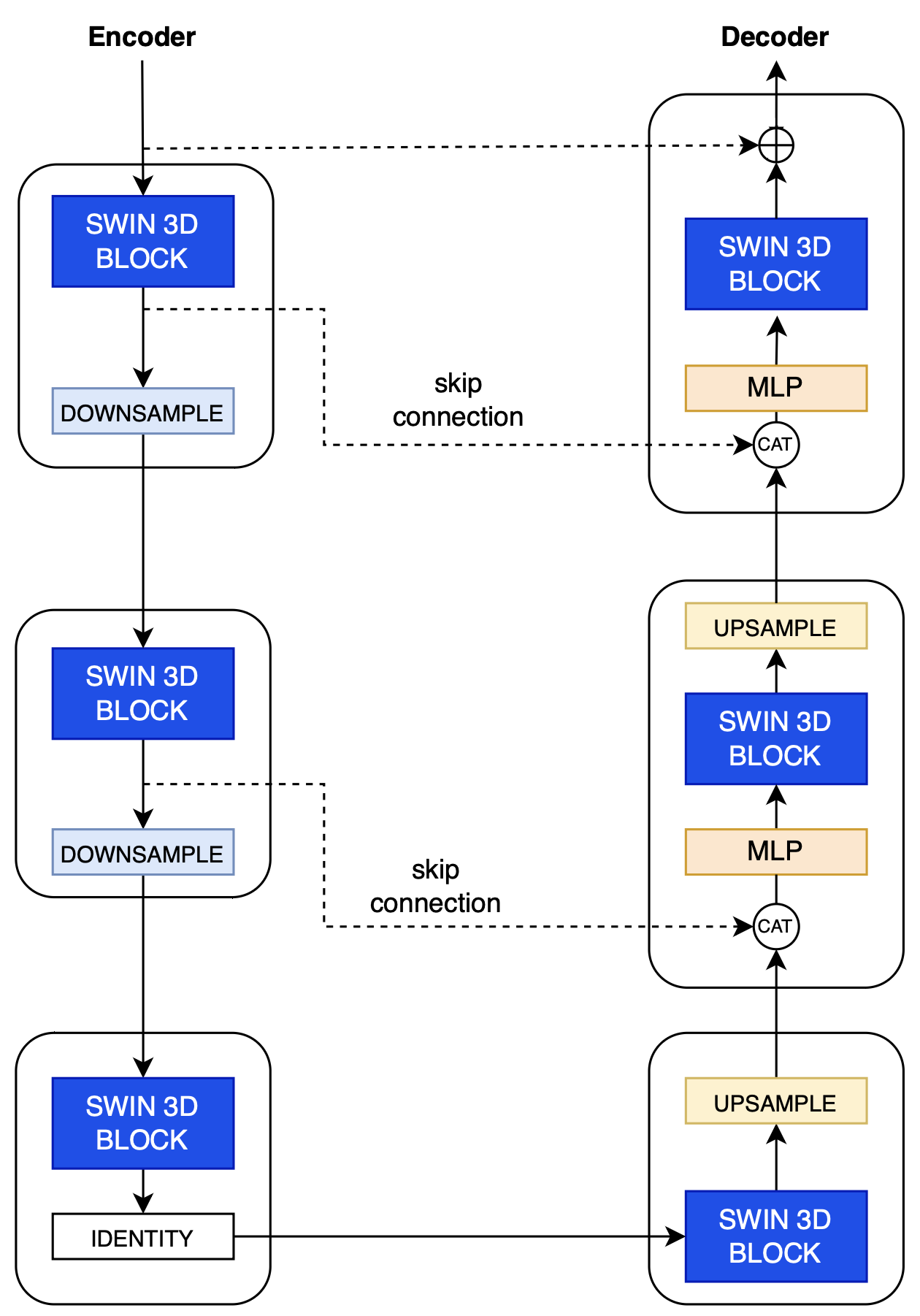} 
        
    \caption{UNet architecture with attention blocks.}
    \label{fig:arch-encoder-decoder}
\end{figure}

The core architecture of MedFormer, shown in Fig.~\ref{fig:arch-encoder-decoder}, is a Multi-scale 3D Swin-UNet Transformer. UNet~\cite{10.1007/978-3-319-24574-4_28} architectures are widely known for image segmentation tasks. An UNet is composed by an encoder and decoder. 
The encoder consists of three Swin3D blocks. Each block includes a series of Swin3D Layers composed by multi-head attention layer, a normalization layer, a multi layer perceptron and a further normalization layer. A max pooling layer follows each Swin3D block. It captures context by progressively reducing spatial resolution and increasing the number of feature channels. It acts like a feature extractor.
The structure of the decoder is mirrored with respect to the encoder. It uses up-sampling (i.e. transposed convolutions) to increase spatial resolution while decreasing the number of channels (in the opposite order done by the encoder). Each up-sampling step is followed by convolutions and concatenation with the corresponding feature maps from the encoder (via skip connections). The skip connections allow the model to reuse high-resolution features lost during down-sampling. Moreover, the skip connections attenuate the gradient vanishing problem for deeper UNet architectures, improving fine-grained performance. Finally, the decoder uses the representation of the ocean state encoded in latent space and the surface atmospheric variables at time $t+1$ to reconstruct the ocean state at time $t+1$.
At each network's layer, the skip connection concatenates the features that come from the encoder to the decoder features, thus doubling the number of channels ($C \rightarrow 2C$). To reduce the number of parameters, the concatenation is followed by a linear layer that halves the channels back to $C$, thus keeping the consistency of the number of channels between encoder and decoder.

\begin{figure}[ht]
    \centering
        \includegraphics[width=0.5\textwidth]{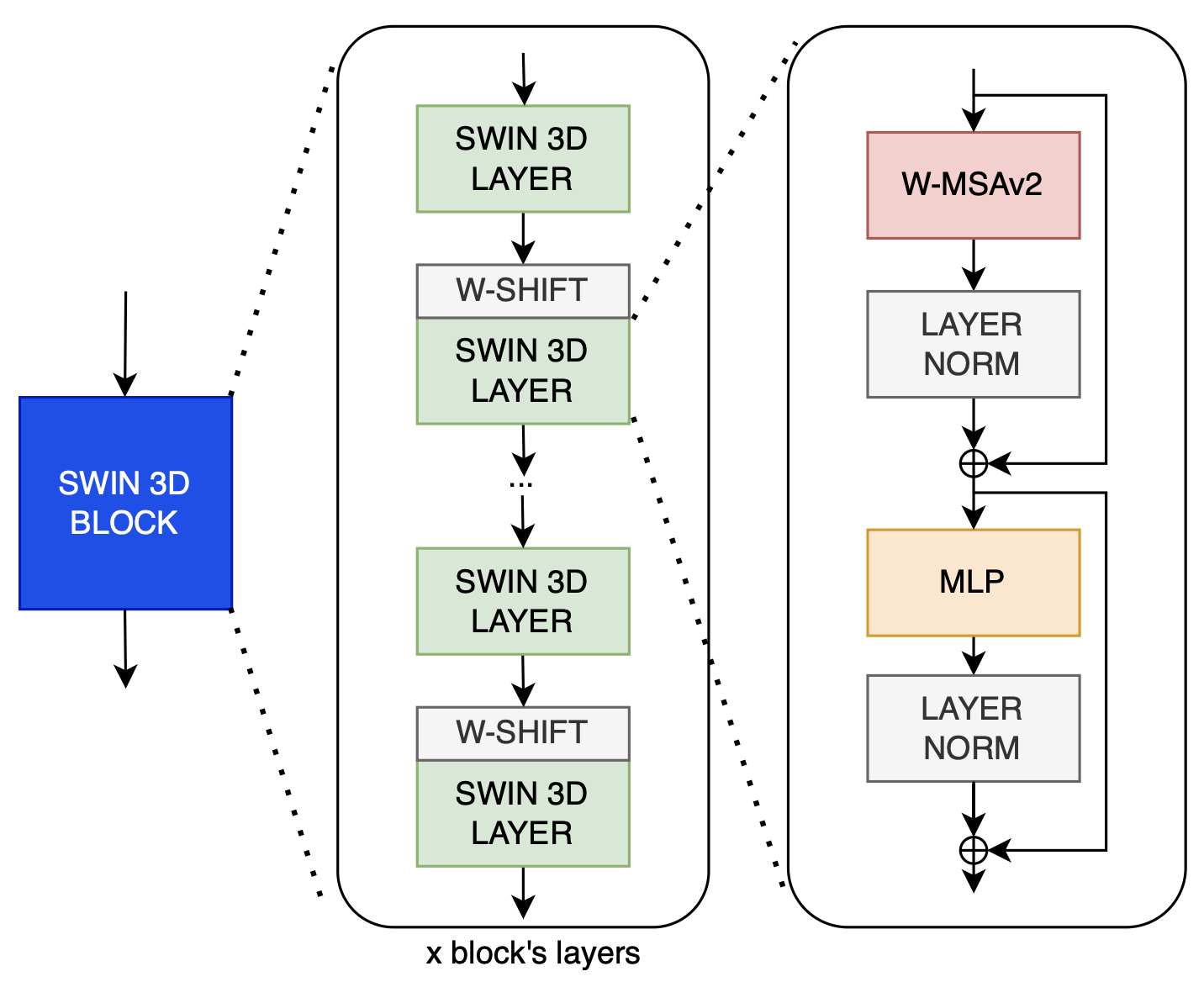} 
        
    \caption{Shifted windows attention block.}
    \label{fig:arch-swin-block}
\end{figure}

Each block of the UNet-shaped architecture includes a 3D SwinTransformer, characterized by the shifted window attention mechanism~\cite{9710580} (see Fig.~\ref{fig:arch-swin-block}). The 3 dimensions managed by MedFormer are time, latitude and longitude, respectively, meaning that it can perform spatio-temporal attention to capture complex correlations in the Mediterranean Sea dynamics. In this work, the Swin Transformer v2~\cite{9879380} has been used, with a window size set to $(2 \times 8 \times 8)$. 

The patch embedding is characterized by a patch size of $(1 \times 4 \times 4)$ with an embedding dimension $C=384$ that doubles after each downsample. To enhance the forecast, the day of the year at time $t$ is used as an auxiliary input and embedded using a Multi Layer Perceptron (MLP). The embedded coefficients are used as scaling factors to perturb the input with an affine transformation.

The number of layers in the encoder and decoder is $[2,8,4]-[4,8,2]$, respectively, resulting in $\sim 400M$ trainable parameters.

\subsection{Training protocol}\label{training}

MedFormer model has been trained in two steps: a pretraining phase, and a fine tuning phase. The AdamW optimizer has been used across the entire training pipeline, with weight decay set to $1e^{-2}$ and $\beta = (0.9, 0.95)$ which are the coefficients for computing running averages of gradient and squared gradient.
The loss function is the Mean Absolute Error (MAE) with weights along the vertical levels ranging linearly from $w_1=1$ for the surface to $w_{18}=0.01$ for the deepest level. Each variable has been weighted differently. Specifically, temperature, salinity, meridional and zonal currents, SSH have weights set to $w_f = [3, 2, 1, 1, 1]$ respectively. The weights have been chosen to balance the contribution of the variables in the loss function. During the pre-training step the loss function is defined as in Eq.~\ref{eq:lossfunction}:

\begin{equation}
\label{eq:lossfunction}
\mathcal{L}(\hat{X}^t, X^t) =  \frac{1}{F}\sum_{f=1}^{F} \frac{w_f}{L_f}\sum_{l=1}^{L_f}\frac{w_l}{N_l}\sum_{i,j\in\mathcal{D}_l} \mid \hat{X}^t_{i,j,l,f} - {X}^t_{i,j,l,f} \mid
\end{equation}

Where $N_l$ is the total number of ocean points at level $l$, $\mathcal{D}_l$ is the set of ocean points at level $l$, $w_l$ is the weight assigned to level $l$ and defined as ${w_l =\frac{1-0.01}{L-1} \cdot (1-l) + 1 }$, $L_f$ is the number of vertical levels for the variable $f$ (the sea surface height has only one level, the other variables have 18 levels), $w_f$ is the weight assigned to fields $f$ and $F$ the number of fields.

The model is fine tuned in the second step of training by using the Mediterranean analysis as initial condition and as target. We used a different loss function aimed at minimizing the error accumulated when the model is used autoregressively; it is defined as in Eq.~\ref{eq:lossfunction_roll} 

\begin{equation}
\label{eq:lossfunction_roll}
\mathbb{L}(\hat{\mathbb{X}}^{t+T}_{t+1}, \mathbb{X}^{t+T}_{t+1}) = \frac{1}{T}\sum_{\tau=t+1}^{t+T} \mathcal{L}(\hat{X}^{\tau}, X^{\tau})
\end{equation}

$T$ is the number of autoregressive steps.

Every experiment has been processed on the CMCC Juno Supercomputer\footnote{CMCC SuperComputing Center web page: https://www.cmcc.it/what-we-do/high-performance-computing-center-hpcc}, using 20 NVidia Ampere A100 40GB GPUs. The pre-trained model has been trained over 160 epochs with a learning rate warm-up for 6 epochs up to $lr=5e^{-4}$ and decreasing to $lr=5e^{-7}$ following cosine annealing schedule. We adopted the Distributed Data Parallel (DDP) approach to parallelize the training in our multinode architecture. Each batch is split among the GPUs and after the forward and back propagation steps, the nodes synchronize by exchanging the gradients. In our configuration, each node processes only an instance, hence the global batch size equals the number of GPUs. 

The fine tuning training  makes use of 4 years of analysis data (2018-2020 for training and 2021 for validation), with a learning rate that starts at $lr=5e^{-5}$ and stops at $lr=5e^{-8}$ with a cosine annealing scheduling. In this case a curriculum learning technique has been used, thus increasing the autoregression steps $T$ by 1 day every $60$ epochs, from 1 to 5. Finally, during the autoregression, the gradients are evaluated only during the last day, to avoid any memory increases and to reduce the training time. The computational cost for training the model was $1200 GPUh$ for the pre-training step and $700 GPUh$ for fine tuning.

\subsection{Evaluation metrics}\label{metrics}

In this section, we report the evaluation metrics used to assess the MedFormer forecasting skills. Specifically, the root mean square difference (RMSD) and the anomaly correlation coefficient (ACC) have been used as main evaluation metrics.

RMSD is a metric that provides the magnitude of the error between the forecast and the ground truth. It is computed using Eq.~\ref{eq:rmse}:

\begin{equation}
\label{eq:rmse}
RMSD = \sqrt{\frac{1}{N} \sum_{i,j\in \mathcal{D}} (\hat{X}_{i,j} - X_{i,j})^2}
\end{equation}

Where $\bar{X}_{i,j}$ and $X_{i,j}$ are, respectively, the forecast and the ground truth values, while $i,j$ are the spatial coordinates of each point in the horizontal domain $\mathcal{D}$. $N=\vert\mathcal{D}\vert$ is the total number of sea points.

The Anomaly Correlation Coefficient is one of the most commonly used metrics for evaluating the accuracy of spatial forecast fields. It measures the spatial correlation between the forecast anomalies and the corresponding anomalies from the analysis. These anomalies are calculated relative to the model climatology derived from MEDREA Reanalysis. A high ACC value indicates that the forecast anomalies closely match those of the analysis, reflecting strong model performance.

\backmatter

\subsection* {Funding}
\subsection* {Competing interests}
The authors declare no competing interests.
\subsection* {Ethics approval and consent to participate}
Not applicable.
\subsection* {Consent for publication}
Not applicable.
\subsection*{Data availability}
Mediterranean Sea Physics Reanalysis available through Copernicus Marine Service.\\
DOI: https://doi.org/10.25423/CMCC/MEDSEA\_MULTIYEAR\_PHY\_006\_004\_E3R1\\

\noindent Mediterranean Sea Analysis and Forecast available through Copernicus Marine Service.\\
DOI: https://doi.org/10.25423/CMCC/MEDSEA\_ANALYSISFORECAST\_PHY\_006\_013\_EAS6
\subsection* {Materials availability}
Not applicable.
\subsection*{Code availability}
MedFormer is available through the GitHub repository:
https://github.com/CMCC-Foundation/MedFormer

\subsection*{Author contribution}
Epicoco and Navarra conceived and designed the study. Navarra and Boccaletti provided the strategic direction of the study. Epicoco, Donno and Accarino defined the methodology and the approach. Clementi, Masina, Navarra, Coppini, Gualdi and Scoccimarro designed the model evaluation process and analyzed the results. Donno, Accarino, Norberti and Elia implemented the Python code, performed the experiments and collected the data. Epicoco, Aloisio and Nassisi supervised the technological development of the model. Grandi and Giurato supported the assessment of the forecast skill. Epicoco, Navarra, Clementi, Masina and Donno wrote the initial draft of the manuscript. All authors reviewed and edited the manuscript. All authors read and approved the final manuscript.

\begin{appendices}

\section{Extended Data}\label{extended_data}

\subsection{Seasonal variability}

\begin{figure}[ht]
    \centering
        \includegraphics[width=0.9\textwidth]{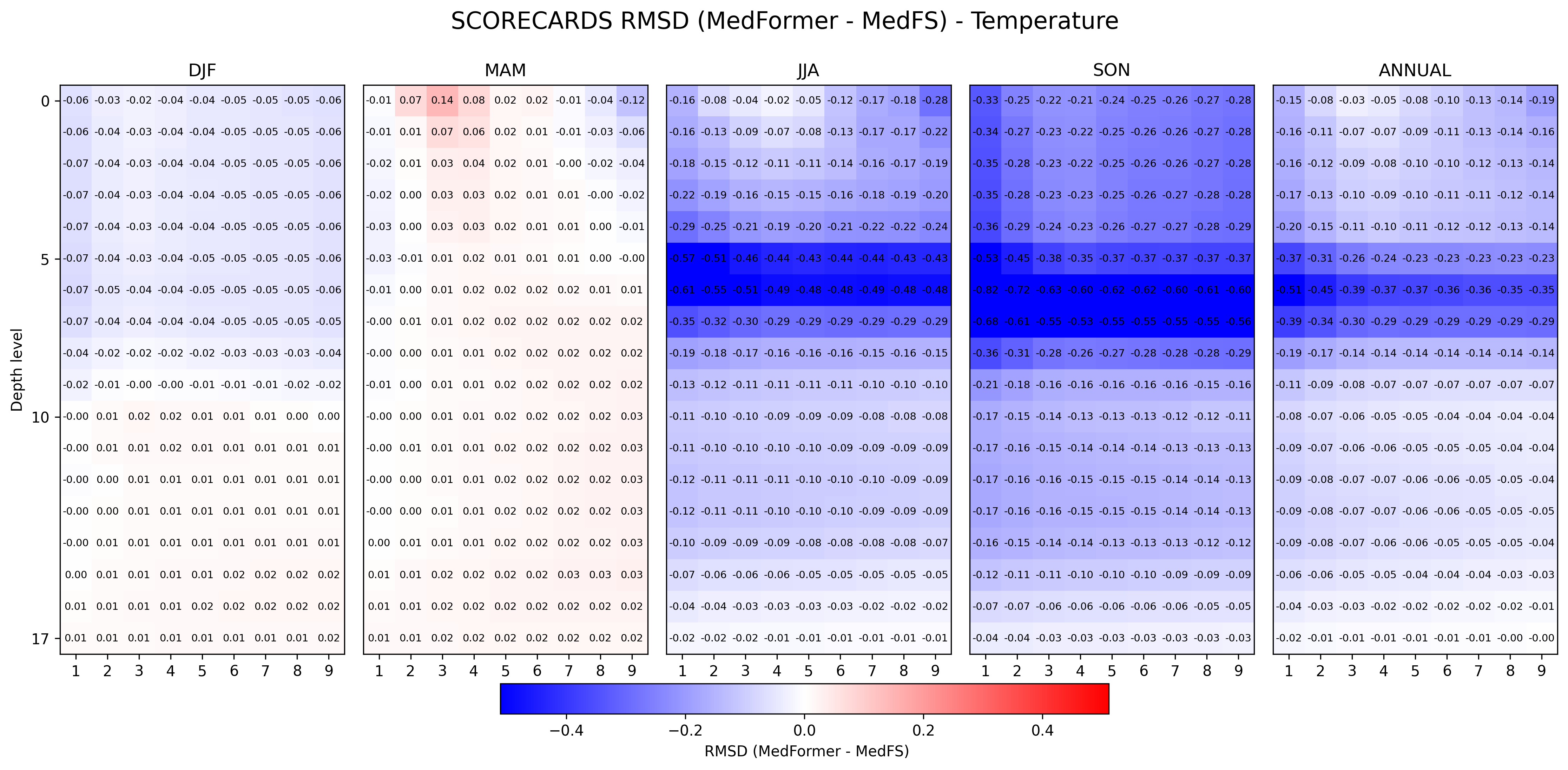}\\ 
        
    \caption{Difference of RMSD between MedFormer and MedFS against Analysis for the 9 forecast lead times for temperature in different seasons. Bluish color means that the MedFormer error is lower than MedFS.}
    \label{fig:rmsd-season-tem}
\end{figure}

\begin{figure}[ht]
    \centering
        \includegraphics[width=0.9\textwidth]{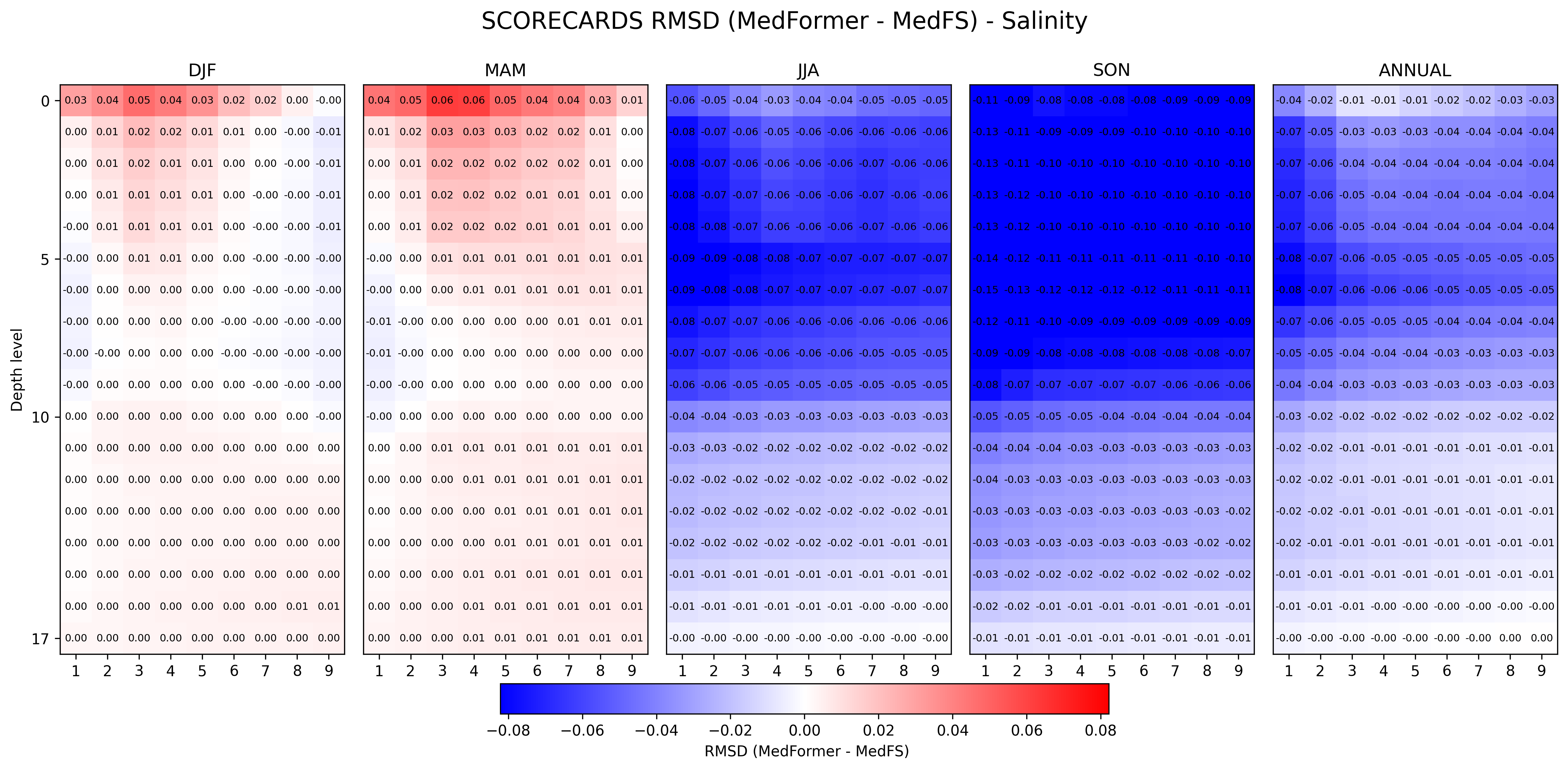}\\ 
        
    \caption{Difference of RMSD between MedFormer and MedFS against Analysis for the 9 forecast lead times for salinity in different seasons. Bluish color means that the MedFormer error is lower than MedFS.}
    \label{fig:rmsd-season-sal}
\end{figure}

\begin{figure}[ht]
    \centering
        \includegraphics[width=0.9\textwidth]{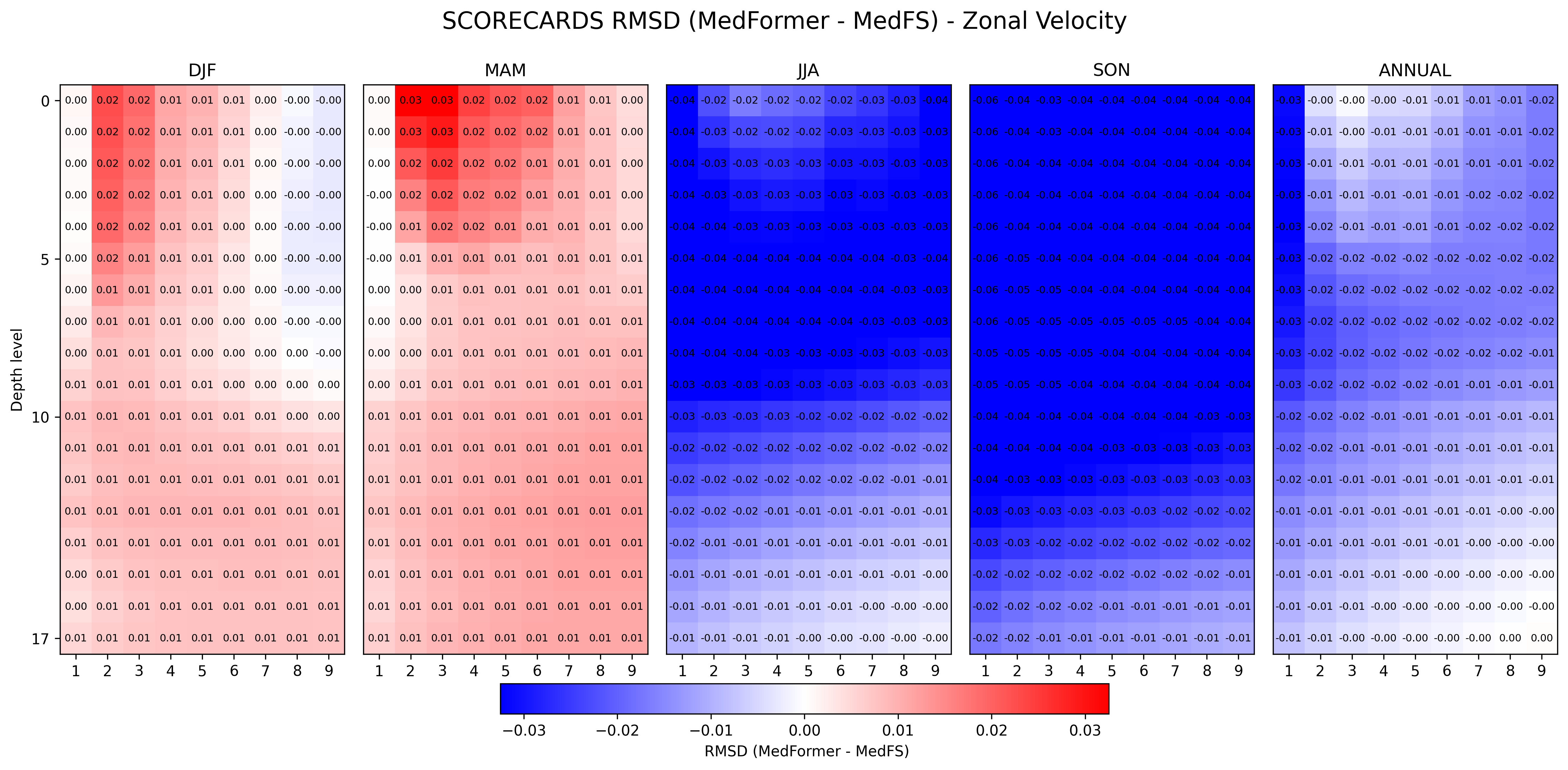}\\
        
    \caption{Difference of RMSD between MedFormer and MedFS against Analysis for the 9 forecast lead times for zonal velocity in different seasons. Bluish color means that the MedFormer error is lower than MedFS.}
    \label{fig:rmsd-season-zon}
\end{figure}

\begin{figure}[ht]
    \centering
        \includegraphics[width=0.9\textwidth]{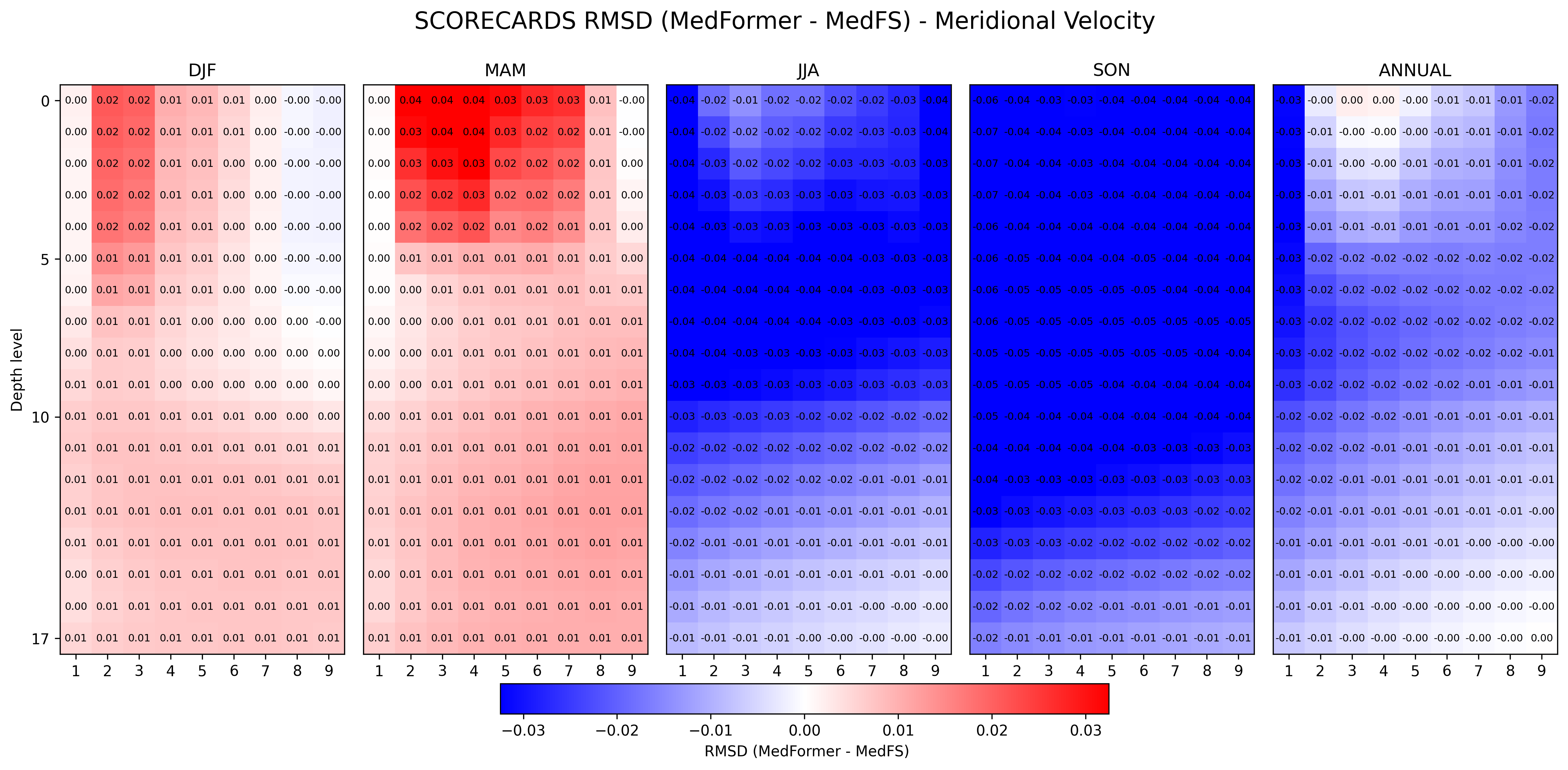}\\
        
    \caption{Difference of RMSD between MedFormer and MedFS against Analysis for the 9 forecast lead times for meridional velocity in different seasons. Bluish color means that the MedFormer error is lower than MedFS.}
    \label{fig:rmsd-season-mer}
\end{figure}

\begin{figure}[ht]
    \centering
        \includegraphics[width=0.5\textwidth]{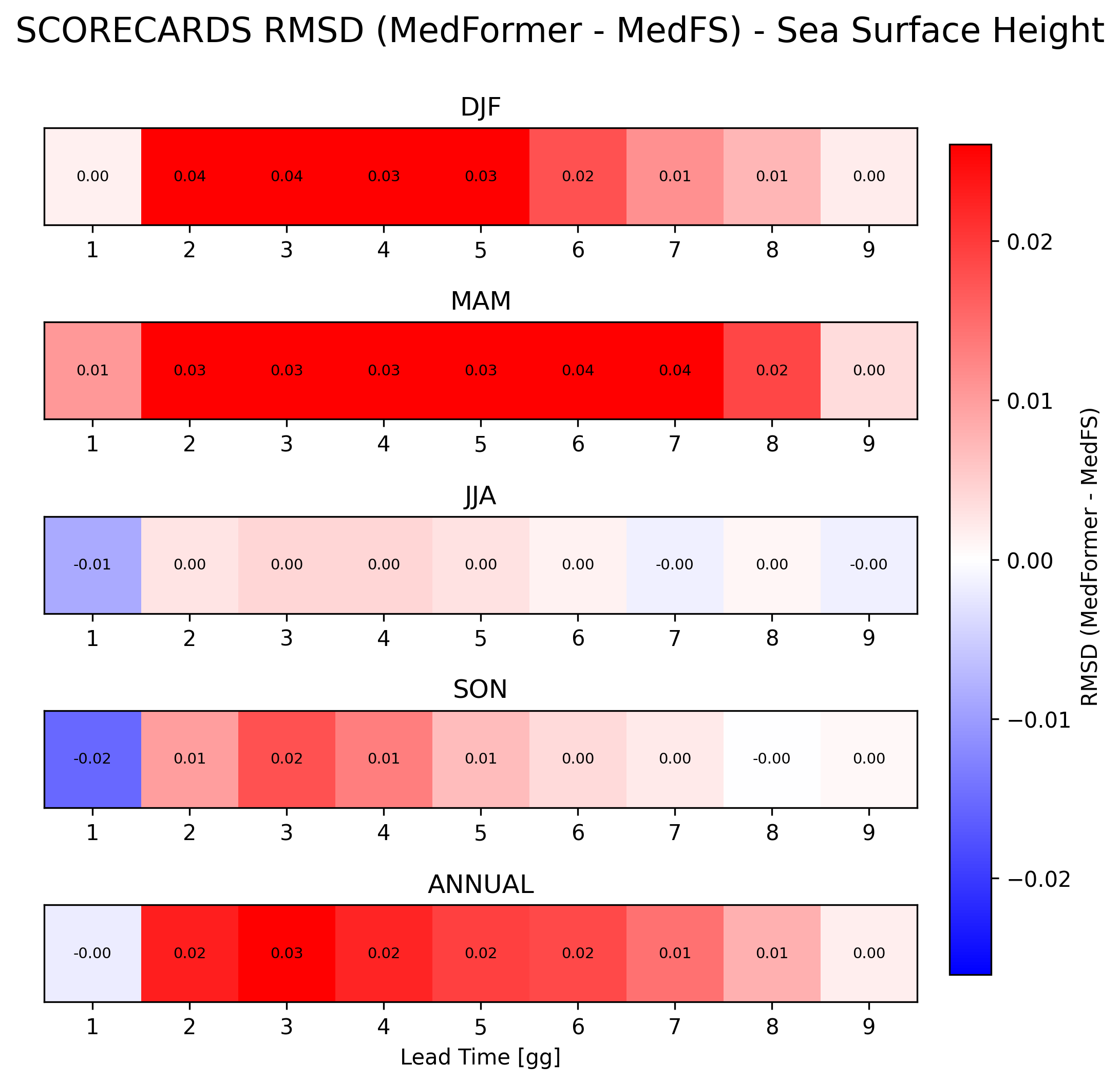}\\
        
    \caption{Difference of RMSD between MedFormer and MedFS against Analysis for the 9 forecast lead times for sea surface height in different seasons. Bluish color means that the MedFormer error is lower than MedFS.}
    \label{fig:rmsd-season-ssh}
\end{figure}

In this section we analyze the forecast skills of MedFormer during different seasons. Figures~\ref{fig:rmsd-season-tem}, \ref{fig:rmsd-season-sal}, \ref{fig:rmsd-season-zon}, \ref{fig:rmsd-season-mer}, \ref{fig:rmsd-season-ssh} 
depict the difference of RMSD between MedFormer and MedFS for salinity, temperature, zonal and meridional velocity and for sea surface height. The last column of each figure reports the annual mean as a reference. Taking into account all variables, MedFormer behaves better than MedFS in summer and autumn, while the skill decreases in winter and spring. 
The seasonal variability between MedFormer and MedFS is more evident for temperature (Fig.~\ref{fig:rmsd-season-tem}) and, in particular, in the thermocline depth where MedFormer exposes greater skill during summer and autumn and behaves similarly to MedFS in winter and spring. 

For salinity (Fig.~\ref{fig:rmsd-season-sal}), the seasonal variability is evident in the upper levels. MedFormer shows less error than MedFS in summer and autumn. In the deepest levels both models show similar skills with no significant differences among the seasons.  

Despite the intensity of the colors in Figures~\ref{fig:rmsd-season-zon}, \ref{fig:rmsd-season-mer}, for both components of the velocity there is no significant difference during the seasons, indeed the scale of the colorbar is narrow ranging between $-0.03$ to $0.03$. Once again MedFormer has better skills in summer and autumn. 

Finally, MedFormer exposes less forecast skills for SSH with respect to MedFS (Fig.~\ref{fig:rmsd-season-ssh}) in all the seasons with more emphasis in spring and summer.

\subsection{Error maps}

\begin{figure}[ht]
    \centering
        \includegraphics[width=0.9\textwidth]{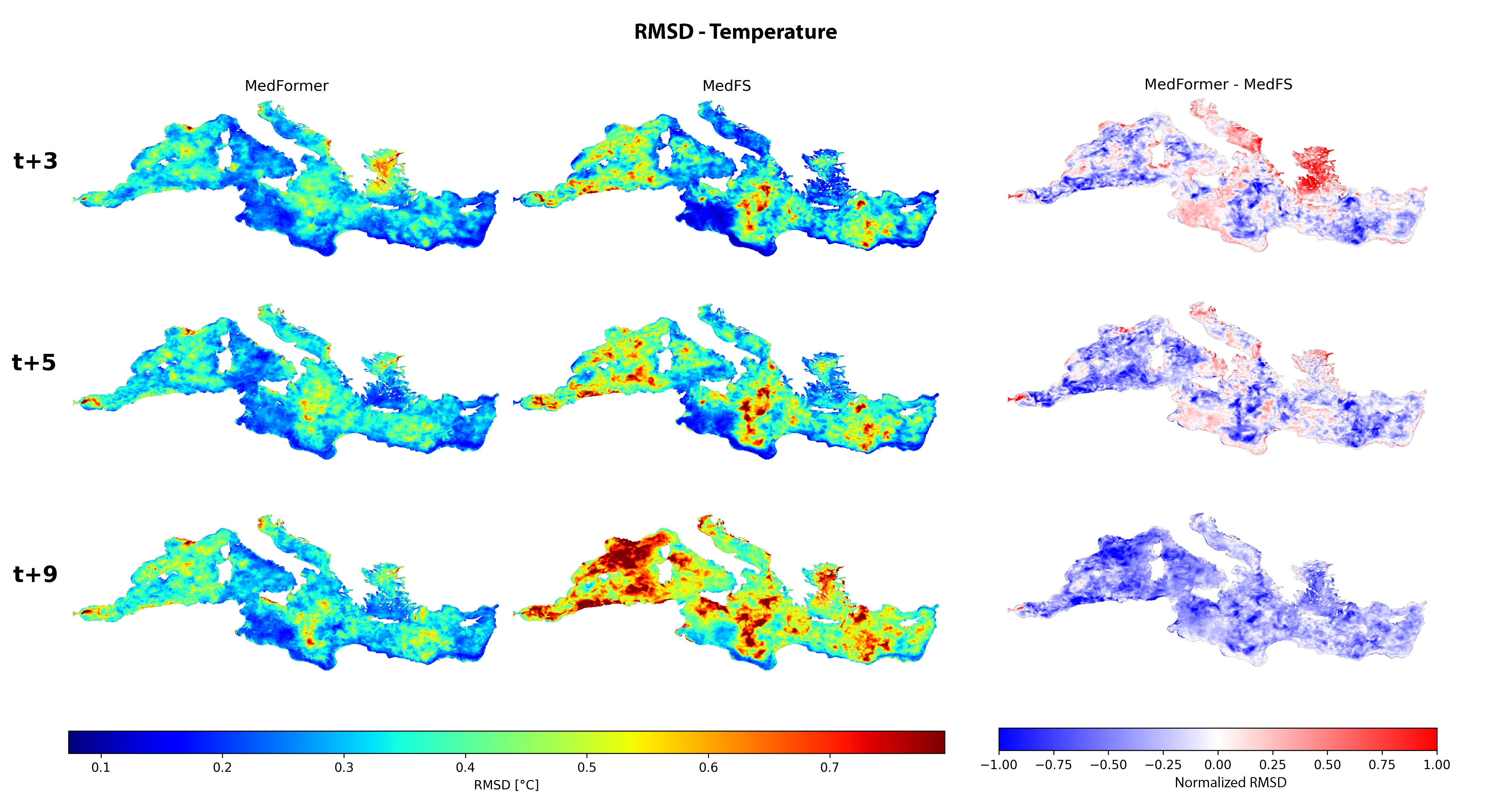}
                
    \caption{RMSD Maps at surface level at forecast lead time 3, 5, and 9 days of temperature against the Analysis for MedFormer (left column) and MedFS (middle column). Difference of RMSD between MedFormer and MedFS in the right column. Bluish color means that the MedFormer error is lower than MedFS.}
    \label{fig:rmsd-errormaps-tem}
\end{figure}

\begin{figure}[ht]
    \centering
        \includegraphics[width=0.9\textwidth]{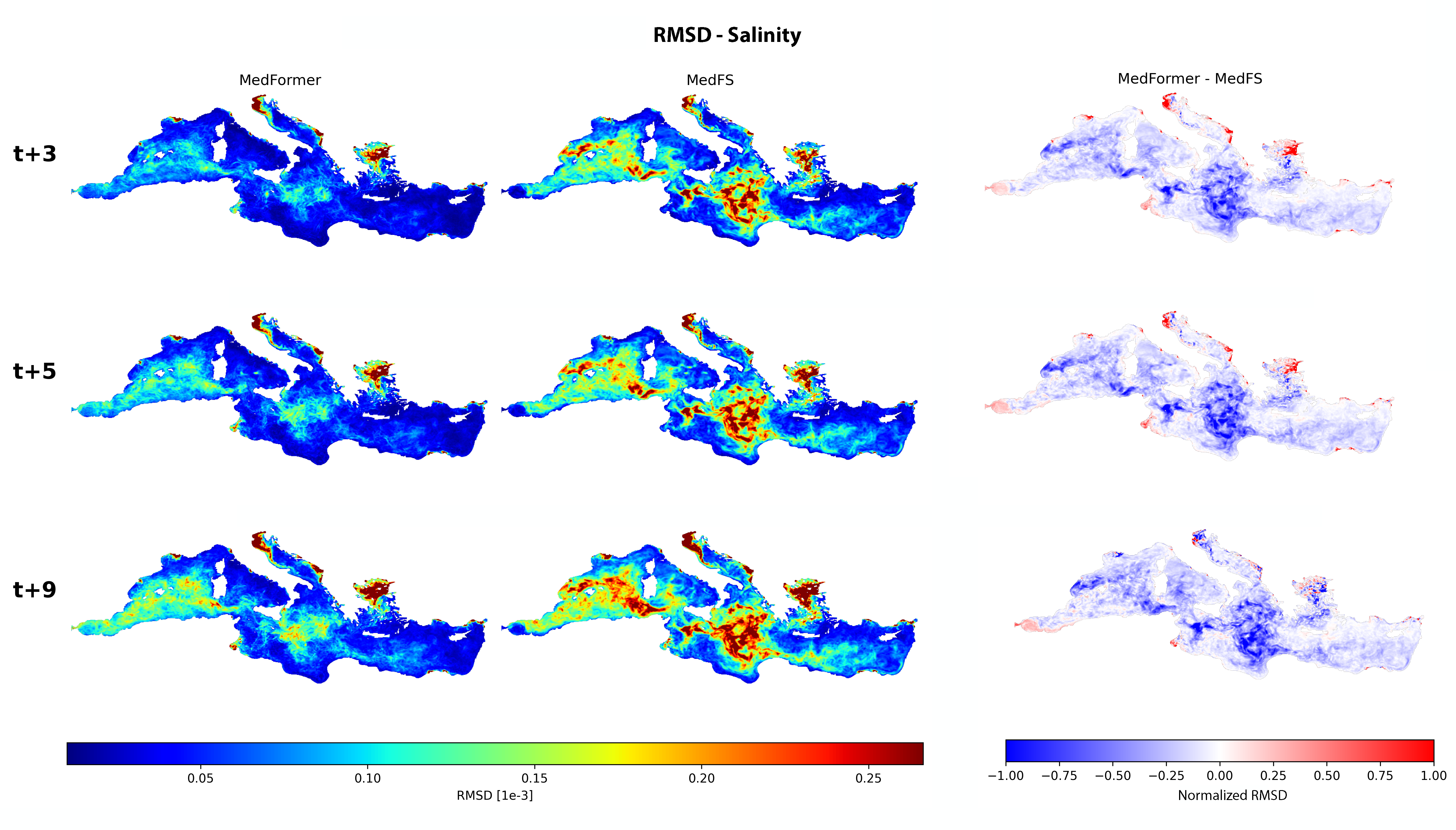}
                
    \caption{RMSD Maps at surface level at forecast lead time 3, 5, and 9 days of salinity against the Analysis for MedFormer (left column) and MedFS (middle column). Difference of RMSD between MedFormer and MedFS in the right column. Bluish color means that the MedFormer error is lower than MedFS.}
    \label{fig:rmsd-errormaps-sal}
\end{figure}

\begin{figure}[ht]
    \centering
        \includegraphics[width=0.9\textwidth]{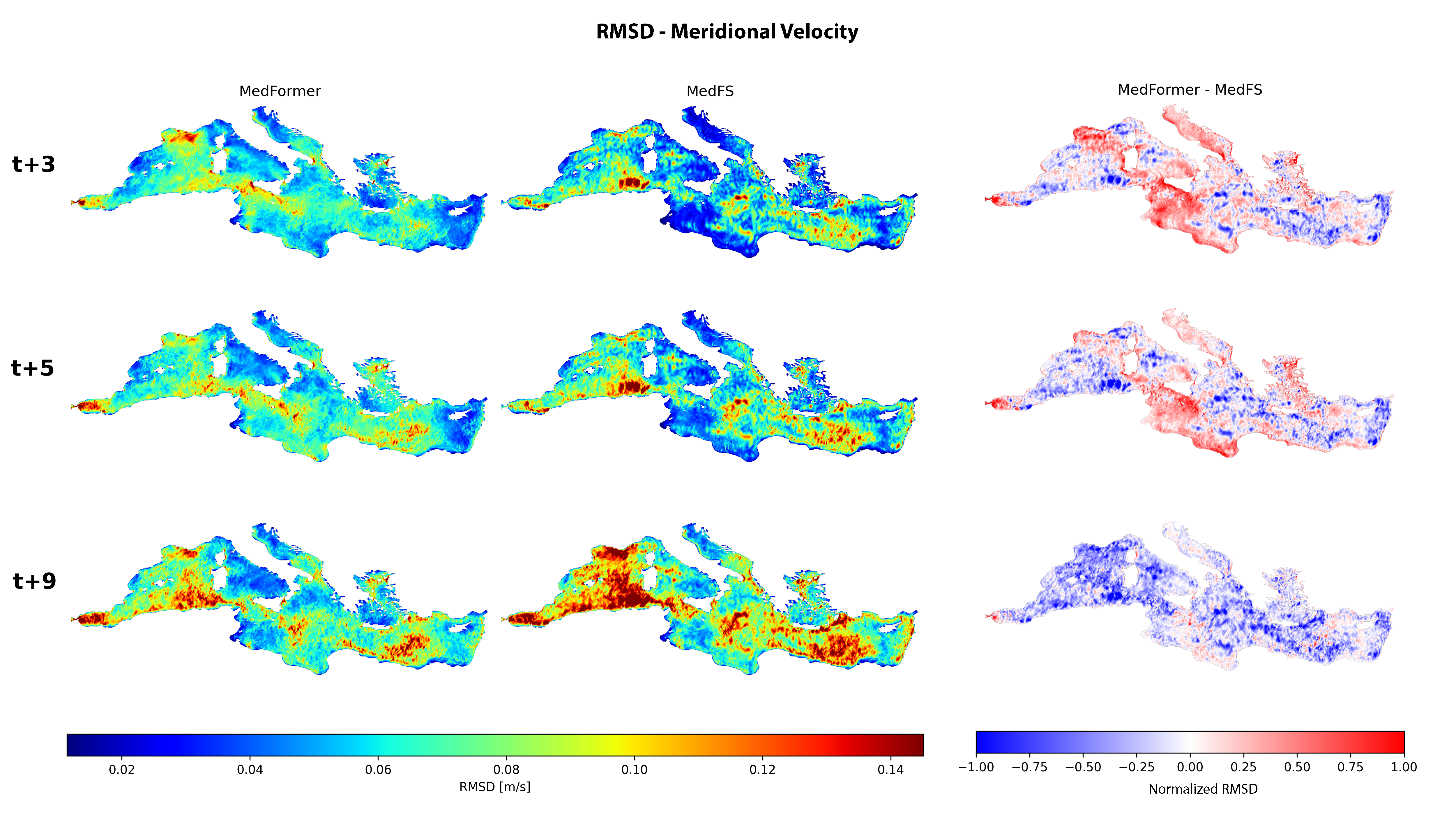}
                
    \caption{RMSD Maps at surface level at forecast lead time 3, 5, and 9 days of meridional velocity against the Analysis for MedFormer (left column) and MedFS (middle column). Difference of RMSD between MedFormer and MedFS in the right column. Bluish color means that the MedFormer error is lower than MedFS.}
    \label{fig:rmsd-errormaps-mer}
\end{figure}

\begin{figure}[ht]
    \centering
        \includegraphics[width=0.9\textwidth]{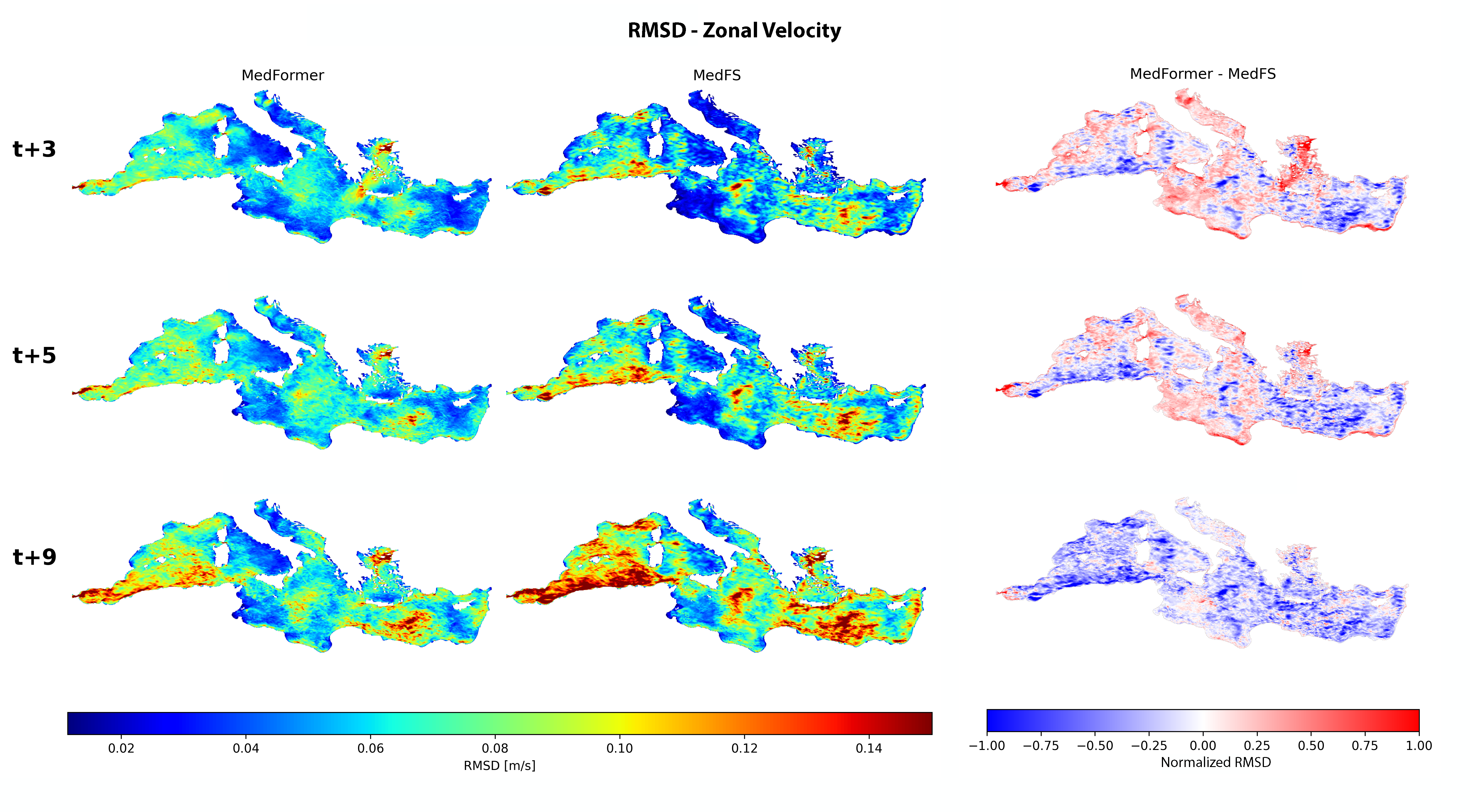}
                
    \caption{RMSD Maps at surface level at forecast lead time 3, 5, and 9 days of zonal velocity against the Analysis for MedFormer (left column) and MedFS (middle column). Difference of RMSD between MedFormer and MedFS in the right column. Bluish color means that the MedFormer error is lower than MedFS.}
    \label{fig:rmsd-errormaps-zon}
\end{figure}

\begin{figure}[ht]
    \centering
        \includegraphics[width=0.9\textwidth]{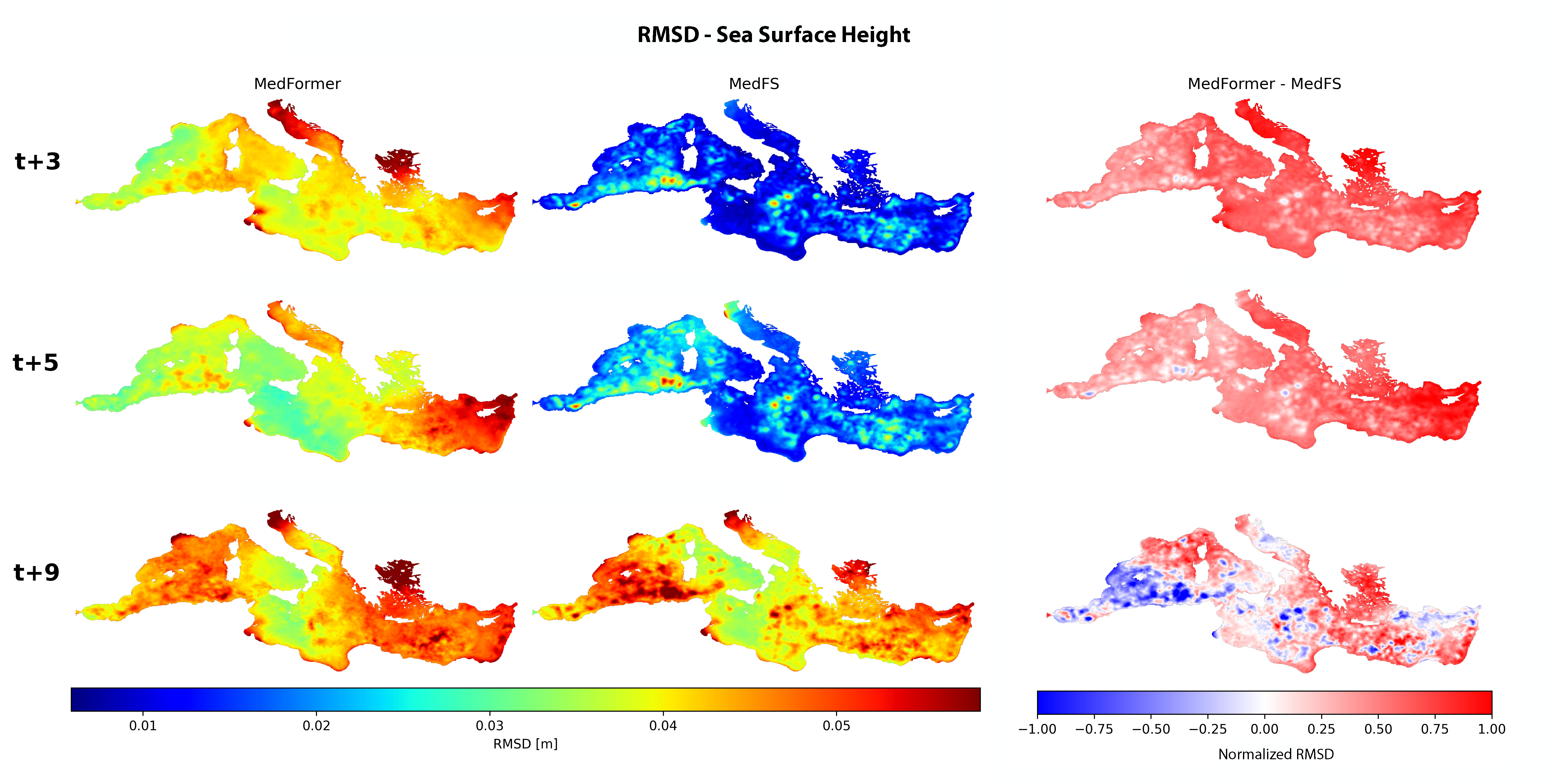}
                
    \caption{RMSD Maps at forecast lead time 3, 5, and 9 days of sea surface height against the Analysis for MedFormer (left column) and MedFS (middle column). Difference of RMSD between MedFormer and MedFS in the right column. Bluish color means that the MedFormer error is lower than MedFS.}
    \label{fig:rmsd-errormaps-ssh}
\end{figure}

The analysis of the RMSD maps provides valuable insight into the spatial distribution of deviations between MedFormer and MedFS against the Mediterranean Analysis. Moreover, the difference of the errors between MedFormer and MedFS highlights the spatial regions mostly responsible for the difference of the forecast skill between the models. Figures~\ref{fig:rmsd-errormaps-tem}, \ref{fig:rmsd-errormaps-sal}, \ref{fig:rmsd-errormaps-mer}, \ref{fig:rmsd-errormaps-zon} and \ref{fig:rmsd-errormaps-ssh} depict the RMSD maps for temperature, salinity, meridional and zonal velocity and for sea surface height at lead times $t+3$, $t+5$, and $t+9$. The fist and second column of each figure show the error maps for MedFormer and for MedFS, while the last column is the difference between MedFormer and MedFS errors measured with a normalized RMSD.

The maps for the temperature (Fig.~\ref{fig:rmsd-errormaps-tem}) highlight an average error of less than $0.8$°C for both models. However, while MedFormer maintains a relatively constant error over the lead time, MedFS shows a steadily increasing error especially in the Ionian Sea and the western Mediterranean Sea

Regarding the salinity (Fig.~\ref{fig:rmsd-errormaps-sal}), MedFormer and MedFS expose a high error in the northern Adriatic Sea (near the Po river) and in the Aegean Sea. But MedFormer is able to forecast the salinity concentration better than MedFS in the Ionian Sea and in the western Mediterranean Sea.   

Both models have a large error when considering the meridional and zonal velocities (Figures~\ref{fig:rmsd-errormaps-mer} and \ref{fig:rmsd-errormaps-zon}) near the coast of Algeria; MedFS has a better skill than MedFormer in the Alboran Sea close to the Gibraltar strait, in the Adriatic Sea, in the Aegean Sea and in the Sicily channel. However, MedFormer is better than MedFS in lead time $t+9$ in almost the whole Mediterranean Sea.

Finally, as already shown by the previous plots, MedFormer is not able to accurately forecast the sea surface height (Fig.~\ref{fig:rmsd-errormaps-ssh}) for the whole Mediterranean basin with errors mostly emphasized in the northern Adriatic and near to the Dardanelles strait.

\subsection{RMSD distribution}
Forecast skills and the comparison between MedFormer and MedFS have been evaluated over a year (2022) used as test dataset. The 9-day forecast has been produced once a week starting from the same initial condition used by MedFS. To provide a better analysis of the distribution of the errors over the testing period, in this Section we analyze the best and the worst forecast at lead time $t+9$ zooming in the maps at the Adriatic Sea, the Ionian Sea and at Western Mediterranean Sea. The best and worst forecasts have been selected as those forecasts with the highest and lowest RMSD for all variables and for all lead times for MedFormer. The results confirm that MedFormer is able to accurately catch the pattern of temperature and salinity for the three regions. Moreover, the maps highlight the presence of artifacts that are small in intensity and almost two orders of magnitude less than the predicted values. Finally, the RMSD does not vary over a wide interval.       

\begin{figure}[ht]
    \centering
        \includegraphics[width=0.45\textwidth]{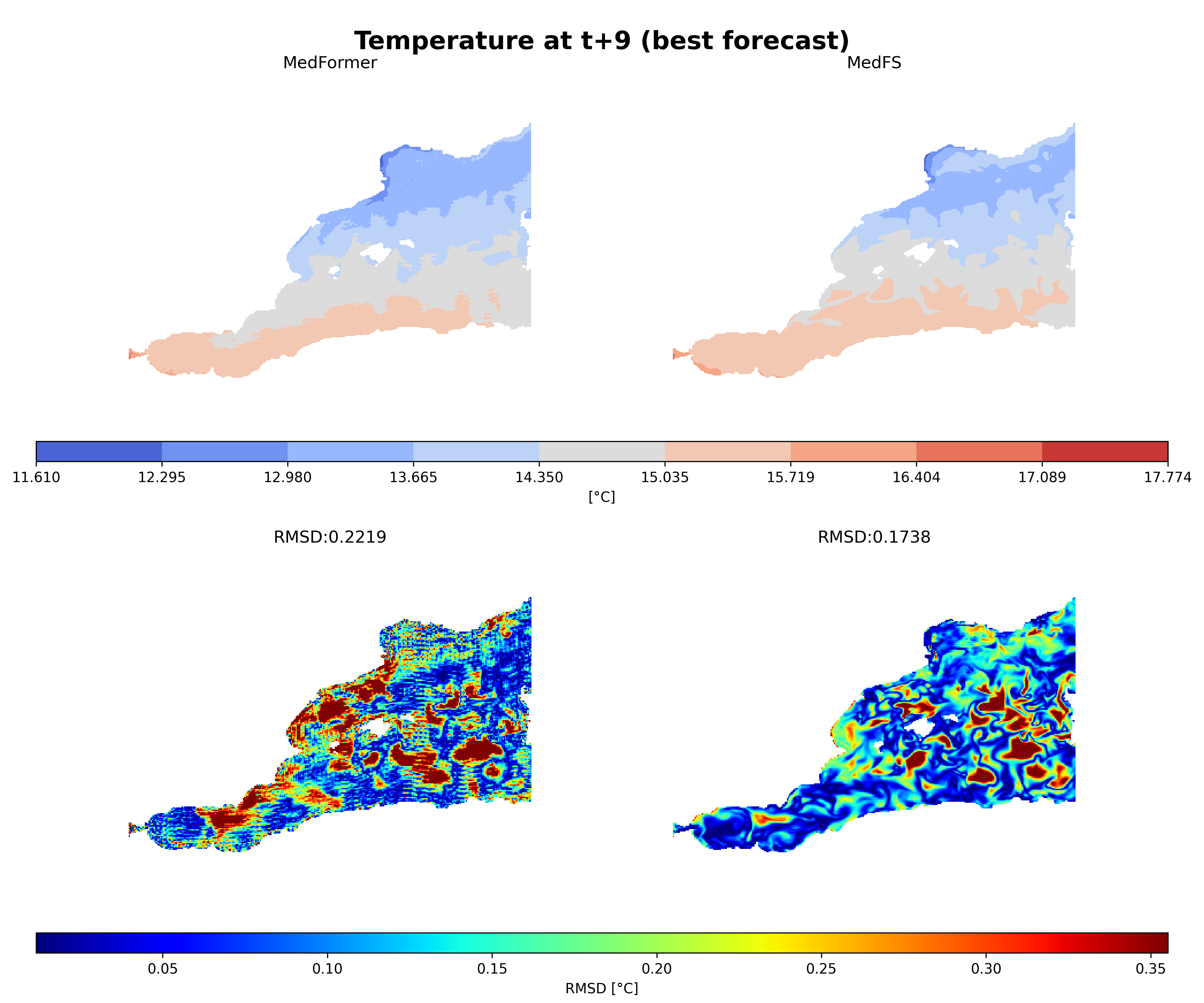}
        \includegraphics[width=0.45\textwidth]{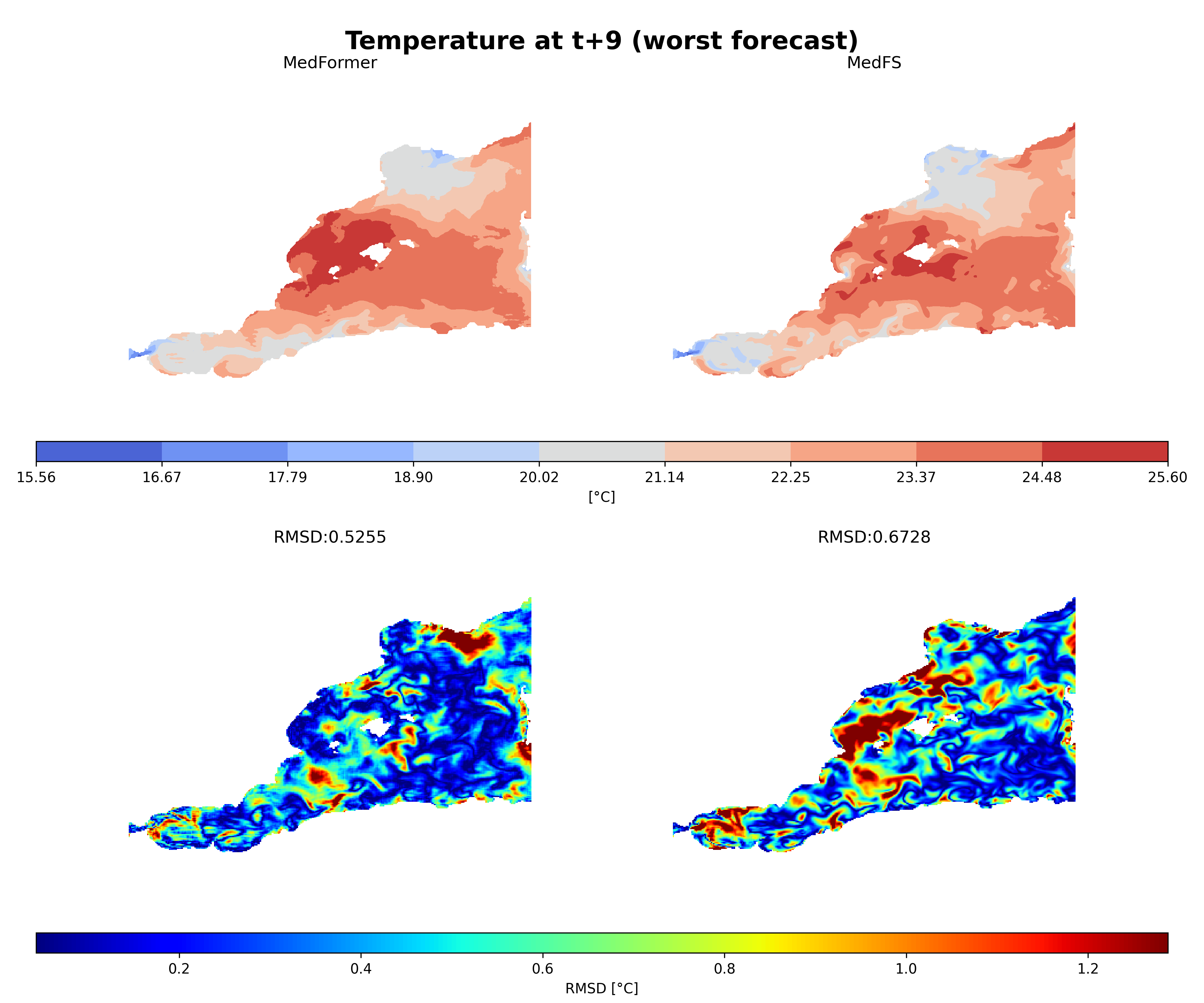}\\

    \caption{Forecast of the temperature in the Western Mediterranean Sea at lead time $t+9$ (first row) in the best (left) and worst (right) case for MedFormer compared with the forecast of MedFS. The second row shows the corresponding RMSD maps against the Analysis.}
    \label{fig:rmsd-bestwort-west-tem}
\end{figure}

\begin{figure}[ht]
    \centering
        \includegraphics[width=0.45\textwidth]{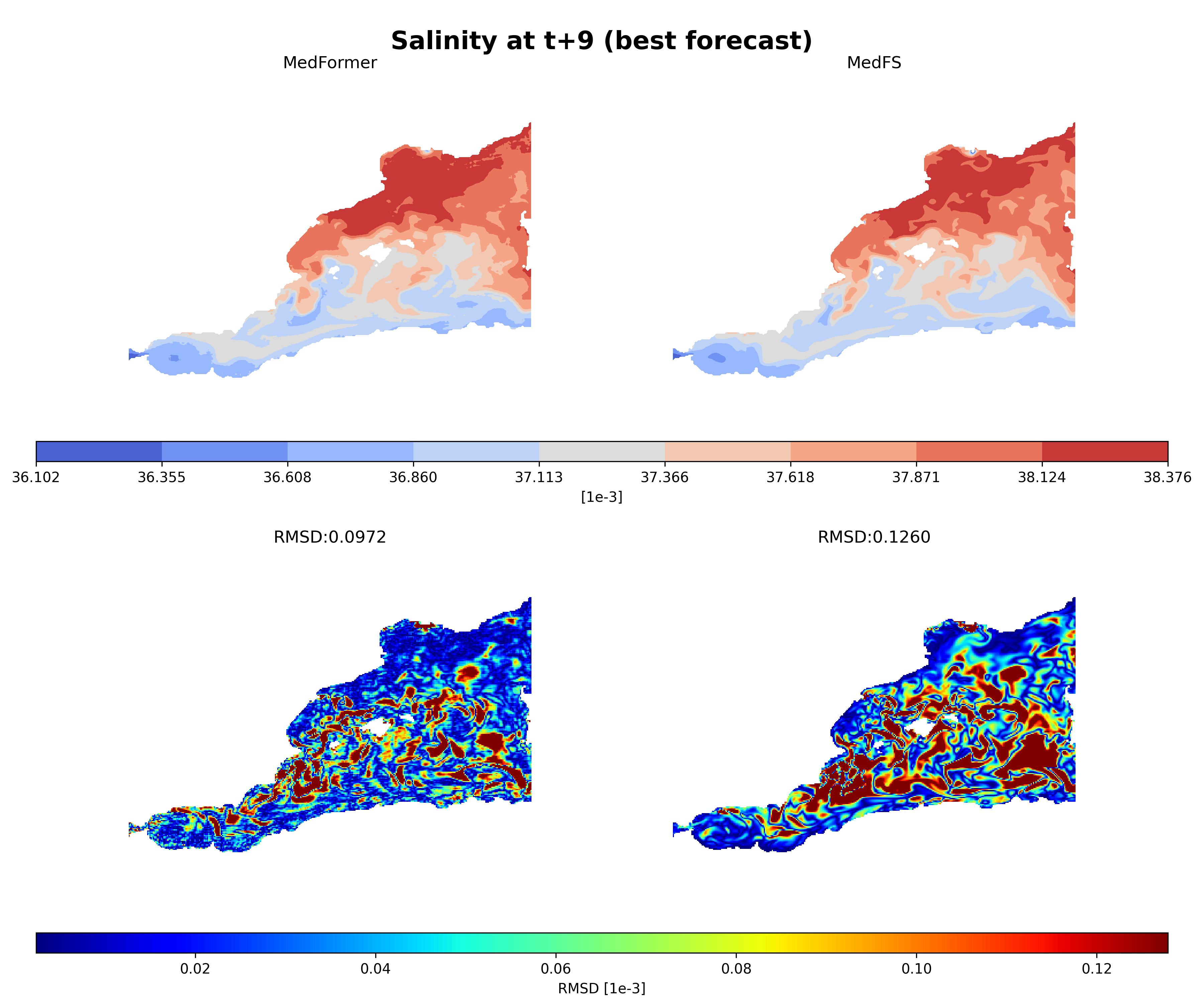}
        \includegraphics[width=0.45\textwidth]{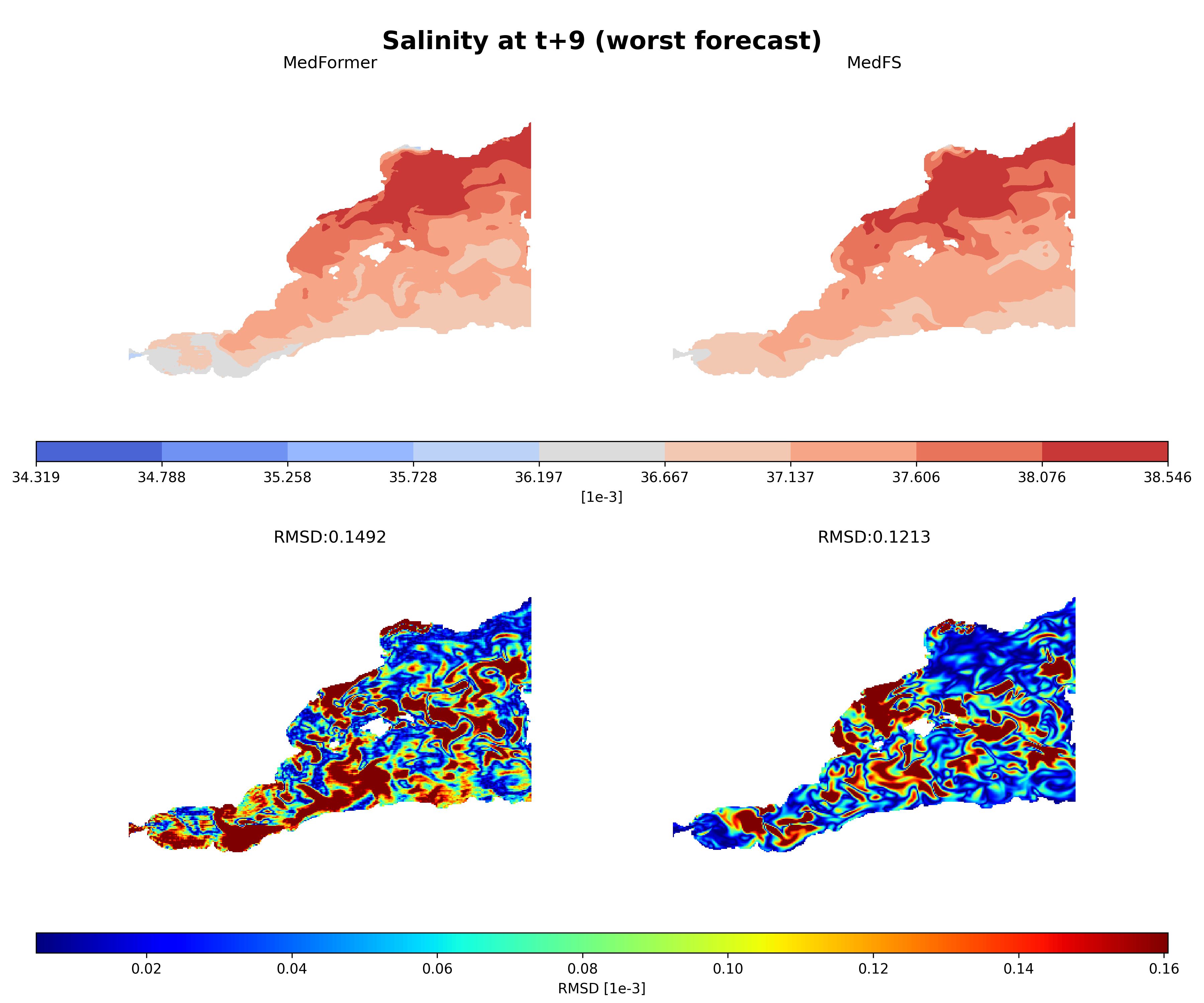}\\

    \caption{Forecast of the salinity in the Western Mediterranean Sea at lead time $t+9$ (first row) in the best (left) and worst (right) case for MedFormer compared with the forecast of MedFS. The second row shows the corresponding RMSD maps against the Analysis.}
    \label{fig:rmsd-bestwort-west-sal}
\end{figure}

\begin{figure}[ht]
    \centering
        \includegraphics[width=0.45\textwidth]{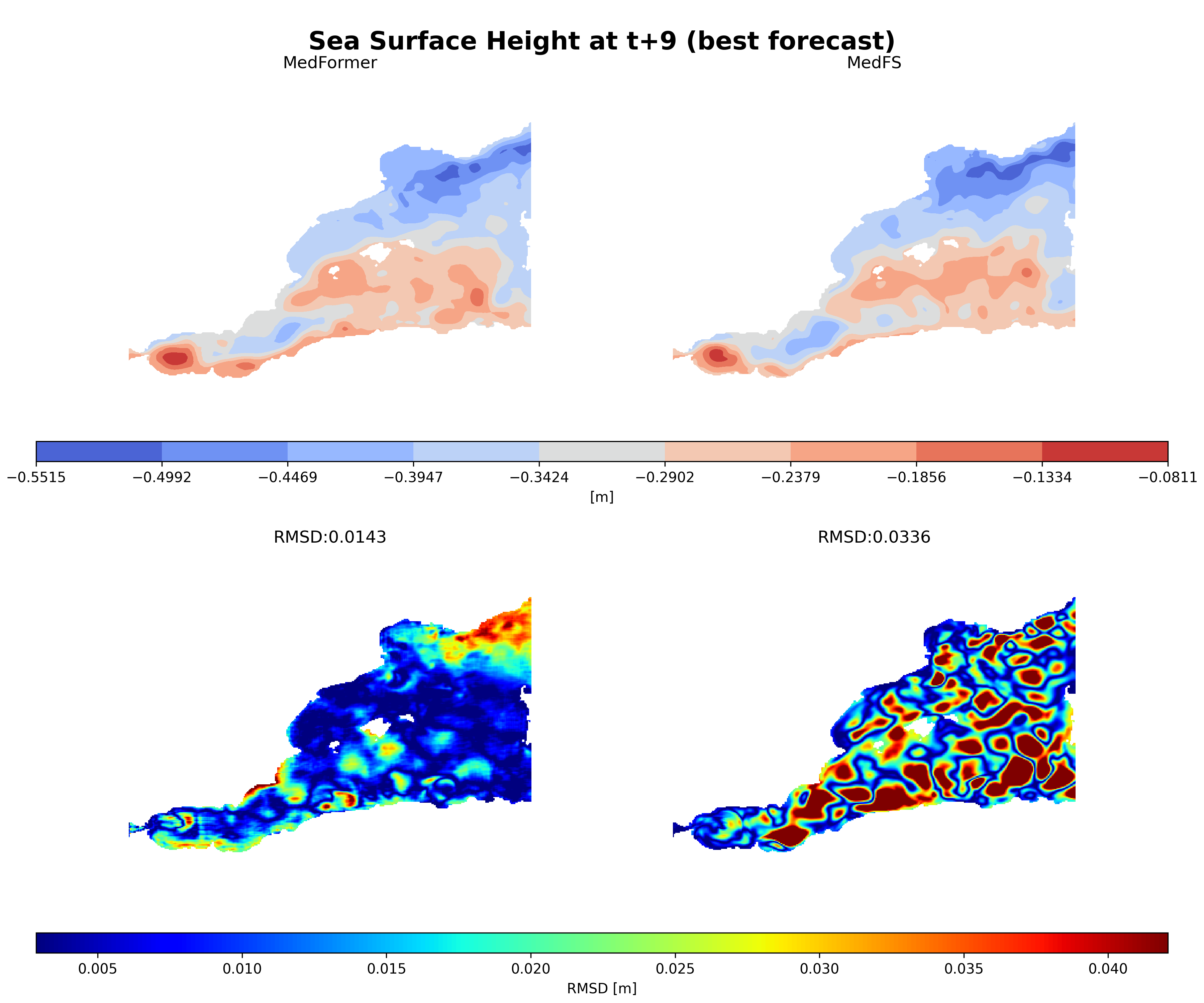}
        \includegraphics[width=0.45\textwidth]{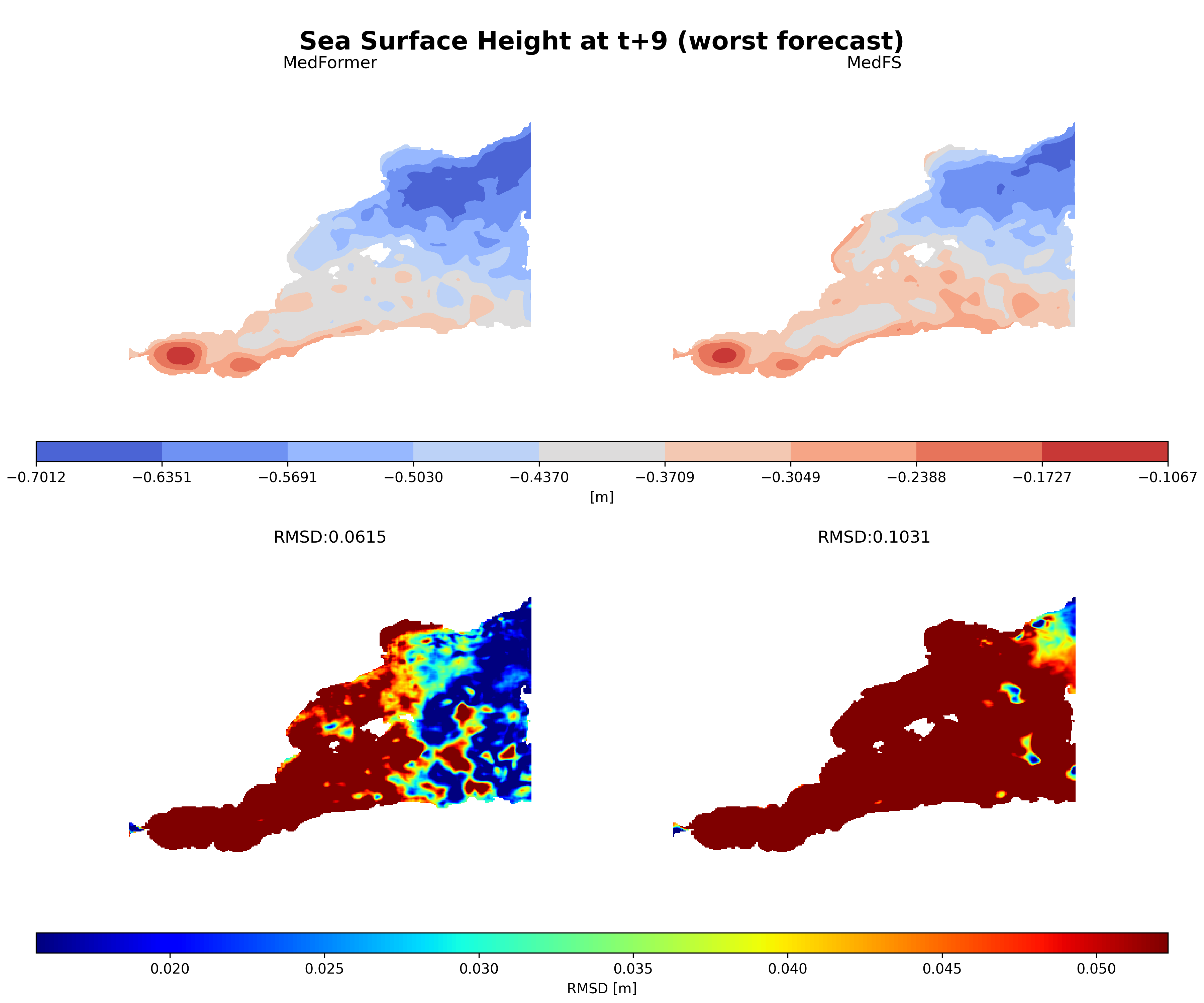}\\

    \caption{Forecast of the sea surface height in the Western Mediterranean Sea at lead time $t+9$ (first row) in the best (left) and worst (right) case for MedFormer compared with the forecast of MedFS. The second row shows the corresponding RMSD maps against the Analysis.}
    \label{fig:rmsd-bestwort-west-ssh}
\end{figure}

\begin{figure}[ht]
    \centering
        \includegraphics[width=0.45\textwidth]{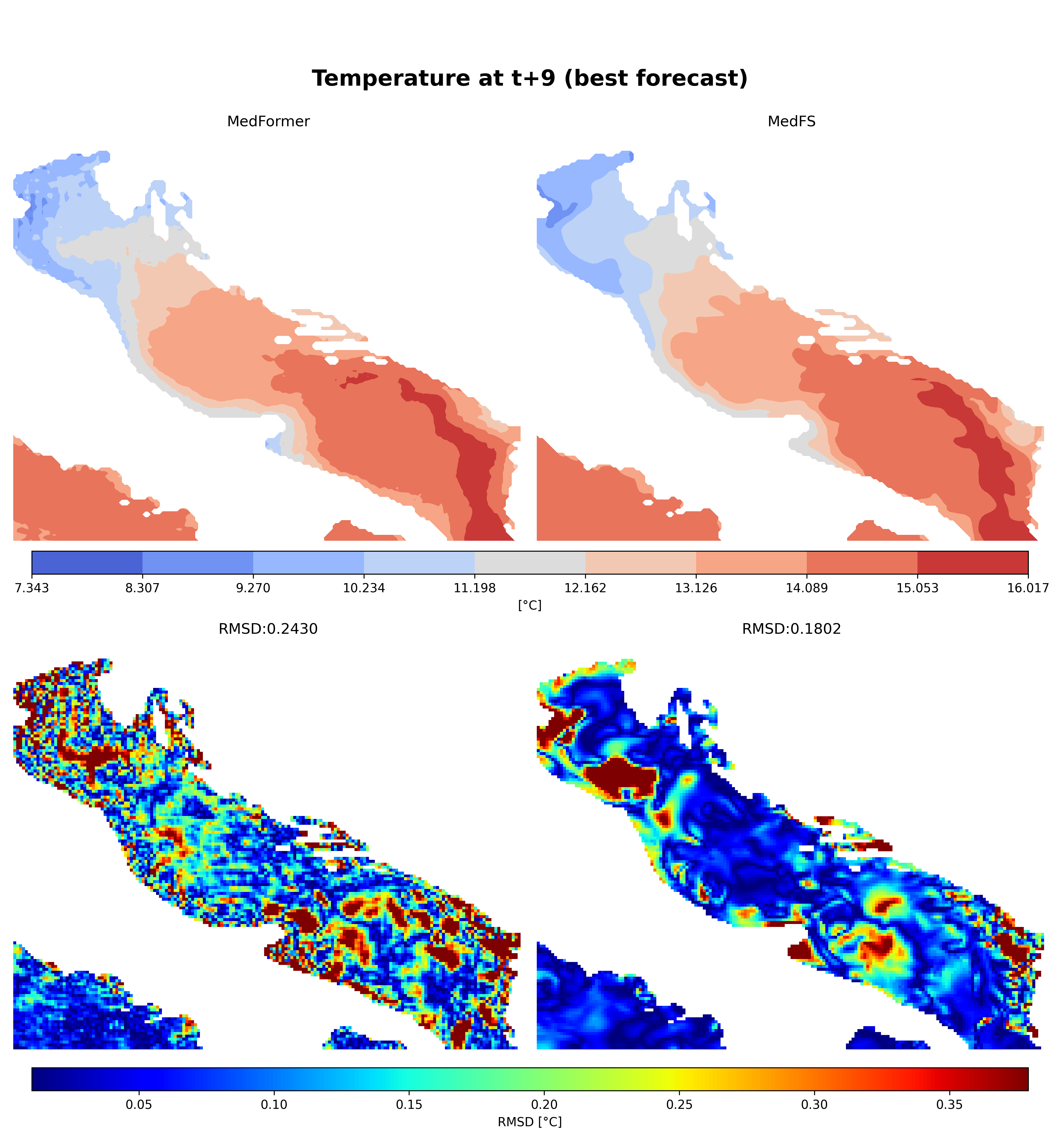}
        \includegraphics[width=0.45\textwidth]{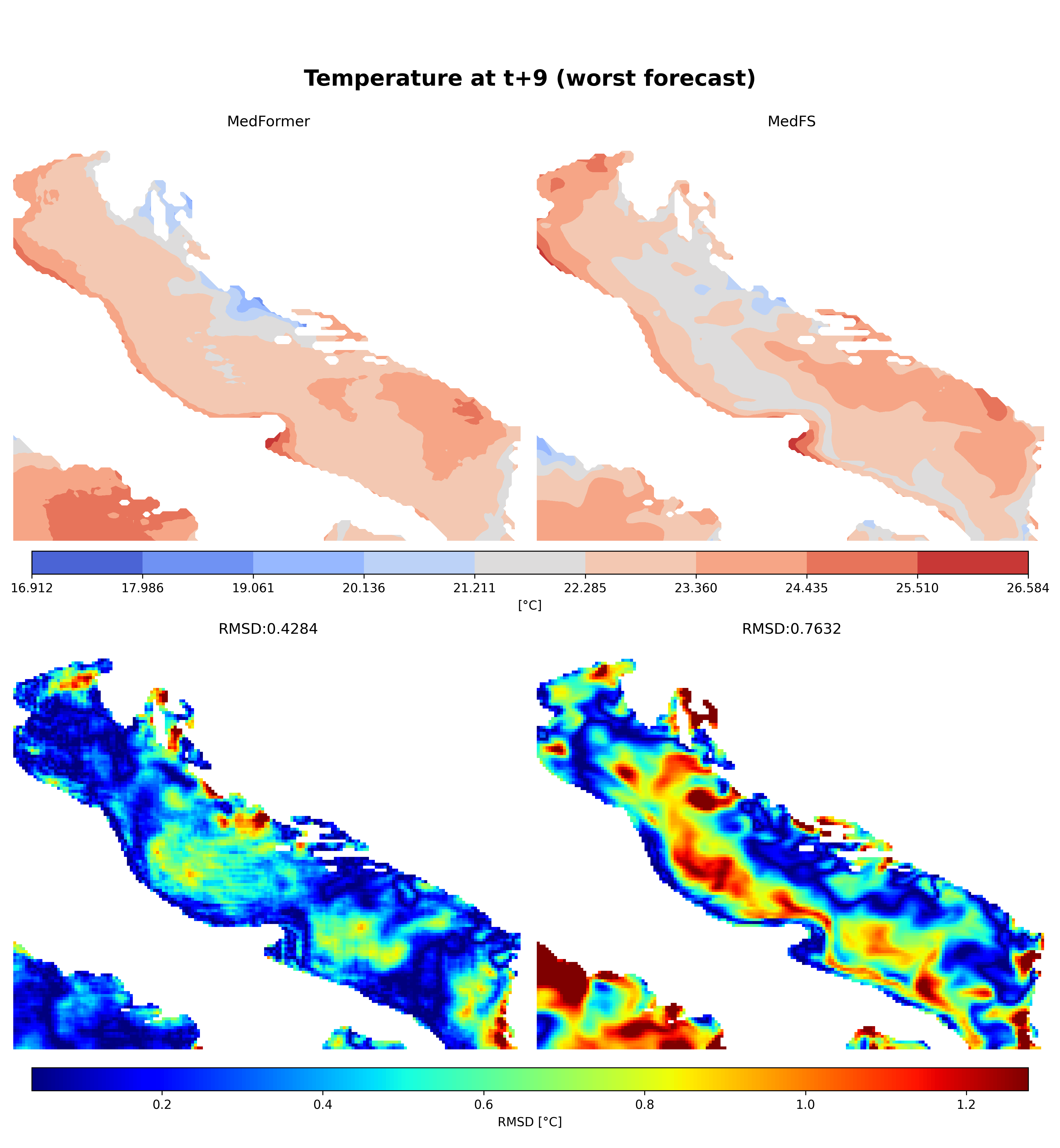}
    \caption{Forecast of the temperature in the Adriatic Sea at lead time $t+9$ (first row) in the best (left) and worst (right) case for MedFormer compared with the forecast of MedFS. The second row shows the corresponding RMSD maps against the Analysis.}
    \label{fig:rmsd-bestworst-adri-tem}
\end{figure}

\begin{figure}[ht]
    \centering
        \includegraphics[width=0.45\textwidth]{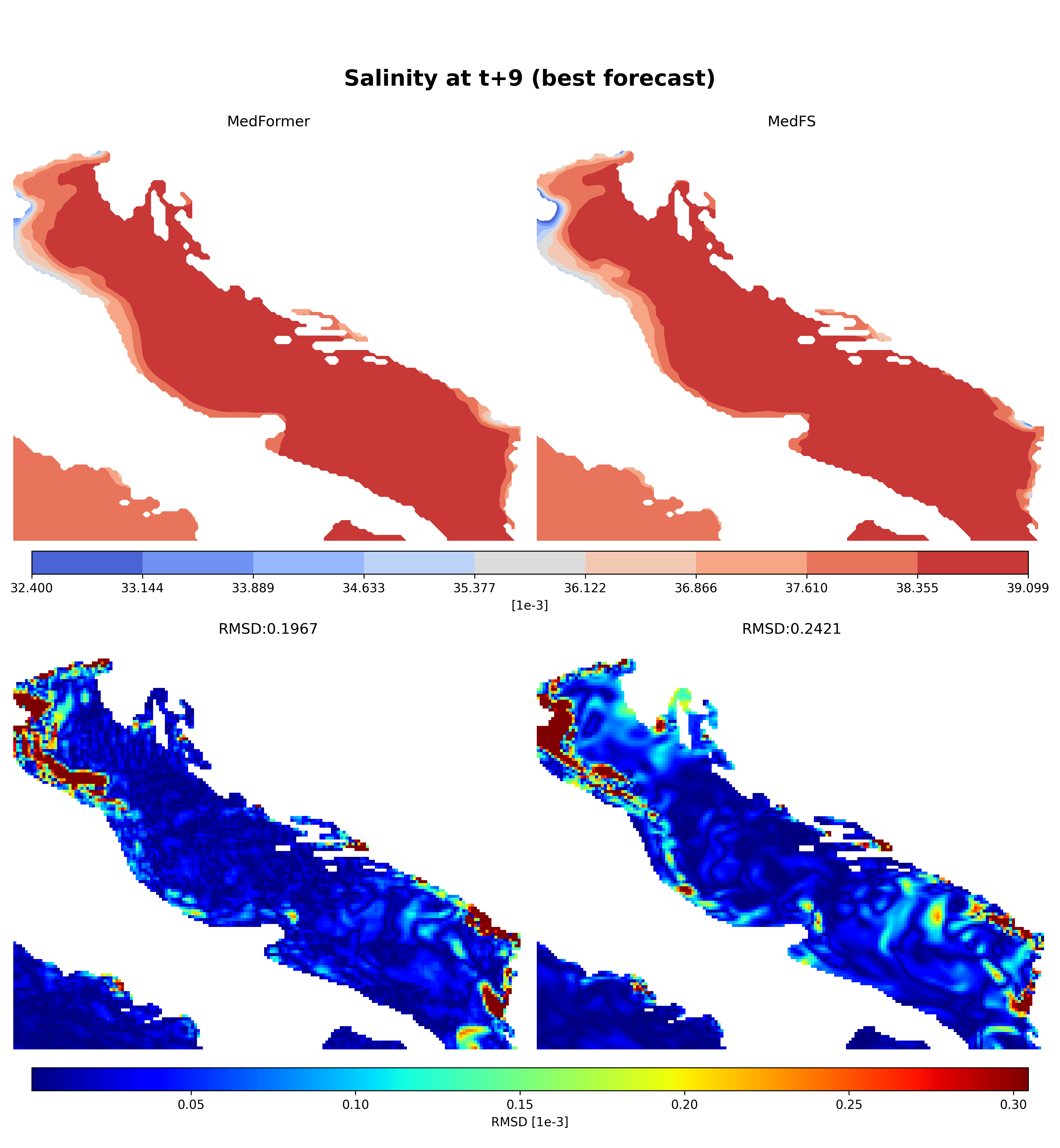}
        \includegraphics[width=0.45\textwidth]{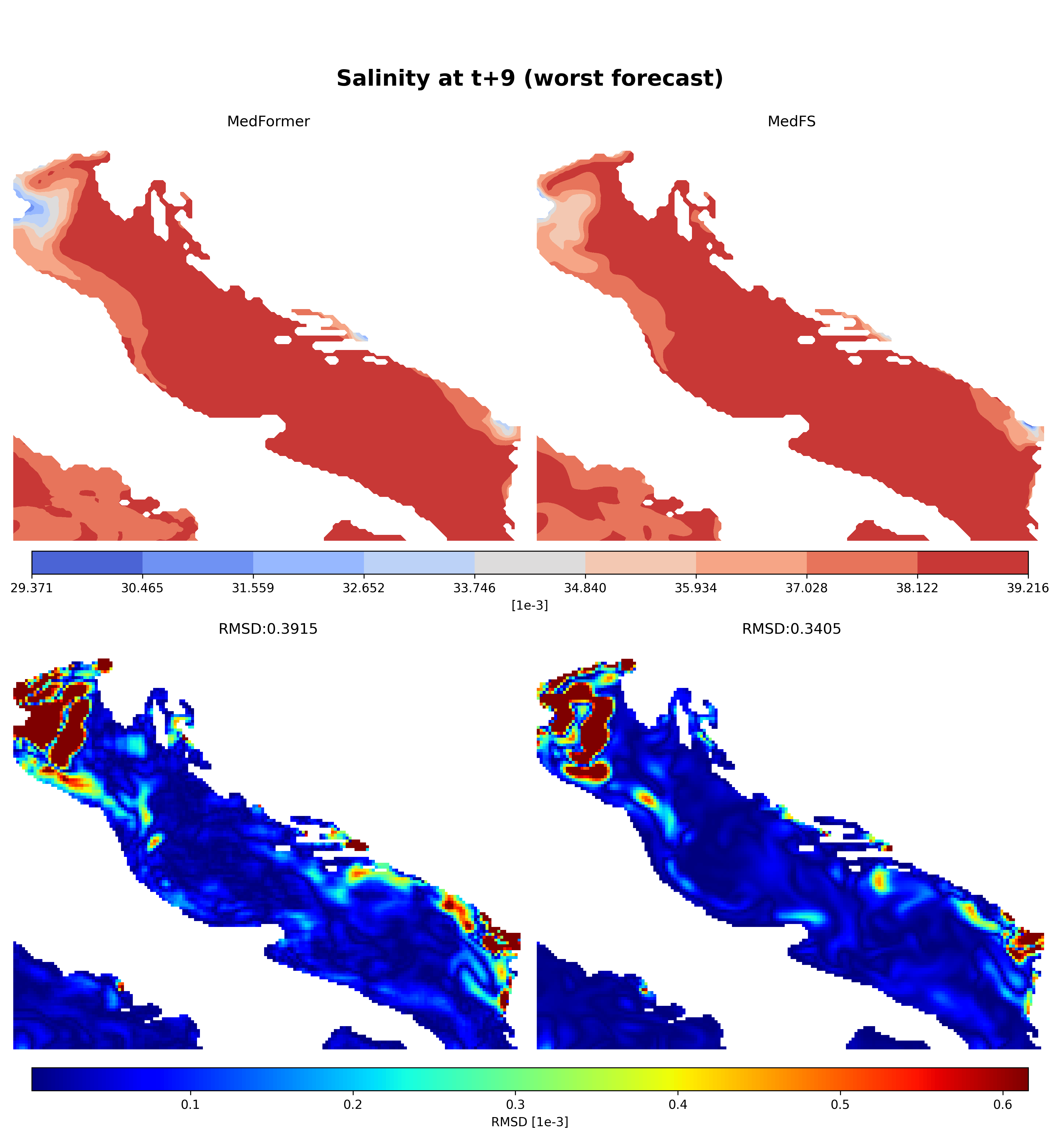}
    \caption{Forecast of the salinity in the Adriatic Sea at lead time $t+9$ (first row) in the best (left) and worst (right) case for MedFormer compared with the forecast of MedFS. The second row shows the corresponding RMSD maps against the Analysis.}
    \label{fig:rmsd-bestworst-adri-sal}
\end{figure}

\begin{figure}[ht]
    \centering
        \includegraphics[width=0.45\textwidth]{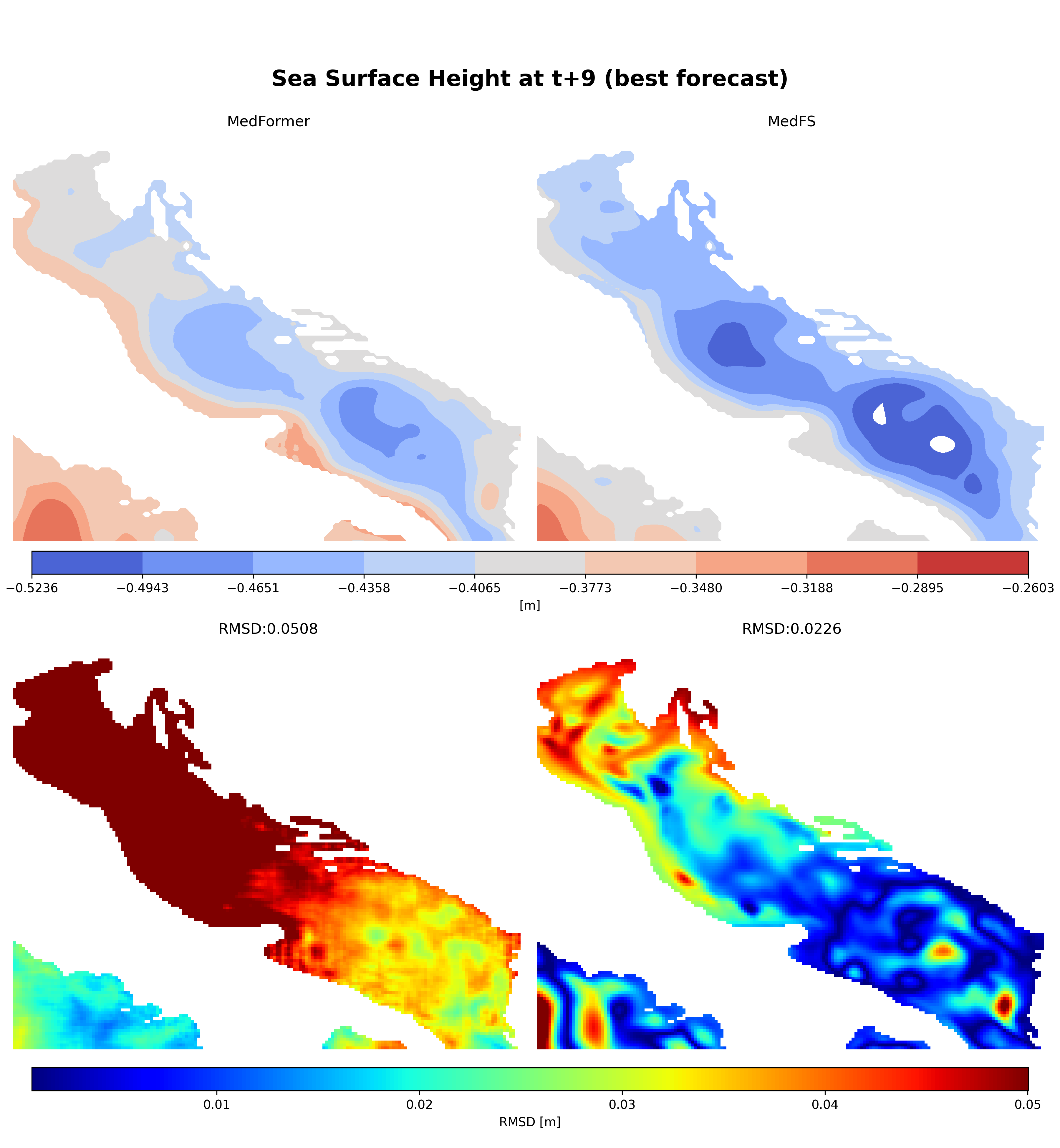}
        \includegraphics[width=0.45\textwidth]{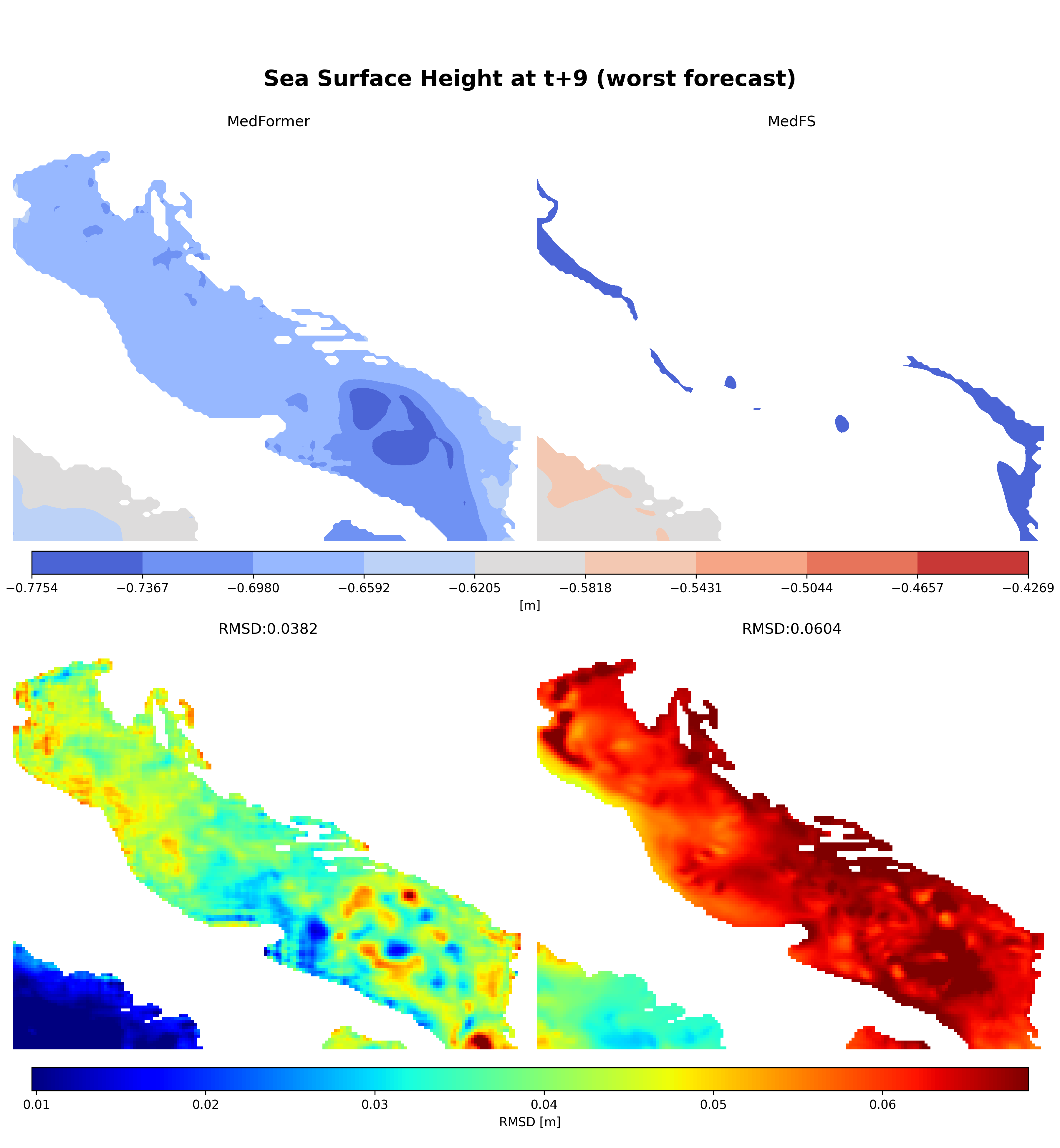}
    \caption{Forecast of the sea surface height in the Adriatic Sea at lead time $t+9$ (first row) in the best (left) and worst (right) case for MedFormer compared with the forecast of MedFS. The second row shows the corresponding RMSD maps against the Analysis.}
    \label{fig:rmsd-bestworst-adri-ssh}
\end{figure}

\begin{figure}[ht]
    \centering
        \includegraphics[width=0.45\textwidth]{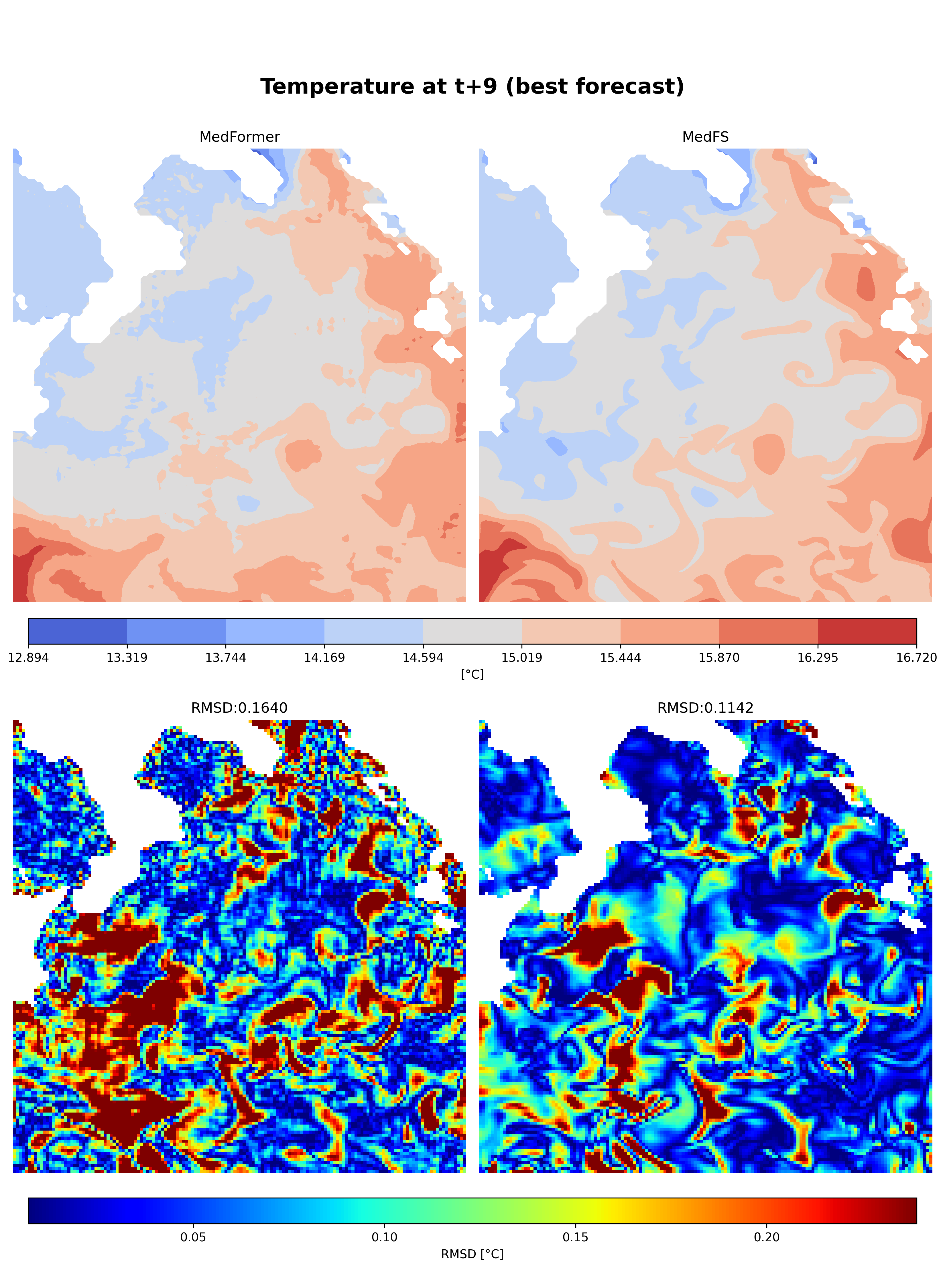}
        \includegraphics[width=0.45\textwidth]{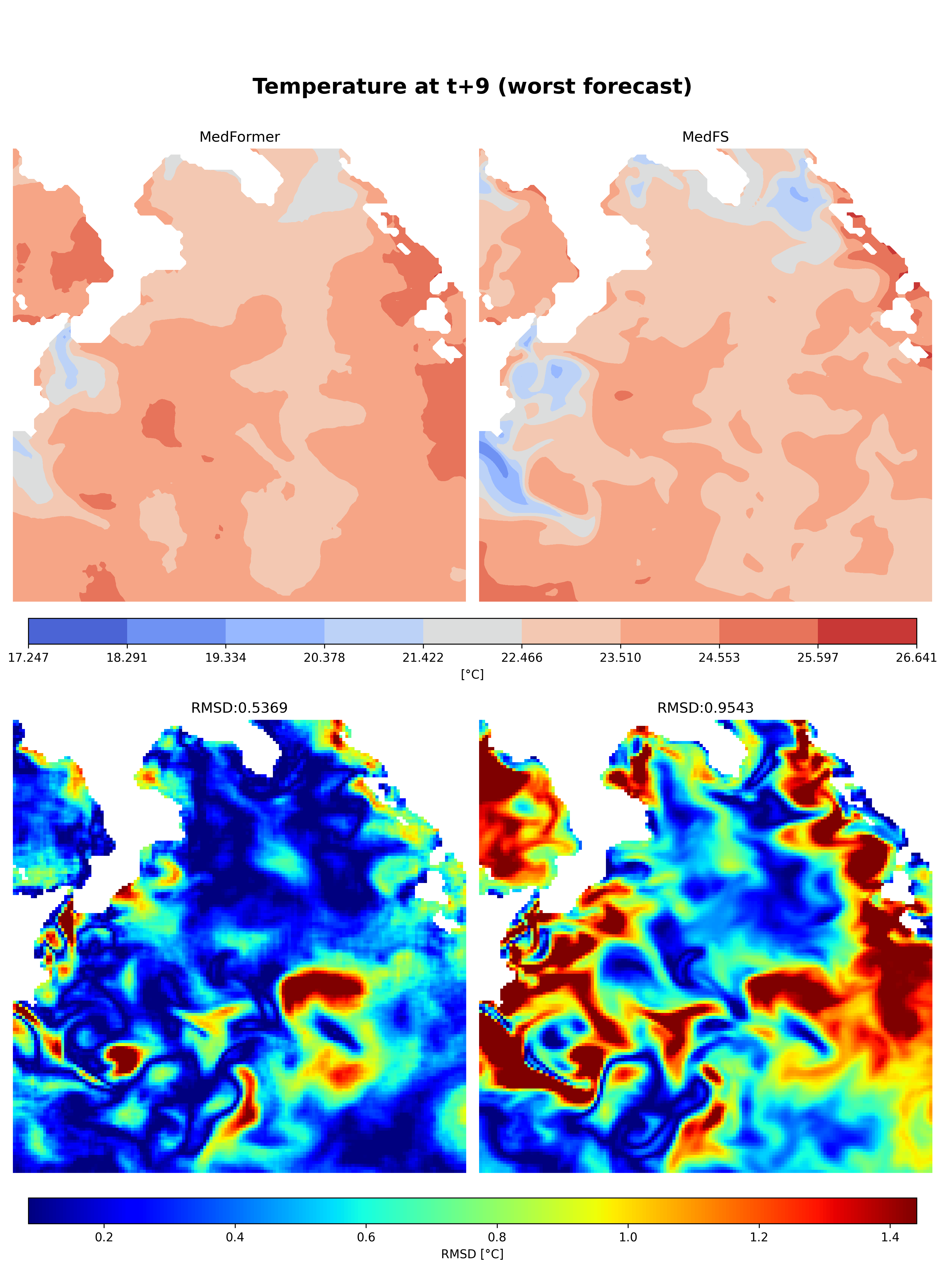}
    \caption{Forecast of the temperature in the Ionian Sea at lead time $t+9$ (first row) in the best (left) and worst (right) case for MedFormer compared with the forecast of MedFS. The second row shows the corresponding RMSD maps against the Analysis.}
    \label{fig:rmsd-bestworst-ioni-tem}
\end{figure}

\begin{figure}[ht]
    \centering
        \includegraphics[width=0.45\textwidth]{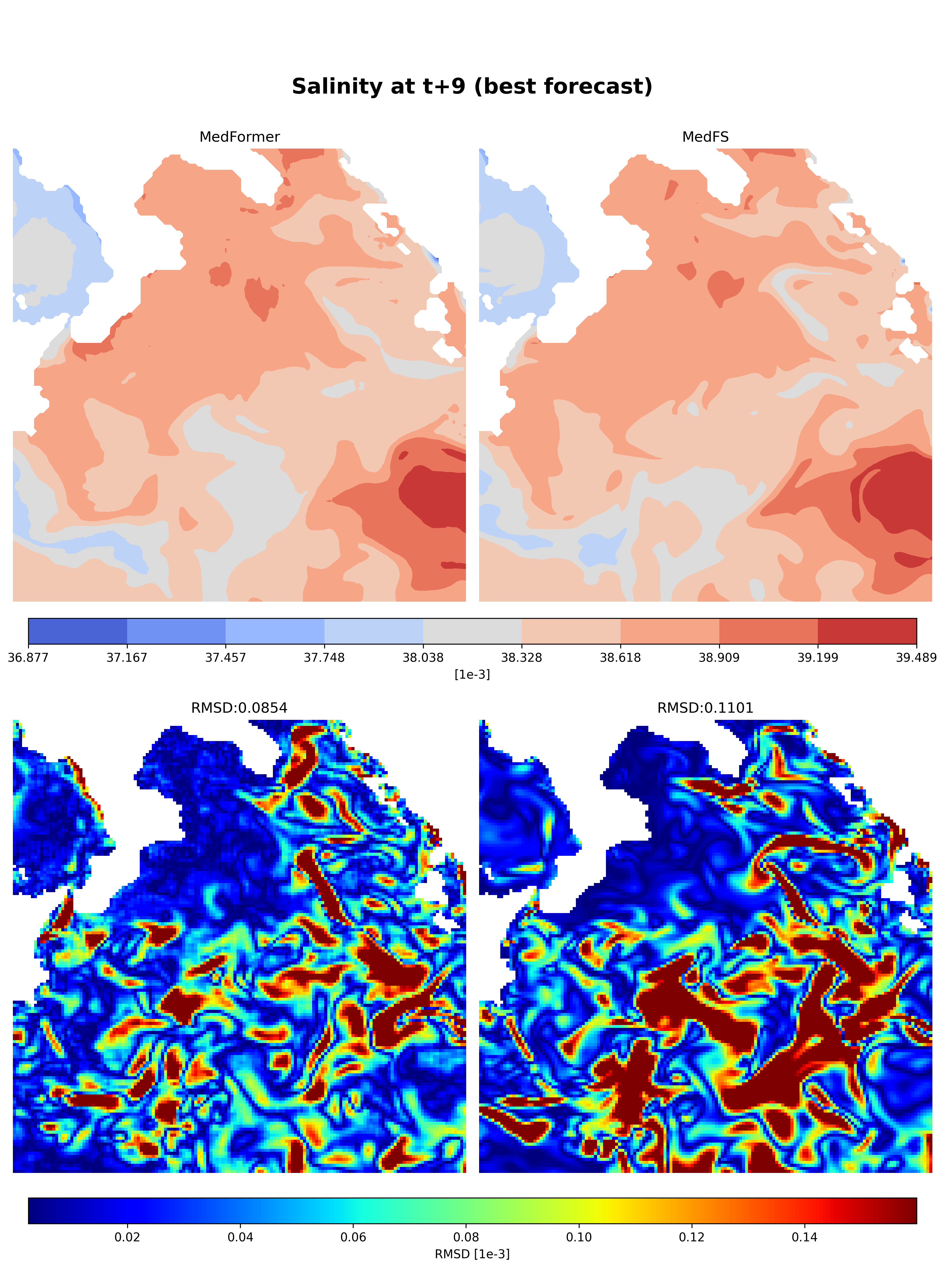}
        \includegraphics[width=0.45\textwidth]{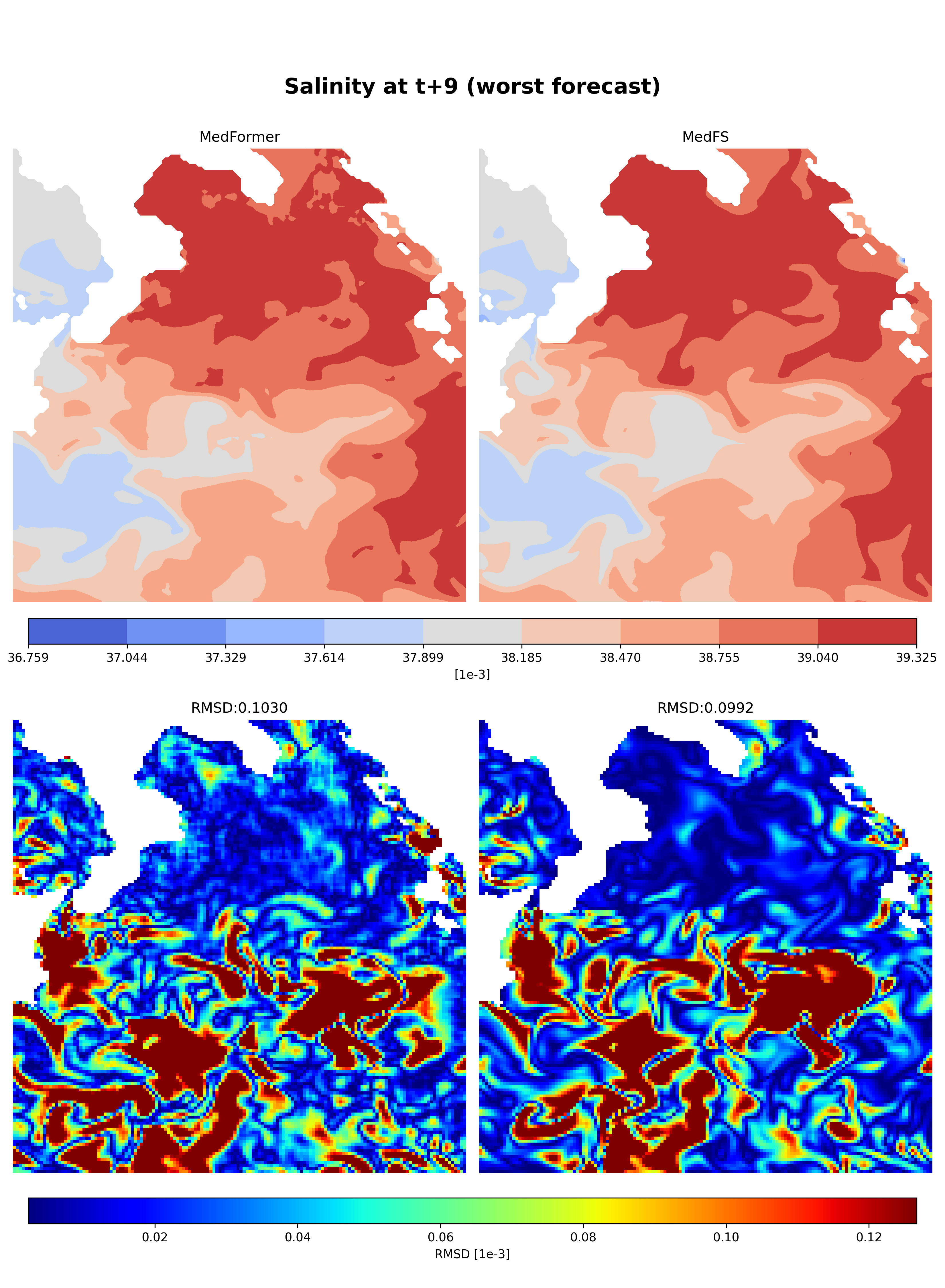}
    \caption{Forecast of the salinity in the Ionian Sea at lead time $t+9$ (first row) in the best (left) and worst (right) case for MedFormer compared with the forecast of MedFS. The second row shows the corresponding RMSD maps against the Analysis.}
    \label{fig:rmsd-bestworst-ioni-sal}
\end{figure}

\begin{figure}[ht]
    \centering
        \includegraphics[width=0.45\textwidth]{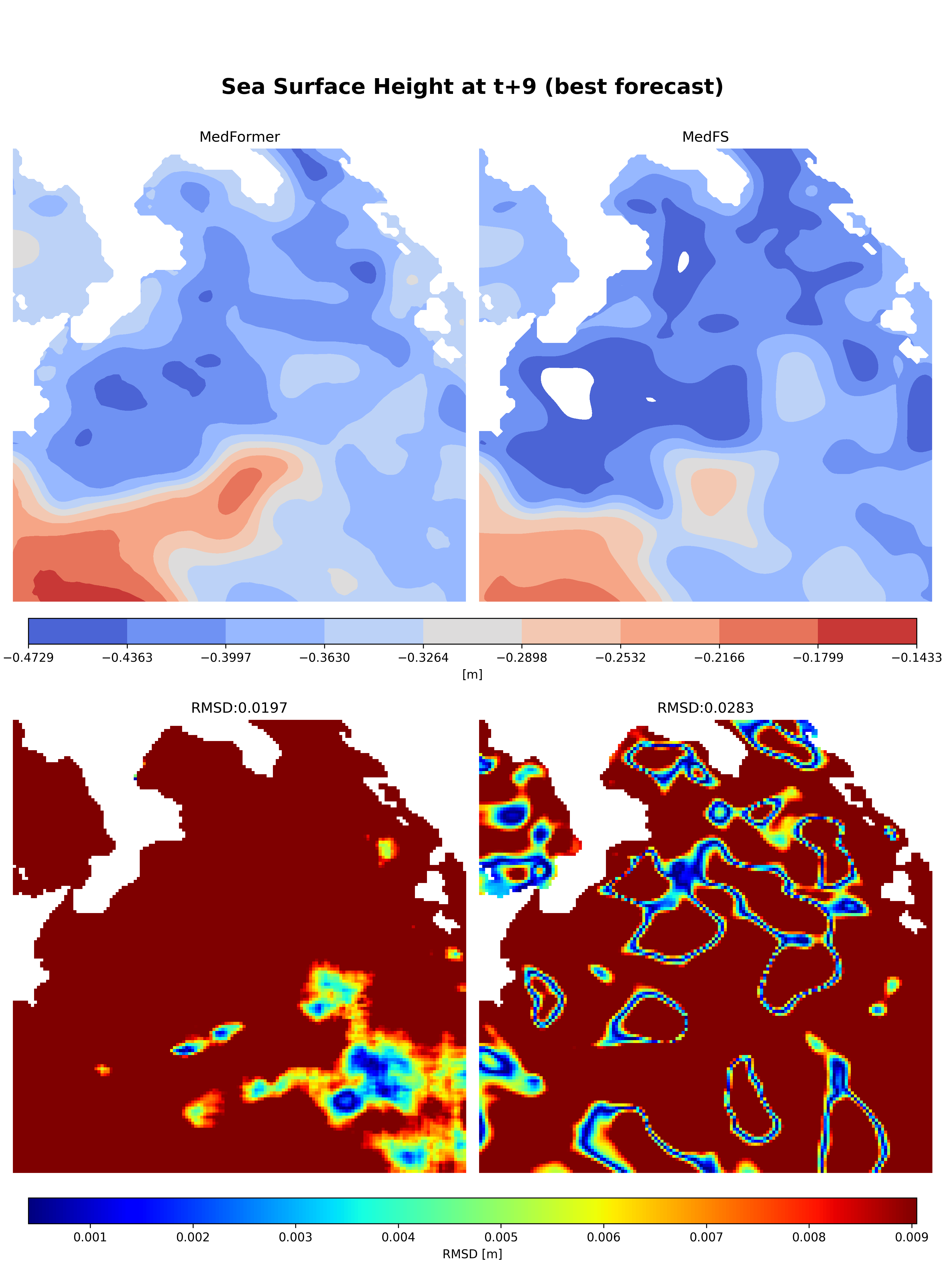}
        \includegraphics[width=0.45\textwidth]{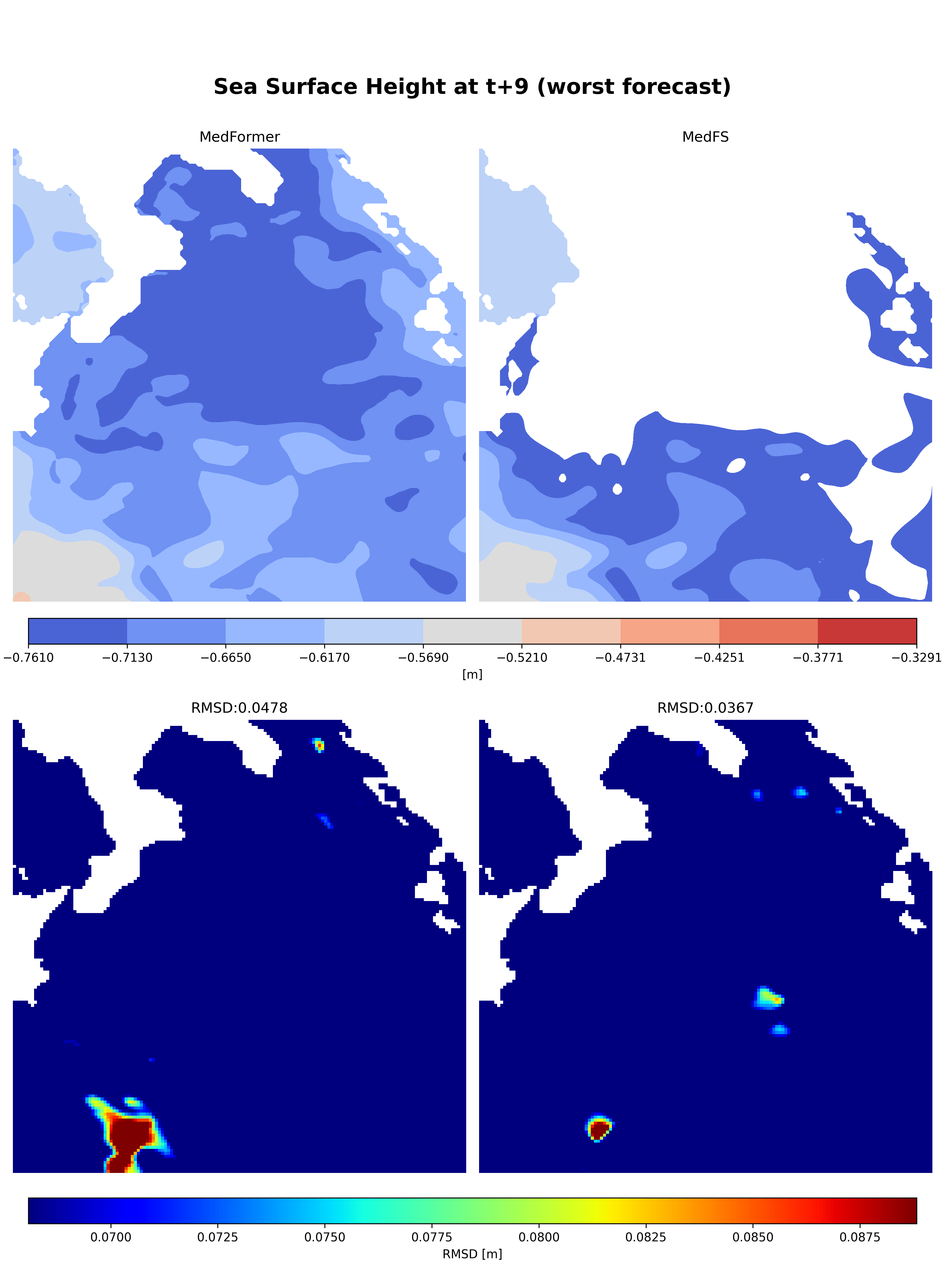}
    \caption{Forecast of the sea surface height in the Ionian Sea at lead time $t+9$ (first row) in the best (left) and worst (right) case for MedFormer compared with the forecast of MedFS. The second row shows the corresponding RMSD maps against the Analysis.}
    \label{fig:rmsd-bestworst-ioni-ssh}
\end{figure}

\subsection{Loss curves}

\begin{figure}[ht]
    \centering
        \includegraphics[width=0.9\textwidth]{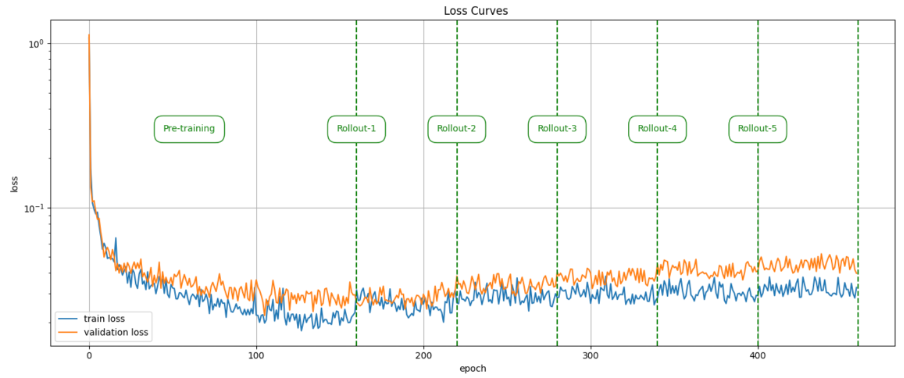}

        \caption{Training and Validation loss curves (respectively blue and orange lines) over the epochs in a logarithmic scale. After 160 epochs the fine tuning starts. Afterwords, a curriculum learning is implemented, i.e. each 60 epochs we increase the number of autoregressive steps.}
    \label{fig:loss_curve_gen}
\end{figure}

The loss function used to train the model computes a weighted average of the error for each variable (see Sec.~\ref{training}).  In this Section we show the behavior of the components of the loss function associated with the prognostic variables. As depicted in Fig.~\ref{fig:loss_curve_gen}, the training loss steadily decreases up to epochs 160 when the pre-trained phase ends. During the fine-tuning phase, we changed both the target dataset and the input dataset by using Analysis data instead of Reanalysis. Indeed, we can notice a small step in the training curve (blue line in Fig.~\ref{fig:loss_curve_gen}) at epoch 160. In the next 60 epochs the training starts accumulating the forecasting error because the loss function is evaluated in autoregressive way, and again a negligible step in the loss curve occurs. After 5 autoregressive steps, we noticed that the distance between the validation loss and the training loss starts to be valuable, and we stop the training for not overfitting the model.   

\begin{figure}[ht]
    \centering
        \includegraphics[width=0.9\textwidth]{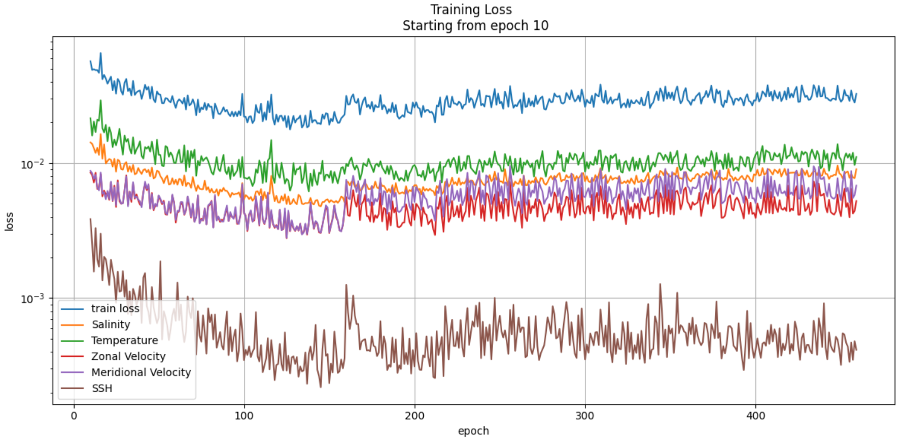}

        \caption{Training loss curves over the epochs in logarithmic scale. In blue the loss function which is obtained as sum of the other colored components. With different colors the components of the loss function due to the variables: SSH (brown); meridional velocity (purple); zonal velocity (red); salinity (orange); temperature (green). The loss curves associated to the variables include also the weight associated to the corresponding variable.}
    \label{fig:loss_curve_var}
\end{figure}

Fig.~\ref{fig:loss_curve_var} shows the contribution of each prognostic variable to the loss function. The contribution of the error made to estimate the SSH is an order of magnitude less than the other contributions. Nevertheless, the brown curve shows a decreasing trend, indicating that the optimization steps are improving all five variables at approximately the same rate. The error in forecasting temperature is the most significant component of the loss function.   

\end{appendices}

\clearpage

\bibliography{bibliography}

\end{document}